\documentclass[graybox]{svmult}

\usepackage{mathptmx}       
\usepackage{helvet}         
\usepackage{courier}        
\usepackage{type1cm}        
\usepackage{makeidx}         
\usepackage{graphicx}        
\usepackage{multicol}        
\usepackage[bottom]{footmisc}

\usepackage{arydshln}

\makeindex             

\begin{document}

\title*{Optimizations of protein force fields}
\author{Yoshitake Sakae and Yuko Okamoto}
\institute{Yoshitake Sakae \at Department of Theoretical and Computational Molecular Science, 
Institute for Molecular Science, Okazaki, Aichi 444-8585, Japan 
\at Department of Physics, Graduate School of Science, 
Nagoya University, Nagoya, Aichi 464-8602, Japan \\ 
\email{sakae@tb.phys.nagoya-u.ac.jp}
\and Yuko Okamoto 
\at Department of Physics, Graduate School of Science, 
Nagoya University, Nagoya, Aichi 464-8602, Japan
\at Structural Biology Research Center, Graduate School of
Science, Nagoya University, Nagoya, Aichi 464-8602, Japan
\at Center for Computational Science, Graduate School of
Engineering, Nagoya University, Nagoya, Aichi 464-8603, Japan
\at Information Technology Center, Nagoya University, 
Nagoya, Aichi 464-8601, Japan \\
\email{okamoto@phys.nagoya-u.ac.jp}}
%
%
\maketitle

\noindent
To be published in {\it Computational Methods to Study the Structure
and Dynamics of}
\noindent
{\it Biomolecules and Biomolecular Processes -- 
from Bioinformatics to Molecular}
\noindent
{\it Quantum Mechanics},  
edited by Adam Liwo, (Springer, Berlin, 2012).

\vspace{1.0cm}


\abstract{ \\
In this Chapter we review our works on force fields for molecular
simulations of protein systems.  We first discuss the functional forms of
the force fields and present some extensions of the conventional ones.
We then present various methods for force-field parameter optimizations.
Finally, some examples of our applications of these parameter optimization
methods are given and they are compared with the results from
the existing force-fields.}


\section{Introduction}
Computer simulations of protein folding into native structures
can be achieved when both of the following two requirements are met:
(1) potential energy functions (or, force fields) for the protein
systems are sufficiently accurate
and (2) sufficiently powerful conformational sampling methods are
available.
Professor Harold A. Scheraga has been one of the most important
pioneers in studies of both of the above 
requirements \cite{ScheragaRev1,Scheragarev2}.
By the developments of the generalized-ensemble algorithms
(for reviews, see, e.g., Refs.~\cite{HOrev,GEA,GEA2,GEA3}) and related
methods, Requirement (2) seems to be almost fulfilled.
In this Chapter, we therefore concentrate our attention on 
Requirement (1). 

There are several well-known all-atom (or united-atom) force fields, 
such as AMBER \cite{parm94, parm96_3, parm99, parm99SB, parm03}, 
CHARMM \cite{charmm,CMAP,CMAP2}, OPLS \cite{opls1,opls2}, 
GROMOS \cite{gromos_v2,gromos_v3}, 
GROMACS \cite{gromacs1,gromacs2}, 
and ECEPP \cite{ECEPP,ECEPP05}.
Generally, the force-field parameters are determined based on 
experimental results for small molecules 
and theoretical results using quantum chemistry calculations of 
small peptides such as alanine dipeptide.

However, the simulations using different force-field parameters will 
give different results.
We have performed detailed comparisons of three version of AMBER  
(ff94 \cite{parm94}, ff96 \cite{parm96_3},
and ff99 \cite{parm99}), CHARMM \cite{charmm},
OPLS-AA/L \cite{opls2}, and GROMOS \cite{gromos_v2}
by generalized-ensemble simulations of two small peptides 
in explicit solvent. \cite{YSO1,YSO2}
We saw that these force fields 
showed clearly different behaviors 
especially with respect to secondary-structure-forming tendencies. 
The folding simulations of the two peptides with implicit solvent model 
also showed similar results \cite{SO1,SO2,SO3}.
For instance, the ff94 \cite{parm94} 
and ff96 \cite{parm96_3} versions of AMBER yield very different
behaviors about the secondary-structure-forming tendencies,
although these force fields differ only in
the main-chain torsion-energy terms.
Many researchers have thus studied the main-chain torsion-energy 
terms and their force-field parameters.
For example, newer force-field parameters for the main-chain 
torsion-energy terms about $\phi$ and $\psi$ angles 
have been developed, which are, e.g., AMBER ff99SB \cite{parm99SB}, 
AMBER ff03 \cite{parm03}, 
CHARMM22/CMAP \cite{CMAP,CMAP2} and OPLS-AA/L \cite{opls2}.
The methods of the force-field optimization thus mainly concentrate on 
the torsion-energy terms.
These modifications of the torsion energy are usually based on 
quantum chemistry calculations \cite{Carlos,Duan,IWA,CMAP,CMAP2,Kamiya} 
or NMR experimental results \cite{Hummer2009, Best_folding}.

We have proposed a new main-chain torsion-energy term, which is 
represented by a double Fourier series in two variables,
the main-chain dihedral angles $\phi$ and $\psi$ \cite{SO4,SO5}.
This expression gives a natural representation of the torsion energy 
in the Ramachandran space \cite{Rama_Sasi}
in the sense that any two-dimensional energy surface periodic 
in both $\phi$ and $\psi$ can be
expanded by the double Fourier series.
We can then easily control secondary-structure-forming tendencies 
by modifying the main-chain torsion-energy surface.
We have presented preliminary results for AMBER ff94 and 
AMBER ff96 \cite{SO4,SO5}.
    
Moreover, we have introduced several optimization methods of 
force-field parameters \cite{SO1,SO2,SO3,SO6,SO7}.
These methods are based on the minimization of some score functions by 
simulations in the force-field parameter space,
where the score functions are derived from the protein coordinate data 
in the Protein Data Bank (PDB).
One of the score functions consists of the sum of the square of 
the force acting on each atom 
in the proteins with the structures from the PDB \cite{SO1,SO2,SO3}. 
Other score functions are taken from the root-mean-square deviations 
between the original PDB structures and 
the corresponding minimized structures \cite{SO6,SO7}.

We have also proposed a new type of the main-chain torsion-energy 
terms for protein systems, 
which can have amino-acid-dependent force-field parameters \cite{SO9}.
As an example of this formulation, we applied this approach to 
the AMBER ff03 force field 
and determined new amino-acid-dependent main-chain torsion-energy parameters 
for $\psi$ (N-C$_{\alpha}$-C-N) and $\psi'$ (C$_{\beta}$-C$_{\alpha}$-C-N)
by using our optimization method in Refs \cite{SO1,SO2,SO3}.

In this Chapter, we review our works on protein force fields.
In section 2 the details of the new main-chain torsion-energy terms 
and the methods for refinements of force-field parameters are given.
In section 3 examples of the applications of these methods
are presented. 
Section 4 is devoted to conclusions.

\section{Methods}

\subsection{General force field for protein systems}

The all-atom force fields for protein systems such as AMBER, CHARMM, OPLS, and ECEPP use 
essentially the same functional forms for the potential energy except for
minor differences.
The commonly used total conformational potential energy $E_{\rm conf}$ is given by

\begin{equation}
E_{\rm conf} = E_{\rm BL} + E_{\rm BA} + E_{\rm torsion} + E_{\rm nonbond}~,
\label{ene_conf_optF1}
\end{equation}
where
\begin{eqnarray}
E_{\rm BL} &=& \sum_{{\rm bond~length}~\ell} K_{\ell} 
(\ell - \ell_{\rm eq})^2~, \\
\label{ene_bond_optF1}
E_{\rm BA} &=& \sum_{{\rm bond~angle}~\theta} K_{\theta} 
( \theta - \theta_{\rm eq})^2~, \\
\label{ene_angle_optF1}
E_{\rm torsion} &=& \sum_{{\rm dihedral~angle}~\Phi} 
\sum_n \frac{V_n}{2} [ 1 + \cos (n \Phi - \gamma_n) ]~, \\
\label{ene_torsion_optF1}
E_{\rm nonbond} &=& \sum_{i<j} \left[ \frac{A_{ij}}{r_{ij}^{12}} - \frac{B_{ij}}{r_{ij}^6} 
+ \frac{332 q_i q_j}{\epsilon r_{ij}} \right] .
\label{ene_nonbond_optF1}
\end{eqnarray}
Here, $E_{\rm BL}$, $E_{\rm BA}$, and $E_{\rm torsion}$
represent the bond-stretching term, the
bond-bending term, and the torsion-energy term, respectively.
The bond-stretching and bond-bending energies are given by harmonic terms
with the force constants, $K_{\ell}$ and $K_{\theta}$, and the equilibrium positions,
$\ell_{\rm eq}$ and $\theta_{\rm eq}$.
The torsion energy is, on the other hand, described by the Fourier
series in Eq.~(\ref{ene_torsion_optF1}),
where the sum is taken over all dihedral angles $\Phi$,
$n$ is the number of waves, $\gamma_n$ is the phase, and $V_n$ is
the Fourier coefficient.
The nonbonded energy in Eq.~(\ref{ene_nonbond_optF1}) is represented by
the  Lennard-Jones and Coulomb terms between pairs of atoms, $i$ and
$j$, separated by the distance $r_{ij}$ (in \AA).
The parameters $A_{ij}$ and $B_{ij}$ in Eq. (\ref{ene_nonbond_optF1}) are the coefficients
for the Lennard-Jones term,
$q_i$ (in units of electronic charges) is the partial charge of
the $i$-th atom,
and $\epsilon$ is the dielectric constant, where we usually
set $\epsilon = 1$ (the value in vacuum).
The factor 332 in the electrostatic term is a constant to express
energy in units of kcal/mol.
Hence, we have five classes of force-field parameters, namely,
those in the bond-stretching term ($K_{\ell}$ and $\ell_{\rm eq}$), those in
the bond-bending term
($K_{\theta}$ and $\theta_{\rm eq}$), those in the torsion term
($V_n$ and $\gamma_n$), those in the Lennard-Jones term
($A_{ij}$ and $B_{ij}$), and those in the electrostatic
term ($q_i$).

Eq.~(\ref{ene_conf_optF1}) represents a standard set of the potential energy terms.
As mentioned above, there are minor differences in the energy functions among different force fields.
For instance, the Urey-Bradley term is used in CHARMM and OPLS, but not in AMBER.
In our parameter refinement methods, we try to optimize a certain set of 
parameters in the existing force fields
without changing the functional forms.
Therefore, if the original force field has non-standard terms, then the optimized one also has them.

\subsection{New torsion-energy terms}

\subsubsection{Representation by a double Fourier series \cite{SO4,SO5}}

Separating the contributions $E(\phi,\psi)$ of the backbone dihedral angles $\phi$ and $\psi$ 
from the rest of the torsion terms $E_{\rm rest}$, we can write
the torsion energy term in Eq.~(\ref{ene_torsion_optF1}) as 
\begin{equation}
E_{\rm torsion} = E(\phi,\psi) + E_{\rm rest}~,
\label{ene_torsionb_Fourier}
\end{equation}
where we have
\begin{equation}
E(\phi,\psi) = \sum_m \frac{V_m}{2} [ 1 + \cos (m \phi - \gamma_m) ]
+\sum_n \frac{V_n}{2} [ 1 + \cos (n \psi - \gamma_n) ]~.
\label{ene_torsion2_Fourier}
\end{equation}
For example, the coefficients for the cases of six force fields 
namely, AMBER parm94, AMBER parm96, AMBER parm99, CHARMM27, 
OPLS-AA, and OPLS-AA/L,
are summarized in Table~\ref{table-org-torsion}, and we can 
explicitly write
$E(\phi,\psi)$ in Eq.~(\ref{ene_torsion2_Fourier}) as follows:
\begin{eqnarray}
\label{ene_torsion_parm94_Fourier}
E_{\rm parm94}(\phi,\psi) 
& = & 2.7 - 0.2 \cos 2 \phi - 0.75 \cos \psi - 1.35 \cos 2 \psi - 0.4 \cos 4 \psi~,\\
\label{ene_torsion_parm96_Fourier}
E_{\rm parm96}(\phi,\psi) 
& = & 2.3 + 0.85 \cos \phi - 0.3 \cos 2 \phi + 0.85 \cos \psi - 0.3 \cos 2 \psi~,\\
\label{ene_torsion_parm99_Fourier}
E_{\rm parm99}(\phi,\psi) 
& = & 5.35 + 0.8 \cos \phi - 0.85 \cos 2 \phi - 1.7 \cos \psi - 2.0 \cos 2 \psi~,\\
\label{ene_torsion_charmm_Fourier}
E_{\rm CHARMM}(\phi,\psi) 
& = & 0.8 - 0.2 \cos \phi + 0.6 \cos \psi~,\\
\label{ene_torsion_oplsaa_Fourier}
E_{\rm OPLS-AA}(\phi,\psi) 
& = & 1.158  - 1.1825 \cos \phi - 0.456 \cos 2 \phi - 0.425 \cos 3 \phi \nonumber\\
& + & 0.908 \cos \psi - 0.611 \cos 2 \psi + 0.7905 \cos 3 \psi~,\\
\label{ene_torsion_oplsaal_Fourier}
E_{\rm OPLS-AA/L}(\phi,\psi) 
& = & 0.81885  -  0.298 \cos \phi - 0.1395 \cos 2 \phi - 2.4565 \cos 3 \phi \nonumber\\
& + & 0.3715 \cos \psi  -  1.254 \cos 2 \psi - 0.4025 \cos 3 \psi~.
\end{eqnarray}

\begin{table}
\caption{Torsion-energy parameters for the backbone dihedral angles $\phi$ and $\psi$ 
for AMBER parm94, AMBER parm96, AMBER parm99, CHARMM27, OPLS-AA, and OPLS-AA/L in Eq.~(\ref{ene_torsion2_Fourier}). }
\label{table-org-torsion}
\vspace{0.3cm}
\begin{center}
\begin{tabular}{lcccccc} \hline
  &   & $\phi$ &            &     & $\psi$ &            \\ \hline
  & $m$ & $\displaystyle{\frac{V_m}{2}}$ (kcal/mol) & $\gamma_m$ (radians)~~~~~ & $n$ & $\displaystyle{\frac{V_n}{2}}$ (kcal/mol) & $\gamma_n$ (radians)\\ \hline
parm94~~~~~ &  2  & 0.2 & $\pi$ & 1  & 0.75 & $\pi$  \\
       &     &       &            &  2  & 1.35 & $\pi$ \\
       &     &       &            &  4  & 0.4 & $\pi$      \\ \hline
parm96~~~~~ &  1  & 0.85 &   0   &  1  & 0.85 &   0   \\ 
       &  2  & 0.3 & $\pi$  &  2  & 0.3 & $\pi$   \\ \hline
parm99~~~~~ &  1  & 0.8  &   0    &  1  & 1.7 & $\pi$   \\ 
            &  2  & 0.85 & $\pi$  &  2  & 2.0 & $\pi$   \\ \hline
charmm~~~~~ &  1  & 0.2  & $\pi$  &  1  & 0.6 &   0     \\ \hline
opls-aa~~~~~ &  1  & $-1.1825$ &   0   &  1  & 0.908    &   0     \\
             &  2  &  0.456    & $\pi$ &  2  & 0.611    & $\pi$   \\
             &  3  & $-0.425$  &   0   &  3  & 0.7905   &   0     \\ \hline
opls-aal~~~~~&  1  & $-0.298$  &   0   &  1  & 0.3715   &   0     \\
             &  2  &  0.1395   & $\pi$ &  2  & 1.254    & $\pi$   \\
             &  3  & $-2.4565$ &   0   &  3  &$-0.4025$ &   0     \\ \hline
\end{tabular}
\end{center}
\end{table}

The backbone torsion-energy term $E(\phi,\psi)$ in Eq.~(\ref{ene_torsion2_Fourier})
is a sum of two one-dimensional Fourier series: 
one is for $\phi$ and the other for $\psi$.
The two variables $\phi$ and $\psi$ are decoupled, and
no correlation of $\phi$ and $\psi$ can be incorporated.
On the other hand, any periodic function of $\phi$ and $\psi$ with period $2\pi$ can be expanded by 
a double Fourier series.
As a simple generalization of $E(\phi,\psi)$, we therefore proposed to express 
this backbone torsion energy by
the following double Fourier series \cite{SO4,SO5}:
\begin{eqnarray}
{\cal E}(\phi,\psi) ~~=~~ a & + & \sum_{m=1}^{\infty}(b_{m} \cos m\phi + c_{m} \sin m\phi) \nonumber\\ 
& + & \sum_{n=1}^{\infty}(d_{n} \cos n\psi + e_{n} \sin n\psi) \nonumber\\ 
& + & \sum_{m=1}^{\infty} \sum_{n=1}^{\infty} (f_{mn} \cos m\phi \cos n\psi + g_{mn} \cos m\phi \sin n\psi \nonumber\\ 
& + & ~~~~~~~~~~~~h_{mn} \sin m\phi \cos n\psi + i_{mn} \sin m\phi \sin n\psi)~.
\label{new_phipsi_Fourier}
\end{eqnarray}
Here, $m$ and $n$ are the numbers of waves, $a$, $b_{m}$, $c_{m}$, $d_{n}$, $e_{n}$, $f_{mn}$, 
$g_{mn}$, $h_{mn}$, and $i_{mn}$ are the Fourier coefficients.
This equation includes cross terms in $\phi$ and $\psi$, while the original term 
in Eq.~(\ref{ene_torsion2_Fourier}) has no mixing of $\phi$ and $\psi$.
Therefore, our new torsion-energy term can represent more complex energy surface 
than the conventional ones.
The Fourier coefficients, by definition, are given by
\begin{eqnarray}
c & = & \frac{1}{\alpha} \int_{-\pi}^{\pi} d\phi \int_{-\pi}^{\pi} d\psi~ 
{\cal E}(\phi,\psi) x(\phi,\psi) \nonumber\\
  & = & \left( \frac{\pi}{180} \right)^2 \frac{1}{\alpha} \int_{-180}^{180} d\tilde{\phi} 
\int_{-180}^{180} d\tilde{\psi}~
{\cal E}\left( \frac{\pi}{180} \tilde{\phi},\frac{\pi}{180} \tilde{\psi} \right) 
x\left( \frac{\pi}{180} \tilde{\phi},\frac{\pi}{180} \tilde{\psi} \right)~,
\label{coe_eqn_Fourier}
\end{eqnarray}
where $\alpha$ are the normalization constants and $x(\phi, \psi)$ are the basis functions for 
the Fourier series.
Table~\ref{coe_table} summarizes these coefficients and functions.
Here, $\phi$ and $\psi$ are given in radians, and $\tilde{\phi}$ and $\tilde{\psi}$ are in degrees 
($\phi = \frac{\pi}{180} \tilde{\phi}$, $\psi = \frac{\pi}{180} \tilde{\psi}$). 
Hereafter, angular quantities without tilde and with tilde are in radians and in degrees,
respectively.

\begin{table}
\caption{Fourier coefficients $c$, normalization constants $\alpha$, and the basis functions 
$x(\phi, \psi)$ for the double Fourier series of the backbone torsion energy ${\cal E}(\phi,\psi)$ 
in Eqs.~(\ref{new_phipsi_Fourier}) and (\ref{coe_eqn_Fourier}).}
\label{coe_table}
\vspace{0.3cm}
\begin{center}
\begin{tabular}{lrr} \hline
$c$      &  ~~$\alpha$  &  ~~~~~~~~~~~~~$x(\phi, \psi)$   \\ \hline
$a$      & $4\pi^2$   &  1                 \\
$b_{m}$  & $2\pi^2$   &  $\cos m\phi$         \\
$c_{m}$  & $2\pi^2$   &  $\sin m\phi$         \\
$d_{n}$  & $2\pi^2$   &  $\cos n\psi$         \\
$e_{n}$  & $2\pi^2$   &  $\sin n\psi$         \\
$f_{mn}$ & $\pi^2$    &  $\cos m\phi \cos n\psi$ \\
$g_{mn}$ & $\pi^2$    &  $\cos m\phi \sin n\psi$ \\
$h_{mn}$ & $\pi^2$    &  $\sin m\phi \cos n\psi$ \\
$i_{mn}$ & $\pi^2$    &  $\sin m\phi \sin n\psi$ \\ \hline
\end{tabular}
\end{center}
\end{table}

Finally, ${\cal E}(\phi,\psi)$ in Eq.~(\ref{new_phipsi_Fourier}) and 
$E_{\rm rest}$ in Eq.~(\ref{ene_torsionb_Fourier}) 
define our torsion-energy term in Eq.~(\ref{ene_conf_optF1}) (instead of Eq.~(\ref{ene_torsion_optF1})):
\begin{equation}
E_{\rm torsion} = {\cal E}(\phi,\psi) + E_{\rm rest}~.
\label{new_torsion_Fourier}
\end{equation}

The double Fourier series in Eq.~(\ref{new_phipsi_Fourier}) is particularly useful,
because it describes the backbone torsion-energy surface in the Ramachandran space.
The Fourier series can express the torsion-energy surface ${\cal E}(\phi,\psi)$ 
that was obtained by any method including quantum chemistry calculations
\cite{opls2,Carlos,Duan,IWA,CMAP,CMAP2,Kamiya}. 

Moreover, one can refine the existing backbone torsion-energy term and 
control the secondary-structure-forming tendencies of the force fields.
For example, $\alpha$-helix is obtained 
for $(\tilde{\phi},\tilde{\psi}) \approx (-57^{\circ},-47^{\circ})$, 
$3_{10}$-helix
for $(\tilde{\phi},\tilde{\psi}) \approx (-49^{\circ},-26^{\circ})$, 
$\pi$-helix
for $(\tilde{\phi},\tilde{\psi}) \approx (-57^{\circ},-70^{\circ})$, 
parallel $\beta$-sheet 
for $(\tilde{\phi},\tilde{\psi}) \approx (-119^{\circ},113^{\circ})$, 
antiparallel $\beta$-sheet 
for $(\tilde{\phi},\tilde{\psi}) \approx (-139^{\circ},135^{\circ})$, 
and so on \cite{Rama_Sasi}.
Hence, if the existing force field gives, say, too little $\alpha$-helix-forming
tendency compared to experimental results, one can lower the backbone torsion-energy 
surface near $(\tilde{\phi},\tilde{\psi}) = (-57^{\circ},-47^{\circ})$ 
in order to enhance
$\alpha$-helix formations.  

We can thus write 
\begin{equation}
{\cal E}(\phi,\psi) = E(\phi,\psi) - f(\phi,\psi)~,
\label{mod_surface_Fourier}
\end{equation}
where
$E(\phi,\psi)$ is the existing backbone torsion-energy term that we want to
refine and $f(\phi,\psi)$ is a function 
that has peaks around the corresponding regions where specific secondary structures 
are to be enhanced.
There are many possible choices for $f(\phi,\psi)$.
For instance, one can use the following function
when one wants to lower the torsion-energy surface
in a single region near $(\phi,\psi) = (\phi_0,\psi_0)$:
\begin{equation}
f(\phi,\psi) = 
\left \{ \begin{array}{ll} 
\displaystyle{A \exp \left( \frac{B}{(\phi - \phi_0)^2 + (\psi - \psi_0)^2 - {r_0}^2} \right) ~,}
& ~{\rm for}~~(\phi - \phi_0)^2 + (\psi - \psi_0)^2 < {r_0}^2~,\\[0.2cm]
                     0~, & ~{\rm otherwise}~,
                     \end{array} \right.
\label{mod_surface2_Fourier}
\end{equation}
where $A$, $B$, and $r_0$ are constants that we adjust for refinement.
In this case, the energy surface is lowered by $f(\phi,\psi)$ in a circular region 
of radius $r_0$, which is centered at $(\phi,\psi) = (\phi_0,\psi_0)$.
Note that we should also impose periodic boundary conditions on $f(\phi,\psi)$.
   
We then express ${\cal E}(\phi,\psi)$ in Eq.~(\ref{mod_surface_Fourier}) in 
terms of 
the double Fourier series in Eq.~(\ref{new_phipsi_Fourier}),  
where the Fourier coefficients are obtained from Eq.~(\ref{coe_eqn_Fourier}).
Hence, we can fine-tune the backbone torsion-energy term by the
above procedure so that it yields correct 
secondary-structure-forming tendencies.

Some remark about the computation time is now in order.
It may appear that we have to expect great increase
in computation time by the introduction
of the double Fourier series, because the number of
terms are much larger.
However, because most of the computation time for the force-field
evaluations is spent in the calculations of distances between
pairs of atoms in the system, the increase in computation
time due to the double Fourier series
is essentially negligible compared to these main computational
efforts.


\subsubsection{Amino-acid-dependent main-chain torsion-energy terms \cite{SO9}}


By writing the dihedral-angle dependence of the parameters explicitly, 
we can rewrite the torsion-energy term in Eq.~(\ref{ene_torsion_optF1}) as
\begin{equation}
E_{\rm torsion} = \sum_{\Phi} \sum_n \frac{V_n\left(\Phi\right)}{2} 
\left\{ 1 + \cos \left[n \Phi - \gamma_n\left(\Phi\right)\right] \right\}~, 
\label{ene_torsion1_eachamino}
\end{equation}
where the first summation 
is taken over all dihedral angles $\Phi$ (both in 
the main chain and in the side chains), 
$n$ is the number of waves, $\gamma_n$ is the phase, and $V_n$ is 
the Fourier coefficient.
Namely, the energy term $E_{\rm torsion}$ has $\gamma_n (\Phi)$ 
and $V_n(\Phi)$ as 
force-field parameters.

We can further write the torsion-energy term as
\begin{equation}
E_{\rm torsion} = E_{\rm torsion}^{(\rm MC)} + E_{\rm torsion}^{(\rm SC)}~,
\label{ene_torsion2_eachamino}
\end{equation}
where
$E_{\rm torsion}^{(\rm MC)}$ and $E_{\rm torsion}^{(\rm SC)}$ are
the torsion-energy terms for dihedral angles around
main-chain bonds and around side-chain bonds, respectively.
Examples of the dihedral angles in
$E_{\rm torsion}^{(\rm MC)}$ are
$\phi$ (C-N-C$_{\alpha}$-C), 
$\psi$ (N-C$_{\alpha}$-C-N), $\phi'$ (C$_{\beta}$-C$_{\alpha}$-N-C), 
$\psi'$ (C$_{\beta}$-C$_{\alpha}$-C-N), and 
$\omega$ (C$_{\alpha}$-C-N-C$_{\alpha}$).
The force-field parameters in
$E_{\rm torsion}^{(\rm SC)}$ can readily depend on amino-acid
residues.
However, those in
$E_{\rm torsion}^{(\rm MC)}$ are usually taken to be
independent of amino-acid residues and the common parameter values
are used for all the amino-acid residues (except for proline).
This is because the amino-acid dependence of the force field
is believed to be taken care of by the very existence of side
chains.  
In Table~\ref{table-org-torsion_eachamino}, we list examples of the
parameter values for
$\psi$ (N-C$_{\alpha}$-C-N)  and $\psi'$ (C$_{\beta}$-C$_{\alpha}$-C-N) 
in general AMBER force fields.

\begin{table}
\caption{Torsion-energy parameters ($V_n$ and $\gamma_n$)
for the main-chain dihedral angles $\psi$ and $\psi'$ 
in Eq.~(\ref{ene_torsion1_eachamino})  
for the original AMBER ff94, ff96, ff99, ff99SB, and ff03 force
fields. 
The values are common among the
amino-acid residues for each force field.
Only the parameters for non-zero $V_n$ are listed.}
\label{table-org-torsion_eachamino}
\vspace{0.3cm}
\begin{center}
\begin{tabular}{lcccccc} \hline
force field  &   & $\psi$ (N-C$_{\alpha}$-C-N) &  &   & $\psi'$ (C$_{\beta}$-C$_{\alpha}$-C-N) &            \\ \hline
  & $n$ & $\displaystyle{V_n/2}$ & $\gamma_n$ & ~~~~~~~ $n$ & $\displaystyle{V_n/2}$  & $\gamma_n$ \\ \hline
ff94~~~~~   &  1  & 0.75 & $\pi$ &  ~~~~~~~ 2  & 0.07 &  0    \\
            &  2  & 1.35 & $\pi$ &  ~~~~~~~ 4  & 0.10 &  0    \\
            &  4  & 0.40 & $\pi$ &  ~~~~~~~    &      &         \\ \hline
ff96~~~~~   &  1  & 0.85 &   0   &  ~~~~~~~ 2  & 0.07 &  0    \\ 
            &  2  & 0.30 & $\pi$ &  ~~~~~~~ 4  & 0.10 &  0    \\ \hline
ff99~~~~~   &  1  & 1.70 & $\pi$ &  ~~~~~~~ 2  & 0.07 &  0    \\ 
            &  2  & 2.00 & $\pi$ &  ~~~~~~~ 4  & 0.10 &  0    \\ \hline
ff99SB~~~   &  1  & 0.45 & $\pi$ &  ~~~~~~~ 1  & 0.20 &  0    \\ 
            &  2  & 1.58 & $\pi$ &  ~~~~~~~ 2  & 0.20 &  0    \\ 
            &  3  & 0.55 & $\pi$ &  ~~~~~~~ 3  & 0.40 &  0    \\ \hline
ff03~~~~~   &  1  & 0.6839 & $\pi$ &  ~~~~~~~ 1  & 0.7784 &  $\pi$    \\ 
            &  2  & 1.4537 & $\pi$ &  ~~~~~~~ 2  & 0.0657 &  $\pi$    \\ 
            &  3  & 0.4615 & $\pi$ &  ~~~~~~~ 3  & 0.0560 &  0    \\ \hline

\end{tabular}
\end{center}
\end{table}

However, this amino-acid independence of the main-chain 
torsion-energy terms is not an absolute requirement, because
we are representing the entire force field by rather a small
number of classical-mechanical terms.  In order to reproduce
the exact quantum-mechanical contributions, one can introduce
amino-acid dependence on any force-field term including
the main-chain torsion-energy terms.
Hence, we can generalize 
$E_{\rm torsion}^{(\rm MC)}$ in Eq.~(\ref{ene_torsion2_eachamino})  
from the expression in Eq.~(\ref{ene_torsion1_eachamino})
to the following amino-acid-dependent form:
\begin{equation}
E_{\rm torsion}^{(\rm MC)} = \sum_{k=1}^{20} 
\sum_{\Phi_{\rm MC}^{(k)}} 
\sum_n \frac{V_n\left(\Phi_{\rm MC}^{(k)}\right)}{2} 
\left\{ 1 + \cos \left[n \Phi_{\rm MC}^{(k)} - 
\gamma_n\left(\Phi_{\rm MC}^{(k)}\right)\right] \right\}~, 
\label{ene_torsion3_eachamino}
\end{equation}
where $k$ ($=1, 2, \cdots, 20$) is the label for the 20 
kinds of amino-acid residues and 
$\Phi_{\rm MC}^{(k)}$ are dihedral angles around the
main-chain bonds in the $k$-th amino-acid
residue.


\subsection{Optimization of force-field parameters}

\subsubsection{Use of force acting on each atom with the PDB coordinates \cite{SO1,SO2,SO3,SO8}}

In the previous subsection, we presented functional forms of the force fields.
Given a fixed set of force-field functions,
we try to optimize a certain set of parameters in the force fields
without changing the functional forms.
Therefore, if the original force field has non-standard terms, then the optimized one also has them.

Our optimization method for these force-field parameters is now described \cite{SO1}.
We first retrieve $N$ native structures (one structure per protein) from PDB.
We try to choose proteins from different folds
(such as all $\alpha$-helix, all $\beta$-sheet, $\alpha$/$\beta$, etc.)
and different homology classes
as much as possible.
If the force-field parameters are of ideal values, then
all the chosen native structures are stable without any force acting on
each atom in the molecules on the average.  Hence, we expect
\begin{equation}
F = 0~,
\label{F0_optF1}
\end{equation}
where
\begin{equation}
F = \sum_{m=1}^N \frac{1}{N_m} \sum_{i_m = 1}^{N_m} \left| \vec{f}_{i_m} \right| ^2~,
\label{F_optF1}
\end{equation}
and
\begin{equation}
\vec{f}_{i_m} = - \frac{\partial E_{\rm tot}^{\{m\}}}{\partial \vec{x}_{i_m}}~. 
\label{F1_optF1}
\end{equation}
Here, $N_m$ is the total number of atoms in molecule $m$, $E_{\rm tot}^{\{m\}}$ is
the total potential energy for molecule $m$, $\vec{x}_i$ is the Cartesian
coordinate vector of atom $i$, and $\vec{f}_{i}$ is 
the force acting on atom $i$.
In reality, $F \ne 0$, and 
because $F \ge 0$, we can optimize the force-field parameters
by minimizing $F$ with respect to these parameters.
In practice, we perform a simulation in the force-field parameter space for this
minimization.

Proteins are usually in aqueous solution, and hence we also have to incorporate
some kind of solvent effects.  Because the more the total number of proteins
($N$) is, the better the force-field parameter optimizations are expected to be, we
want to minimize our efforts in the calculations of the solvent effects.
Here, we employ the generalized-Born/surface area (GB/SA) terms
for the solvent contributions \cite{gb1,gb2}.  
Hence, we use in Eq.~(\ref{F1_optF1}) (we suppress the label $m$ for each
molecule)
\begin{equation}
E_{\rm tot} = E_{\rm conf} + E_{\rm solv}~,
\label{ene_tot_optF1}
\end{equation}
where
\begin{equation}
E_{\rm solv} = E_{\rm GB} + E_{\rm SA}~,
\label{ene_gbsa1_optF1}
\end{equation}
\begin{equation}
E_{\rm GB} = - 166 \left( 1 - \frac{1}{\epsilon_{s}} \right) \sum_{i,j} \frac{q_i q_j}
{\sqrt{r_{ij}^2 + \alpha_{ij}^2 e^{-D_{ij}}}}~,
\label{ene_gbsa3_optF1}
\end{equation}
\begin{equation}
E_{\rm SA} = \sum_k \sigma_k A_k~.
\label{ene_gbsa2_optF1}
\end{equation}
Namely, in the GB/SA model, the total solvation free energy 
in Eq.~(\ref{ene_gbsa1_optF1}) is given by the sum 
of a solute-solvent electrostatic 
polarization term,
a solvent-solvent 
cavity term, and a solute-solvent van der Waals term.
A solute-solvent electrostatic polarization term can be calculated by the generalized Born 
equation in Eq. (\ref{ene_gbsa3_optF1}), where
$\alpha_{ij} = \sqrt{\alpha_i \alpha_j}$,
$\alpha_i$ is the so-called Born radius of atom $i$,
$D_{ij} = r_{ij}^2/(2\alpha_{ij})^2$, and
$\epsilon_s$ is the dielectric constant of bulk water (we take $\epsilon_s = 78.3$).
A solvent-solvent cavity term and a solute-solvent van der Waals term can 
be approximated by the term in Eq.~(\ref{ene_gbsa2_optF1})
that is proportional to the 
solvent accessible surface area.
Here,
$A_k$ is the total solvent-accessible surface area of atoms of type $k$ and $\sigma_k$ is 
an empirically determined proportionality constant
\cite{gb1,gb2}.  

\begin{figure}
\begin{center}
\resizebox*{12cm}{!}{\includegraphics{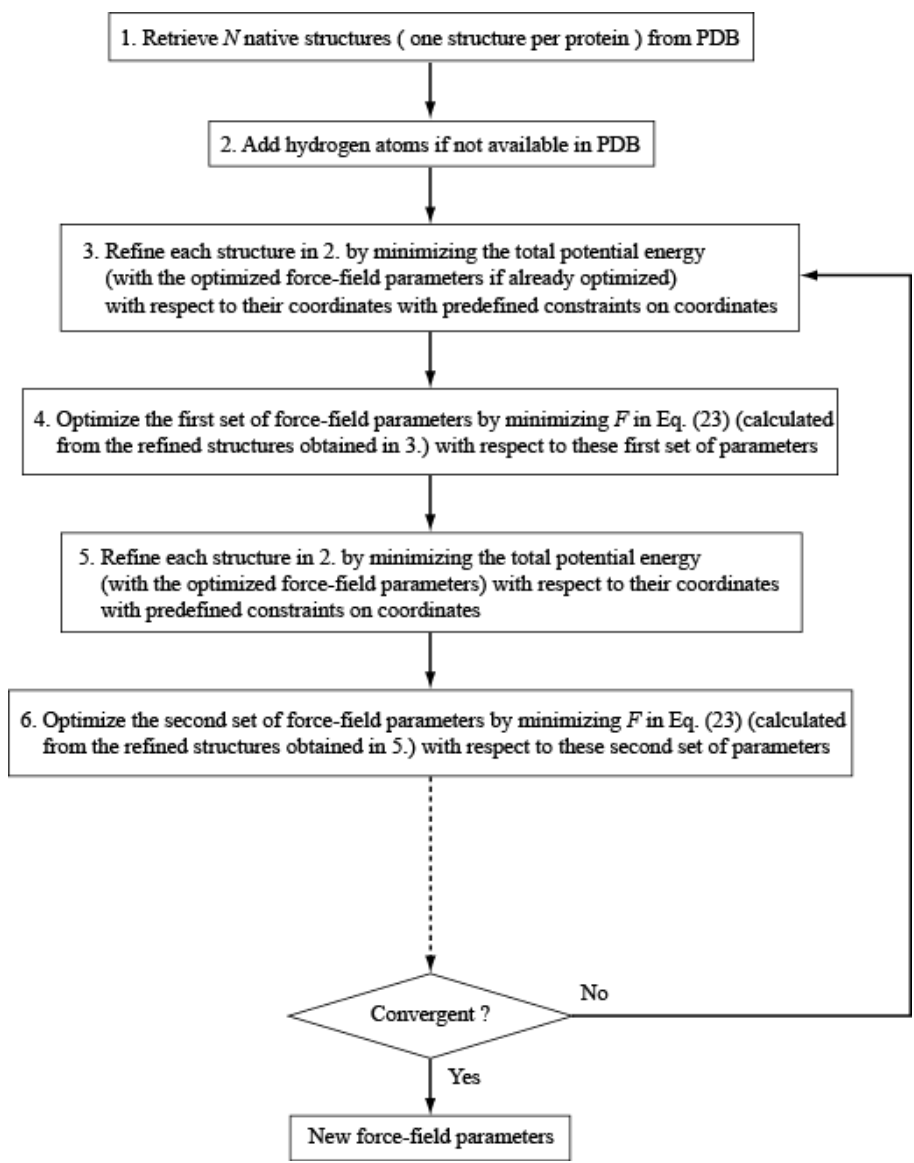}}%
\caption{The flowchart of our method for the optimization of force-field 
parameters.}
\label{fig_flow}
\end{center}
\end{figure}

The flowchart of our method for the optimization of force-field parameters 
is shown in Fig.~\ref{fig_flow}.

In Step 1 of the flowchart we try to obtain as many structures as possible 
from PDB.
The number is limited by the computer power that we have available in our laboratory.
We want to choose proteins with different sizes (numbers of amino acids), 
different folds, and different homology classes as much as possible.
We also want to use only those with high experimental resolutions.
Note that only atomic coordinates of proteins are extracted from PDB
(and coordinates from other molecules such as crystal water are neglected).

If we use data from X-ray experiments, hydrogen atoms are missing, and thus in
Step 2 we have to add hydrogen coordinates.
Many protein simulation software packages provide with routines that add hydrogen atoms to
the PDB coordinates, and one can use one of such routines.  

We now have $N$ protein coordinates ready, but usually such ``raw data''
result in very high total potential energy and strong forces will be
acting on some of the atoms in the molecules. 
This is because the hydrogen coordinates that we added as above are not based on
experimental results and have rather large uncertainties.
The coordinates of heavy atoms from PDB also have experimental errors.
We take the position that we leave the coordinates of heavy atoms as they are
in PDB as much as possible, and adjust the hydrogen coordinates to
reduce this mismatch.
This is why we want to include as many PDB data as possible with high
experimental resolutions (so that the effects of experimental errors
in PDB may be minimal).
We thus minimize the total potential energy 
$E_{\rm tot} = E_{\rm conf} + E_{\rm solv} + E_{\rm constr}$
with respect to the coordinates for each protein conformation,
where $E_{\rm constr}$ is the constraint energy term that is imposed on
the heavy atoms in PDB (it is referred to as the ``predefined constraints''
in Steps 3 and 5 in Fig.~\ref{fig_flow}):
\begin{equation}
E_{\rm constr} = \sum_{\rm heavy~atom} K_x (\vec{x} - \vec{x}_0)^2~.
\label{ene_constr_optF1}
\end{equation}
Here,
$K_x$ is the force constant of the restriction, and
$\vec{x}_0$ are the original coordinate vectors of heavy atoms in PDB.
Because we are searching for the nearest local-minimum states,
usual minimization routines such as the conjugate-gradient method and 
Newton-Raphson method can be employed here.
As one can see from Eq.~(\ref{ene_constr_optF1}), the coordinates of hydrogen
atoms will be mainly adjusted, but unnatural heavy-atom coordinates
will also be modified.
We perform this minimization for all $N$ protein structures separately, and
obtain $N$ refined structures.

Given $N$ set of ``ideal'' reference coordinates in Step 3 of the flowchart,
we now optimize the first set of force-field parameters in Step 4.
In Eq.~(\ref{ene_conf_optF1}) we have five classes of force-field parameters as mentioned
above.
Namely, the force-field parameters are
those in the bond-stretching term ($K_{\ell}$ and $\ell_{\rm eq}$), those in 
the bond-bending term
($K_{\theta}$ and $\theta_{\rm eq}$), those in the torsion term
($V_n$ and $\gamma_n$), those in the Lennard-Jones term
($A_{ij}$ and $B_{ij}$), and those in the electrostatic
term ($q_i$).
Because they are of very different nature, we believe that it is better to
optimize these classes of force-field parameters separately (as in
Steps 4, 6, and so on in Fig.~\ref{fig_flow}).
Note also that if we optimize all the parameters simultaneously, the null result 
(with all the parameter values equal to zero) is a solution to Eq.~(\ref{F0_optF1}).
This is the main reason why 
we optimize each class of parameters separately.

For each set of force-field parameters, the optimization is carried out
by minimizing $F$ in Eq.~(\ref{F_optF1}) with respect to these parameters.
Here, $E_{\rm tot}$ in Eq.~(\ref{F1_optF1}) is given by Eq.~(\ref{ene_tot_optF1}).
For this purpose usual minimization routines such as the conjugate-gradient
method are not adequate, because we need a global optimization.
One should employ more powerful methods such as simulated annealing \cite{SA}
and generalized-ensemble algorithms \cite{GEA}.
We perform this minimization simulation
in the above parameter space to obtain the parameter
values that give the global minimum of $F$.

These processes are repeated until the optimized force-field parameters
converge.
We can, in principle, optimize all the force-field parameters following the flowchart
in Fig.~\ref{fig_flow}.
In the examples given below, however, we just optimize two classes
of the force-field parameters for simplicity; namely, the partial charges and the
backbone torsion-energy parameters.
For the optimization of the partial charges ($q_i$), we impose a condition that the 
total charge of each amino acid remains constant, which is the usual assumption 
adopted by the force fields of Eq.~(\ref{ene_conf_optF1}) based on classical mechanics.
As for the main chain torsion-energy parameters, we use the following functional form 
for each backbone dihedral angle $\phi$ and $\psi$ (see Eq.~(\ref{ene_torsion_optF1})):
\begin{equation}
E_{\Phi=\phi,\psi} = \frac{V_a}{2} \left[ 1+\cos (n_a\Phi - \gamma_a) \right] 
+ \frac{V_b}{2} \left[ 1+\cos (n_b\Phi - \gamma_b) \right] + \frac{V_c}{2} 
\left[ 1+\cos (n_c\Phi - \gamma_c) \right]~.
\label{ene_main_optF1}
\end{equation}
We optimize only the parameters ($V_a$, $V_b$, and $V_c$) and fix the number of waves 
($n_a$, $n_b$, and $n_c$) and the phases ($\gamma_a$, $\gamma_b$, and $\gamma_c$) as in the 
original force field.
This torsion-energy parameter optimization strongly depends on the values of the force 
constant $K_x$ of the constraint energy in Eq.~(\ref{ene_constr_optF1}):
The larger the values of $K_x$ are, the larger those of $V_a$, $V_b$, and $V_c$ tend to be.
In order to minimize such dependences, we impose the constraint that the total area 
enclosed by the curve of $|E_\Phi|$ (from $\Phi=-180^\circ$ to $180^\circ$) remains 
less than or equal to the original value during the optimization.

We believe that these two classes of parameters have the most uncertainty among 
all the force-field parameters. 
This is because partial charges are usually obtained by quantum chemistry
calculations of an isolated amino acid in vacuum separately, which is a very
different condition from that in amino acids of proteins in aqueous solution, 
and because the torsion-energy term is the most problematic (for instance, the parm94,
parm96, and parm99 versions of AMBER differ mainly in backbone torsion-energy parameters).



Moreover, when we perform the optimizations of force-field parameters by using $F$ in Eq.~(\ref{F_optF1}), 
we can neglect unnaturally large forces acting on atoms
in order to remove the errors of PDB structures.
Namely, we can exclude the term for $\vec{f}_{i_m}$ in Eq.~(\ref{F_optF1}) that satisfies 
\begin{equation}
\left| \vec{f}_{i_m} \right| > f_{\rm cut}.
\label{f_cut_rmsd3}
\end{equation}
We determine the cutoff value $f_{\rm cut}$ by using the following function:
\begin{equation}
\Phi {\rm RMSD}=\sqrt{\frac{1}{n}\sum_{i=1}^n{(\Phi_i^{\rm native}-\Phi_i^{\rm min})^2}}.
\label{eq_phi_rmsd_rmsd3}
\end{equation}
Here, $n$ is the total number of backbone dihedral angles ($\phi$ and $\psi$ angles) in all molecules, 
$\Phi_i^{\rm native}$ is the $i$-th backbone dihedral angle of the native structures 
and $\Phi_i^{\rm min}$ is the corresponding $i$-th backbone dihedral angle of the minimized structures using the trial force-field parameters.
The optimal value of $f_{\rm cut}$ is chosen so that $\Phi {\rm RMSD}$ is the minimal value with 
$f_{\rm cut} \leq f^{\rm max}_{\rm cut}$, where $f^{\rm max}_{\rm cut}$ is obtained in an appropriate way (see an example below).


\subsubsection{Use of RMSD I \cite{SO7}}

We now describe another second method for optimizing the force-field parameters.
We use $N$ proteins again from PDB, which can be the same proteins as those that we used 
in the previous optimization method.
If the force-field parameters are of ideal values, we expect that all the chosen native structures minimized 
by the ideal force field do not change after minimizations.
Namely, we believe that force-field parameters are better, if they have 
smaller deviations 
obtained by minimizations of protein structures.
Hence, we expect
\begin{equation}
R = 0,
\end{equation}
where
\begin{equation}
R = \frac{{\displaystyle \sum_{i=1}^N RMSD_i}}{N}.
\label{eq_R_rmsd1}
\end{equation}
Here, $RMSD_i$ is the root-mean-square deviation 
between the native structure of protein $i$ and 
the corresponding minimized structure using the trial force-field parameters.
In reality, $R \ne 0$, and because $R \geq 0$, we expect that we can optimize 
the force-field parameters by minimizing $R$ with respect to these force-field parameters.
In practice, we perform a simulation in the force-field parameter space for this minimization.
Namely, in the previous method we minimize $F$ in Eq.~(\ref{F_optF1}), 
and in the present method we minimize $R$ in Eq.~(\ref{eq_R_rmsd1}) instead.


\subsubsection{Use of RMSD II \cite{SO6}}

We now describe our third method for optimizing the force-field parameters.
We first select $N$ proteins from PDB
as in the previous two methods.
If the force-field parameters are of ideal values, we expect that all the chosen native structures minimized 
by the ideal force field do not change.
Namely, we believe that force-field parameters are better, if they have lower deviations obtained from minimizations of protein structures.
Hence, we expect
\begin{equation}
\Phi {\rm RMSD}=0,
\end{equation}
where
\begin{equation}
\Phi {\rm RMSD}=\sqrt{\frac{1}{n}\sum_{i=1}^n{(\Phi_i^{\rm native}-\Phi_i^{\rm min})^2}}.
\label{eq_e_dif_rmsd2}
\end{equation}
Here, $n$ is the total number of backbone dihedral angles ($\phi$ and $\psi$ angles) in all molecules, 
$\Phi_i^{\rm native}$ is the $i$-th backbone dihedral angle of the native structures 
and $\Phi_i^{\rm min}$ is the corresponding $i$-th backbone dihedral angle of the minimized structures using the trial force-field parameters.
In reality, $\Phi {\rm RMSD} \ne 0$, because $\Phi {\rm RMSD} \geq 0$, we expect that we can optimize 
the force-field parameters by minimizing $\Phi {\rm RMSD}$ with respect to these force-field parameters.
In practice, we perform a simulation in the force-field parameter space for this minimization.

However, our first aim is to determine the balance of secondary-structure-forming tendencies such as 
helix structure and $\beta$-sheet structure.
Additionally, it is difficult to perform the minimization of $\Phi {\rm RMSD}$ in wider force-field paramter space 
until $\Phi {\rm RMSD}$ is close to $0$
because of the computational cost.
Therefore, we only focus on secondary-structure regions of helix structure and $\beta$-sheet structure in the amino-acid sequence.
Namely, we only consider the backbone dihedral angles of residues in the 
native structures which are 
identiffied by the DSSP program \cite{DSSP} that they constitute one of $\alpha$-helix, $3/10$-helix, $\pi$-helix, and $\beta$-sheet structures.
We calculate two kinds of $\Phi {\rm RMSD}$ for secondary structures, namely, $\Phi {\rm RMSD}_{\rm helix}$ and 
$\Phi {\rm RMSD}_{\rm \beta}$.
Here, $\Phi {\rm RMSD}_{\rm helix}$ stands for $\Phi {\rm RMSD}$ of 
backborn dihedral angles of residues which have helix structures 
in the native structures, and $\Phi {\rm RMSD}_{\rm \beta}$ means that of only $\beta$-sheet structures in the native structures.
Using these two $\Phi {\rm RMSD}$s, we want to optimize the torsion-energy parameters, which will have better balance of 
secondary-structure-forming tendencies.
We propose the following combination: 
\begin{equation}
\Phi {\rm RMSD}_{\rm 2ndly} = \lambda \Phi {\rm RMSD}_{\rm helix} + \Phi {\rm RMSD}_{\rm \beta},
\label{helix_plus_beta_rmsd2}
\end{equation}
where we have introduced a fixed scaling factor $\lambda$. 
 
Finally, by minimizing $\Phi {\rm RMSD}_{\rm 2ndly}$ with respect to 
the force-field parameters, we can obtain 
the optimized force-field parameters.





\subsubsection{Use of short MD simulations \cite{SO10}}

We now describe our fourth method for optimizing the force-field parameters.
In this method, we prepare $M$ protein structures, which are some
experimentally determined conformations.
For these proteins, we perform MD simulations, which start from
the experimental conformations, 
by using a trial force field.
We try to perform MD simulations with varied values of force-field parameters. 
After that, we estimate the ``$S$'' value defined by the following function 
for the trajectories of the $M$ proteins 
obtained from  the trial MD simulations:
\begin{equation}
S = \sum_{i=1}^{M} \left( \frac{n_i^{\rm S \to U}}{N_i^{\rm S}} + \frac{n_i^{\rm U \to S}}{N_i^{\rm U}}  \right).
\label{su_num_MD}
\end{equation}
Here, 
$n_i^{\rm S \to U}$ is the number of the amino acids in protein $i$ where their
structures in PDB (initial conformation) had some secondary structures
(such as $\alpha$-helix, $3_{10}$-helix, $\pi$-helix, and $\beta$ structures)
but transformed into unstructured, coil structures without any secondary
structures after a short MD simulation.
Likewise, 
$n_i^{\rm U \to S}$ is is the number of amino acids in protein $i$
where their structures in PDB had coil structures but transformed
to have some secondary structures after a MD simulation.
$N_i^{\rm S}$ is the total number of amino acids in protein $i$
which have some secondary structures in PDB, and 
$N_i^{\rm U}$ is the total number of amino acids in protein $i$
which have coil structures in PDB.

When we calculate the $S$ values for the conformations obtained from MD simulations by using trial force-field parameters, 
the parameter set, which yields the minimum $S$ value, is considered to
give the optimized force field.


\section{Examples of Optimizations of Force-Field Parameters}
\label{results}

\subsection{New torsion-energy terms}

\subsubsection{Representation by a double Fourier series \cite{SO4,SO5}}


We now present various examples of our refinements of force-field parameters.
We first consider the following truncated double Fourier series
(see Eq.~(\ref{new_phipsi_Fourier}):
\begin{eqnarray}
{\cal E}(\phi,\psi) = a & + & b_{1} \cos \phi + c_{1} \sin \phi + b_{2} \cos 2\phi + c_{2} \sin 2\phi + b_{3} \cos 3\phi  + c_{3} \sin 3\phi \nonumber\\
& + & d_{1} \cos \psi + e_{1} \sin \psi + d_{2} \cos 2\psi + e_{2} \sin 2\psi + d_{3} \cos 3\psi + e_{3} \sin 3\psi \nonumber\\ 
& + & f_{11} \cos \phi \cos \psi + g_{11} \cos \phi \sin \psi + h_{11} \sin \phi \cos \psi + i_{11} \sin \phi \sin \psi \nonumber\\ 
& + & f_{21} \cos 2\phi \cos \psi + g_{21} \cos 2\phi \sin \psi + h_{21} \sin 2\phi \cos \psi + i_{21} \sin 2\phi \sin \psi \nonumber\\
& + & f_{12} \cos \phi \cos 2\psi + g_{12} \cos \phi \sin 2\psi + h_{12} \sin \phi \cos 2\psi + i_{12} \sin \phi \sin 2\psi \nonumber\\
& + & f_{22} \cos 2\phi \cos 2\psi + g_{22} \cos 2\phi \sin 2\psi \nonumber\\
& + & h_{22} \sin 2\phi \cos 2\psi + i_{22} \sin 2\phi \sin 2\psi~.   
\label{use_torsion_Fourier}
\end{eqnarray}
This function has 29 Fourier-coefficient parameters.
We will see below that this number of Fourier terms is sufficient for
most of our purposes.

We first check how well the truncated Fourier series in Eq.~(\ref{use_torsion_Fourier})
can reproduce the six original backbone torsion-energy terms 
in Eqs.~(\ref{ene_torsion_parm94_Fourier})--(\ref{ene_torsion_oplsaal_Fourier}).
Because these functions are already the sum of one-dimensional Fourier series
and subsets of the double Fourier series 
in Eq.~(\ref{new_phipsi_Fourier}), the Fourier coefficients in Eq.~(\ref{coe_eqn_Fourier}) can be
analytically calculated and agree with those 
in Eqs.~(\ref{ene_torsion_parm94_Fourier})--(\ref{ene_torsion_oplsaal_Fourier})
except for the last one (that for $\cos 4 \psi$) 
in Eq.~(\ref{ene_torsion_parm94_Fourier}).  This term is missing in Eq.~(\ref{use_torsion_Fourier}).
These cases thus give us good test of numerical integrations in 
Eq.~(\ref{coe_eqn_Fourier}).  The numerical integrations were evaluated as follows.
We divided the Ramachandran space 
($-180^{\circ} < \tilde{\phi} < 180^{\circ}$, 
$-180^{\circ} < \tilde{\psi} < 180^{\circ}$) 
into unit square cells of side length $\tilde{\epsilon}$ (in degrees).
Hence, there are $(360/\tilde{\epsilon})^2$ unit cells altogether.
The double integral on the right-hand side of Eq.~(\ref{coe_eqn_Fourier}) was approximated by the sum of
$\left[{\cal E}\left( \frac{\pi}{180} \tilde{\phi},\frac{\pi}{180} \tilde{\psi} \right) 
x\left( \frac{\pi}{180} \tilde{\phi},\frac{\pi}{180} \tilde{\psi} \right)\right]
\times \left(\tilde{\epsilon}\right)^2$, 
where each 
${\cal E}\left( \frac{\pi}{180} \tilde{\phi},\frac{\pi}{180} \tilde{\psi} \right) 
x\left( \frac{\pi}{180} \tilde{\phi},\frac{\pi}{180} \tilde{\psi} \right)$ was
evaluated at one of the four corners of each unit cell.
We tried two values of $\tilde{\epsilon}$ ($1^{\circ}$ and $10^{\circ}$).
Both cases gave almost complete agreement of Fourier coefficients with the resutls of 
the analytical integrations (see, for example,  Tables~\ref{coe_table2} below).

\begin{table}
\caption{Fourier coefficients in Eq.~(\ref{use_torsion_Fourier}) obtained from
the numerical evaluations of the integrals in Eq.~(\ref{coe_eqn_Fourier}).
``org94'' stands for the original AMBER parm94 force field.
``mod94($\alpha$)'' and ``mod94($\beta$)'' stand for AMBER parm94 force fields 
that were modified to
enhance $\alpha$-helix structures and $\beta$-sheet structures, respectively,
by Eqs.~(\ref{mod_surface_Fourier}) and (\ref{mod_surface2_Fourier}).
The bin size $\tilde{\epsilon}$ is the length of the sides of each unit
square cell for the numerical integration in Eq.~(\ref{coe_eqn_Fourier}).}
\label{coe_table2}
\vspace{0.3cm}
\begin{center}
\begin{tabular}{lrrrrrr} \hline
bin size $\tilde{\epsilon}$   &             & $1^\circ$       &                &             & $10^\circ$      &                 \\ \hline
coefficient & org94    &  mod94($\alpha$) &  mod94($\beta$) &  org94   &  mod94($\alpha$) &  mod94($\beta$)   \\ \hline
$a$  &  2.700000 & 2.308359   & 1.916719  & 2.700000  & 2.308370   & 1.916742  \\ 
$b_1$ &  0.000000 &$-0.330937$ & 0.781150  & 0.000000  &$-0.331053$ & 0.781041  \\ 
$c_1$ &  0.000000 & 0.509599   & 0.930938  & 0.000000  &  0.509517  & 0.930809  \\ 
$b_2$ &$-0.200000$&$-0.101549$ &$-0.115937$&$-0.200000$&$-0.101513$ &$-0.115970$\\ 
$c_2$ &  0.000000 & 0.221123   &$-0.476745$& 0.000000  &  0.221100  &$-0.476558$\\ 
$b_3$ &  0.000000 &$-0.018073$ & 0.031693  & 0.000000  &$-0.018084$ &  0.031714 \\
$c_3$ &  0.000000 &$-0.002862$ &$-0.018298$& 0.000000  &$-0.003036$ &$-0.018310$\\ 
$d_1$ &$-0.750000$&$-1.164401$ &$-0.052959$&$-0.750000$&$-1.164500$ &$-0.052874$\\ 
$e_1$ &  0.000000 &  0.444390  &$-0.995478$& 0.000000  &  0.444289  &$-0.995599$\\ 
$d_2$ &$-1.350000$&$-1.333115$ &$-1.184428$&$-1.350000$&$-1.333073$ &$-1.184340$\\ 
$e_2$ &  0.000000 &  0.241460  &  0.454905 & 0.000000  &  0.241451  &  0.455147 \\ 
$d_3$ &  0.000000 &$-0.014220$ &  0.035349 & 0.000000  &$-0.014143$ &  0.035324 \\ 
$e_3$ &  0.000000 &$-0.011515$ &  0.009472 & 0.000000  &$-0.011671$ &  0.009465 \\ 
$f_{11}$&  0.000000 &$-0.342789$ &$-0.680493$& 0.000000  &$-0.343087$ &$-0.680497$\\ 
$g_{11}$&  0.000000 &  0.367596  &  0.971845 & 0.000000  &  0.367697  &  0.971851 \\ 
$h_{11}$&  0.000000 &  0.527849  &$-0.810980$& 0.000000  &  0.527949  &$-0.810985$\\ 
$i_{11}$&  0.000000 &$-0.566049$ &  1.158199 & 0.000000  &$-0.565751$ &  1.158206 \\ 
$f_{21}$&  0.000000 &  0.090016  &$-0.064642$& 0.000000  &  0.090168  &$-0.064636$\\ 
$g_{21}$&  0.000000 &$-0.096530$ &  0.092318 & 0.000000  &$-0.096472$ &  0.092309 \\ 
$h_{21}$&  0.000000 &  0.202178  &  0.366601 & 0.000000  &  0.202421  &  0.366565 \\
$i_{21}$&  0.000000 &$-0.216810$ &$-0.523561$& 0.000000  &$-0.216596$ &$-0.523509$\\ 
$f_{12}$&  0.000000 &  0.012329  &$-0.142682$& 0.000000  &  0.012385  &$-0.142712$\\ 
$g_{12}$&  0.000000 &  0.176308  &$-0.392017$& 0.000000  &  0.176622  &$-0.392098$\\ 
$h_{12}$&  0.000000 &$-0.018984$ &$-0.170042$& 0.000000  &$-0.019013$ &$-0.170077$\\ 
$i_{12}$&  0.000000 &$-0.271490$ &$-0.467187$& 0.000000  &$-0.271321$ &$-0.467284$\\ 
$f_{22}$&  0.000000 &$-0.000586$ &$-0.002453$&$-0.000001$&$-0.000585$ &$-0.002451$\\ 
$g_{22}$&  0.000000 &$-0.008378$ &$-0.006738$&  0.000000 &$-0.008397$ &$-0.006733$\\ 
$h_{22}$&  0.000000 &$-0.001316$ &  0.013909 &  0.000000 &$-0.001317$ &  0.013897 \\ 
$i_{22}$&  0.000000 &$-0.018817$ &  0.038215 &  0.000000 &$-0.018867$ &  0.038183 \\ \hline
\end{tabular}
\end{center}
\end{table}

\begin{figure}
\begin{center}
\includegraphics[width=12cm,keepaspectratio]{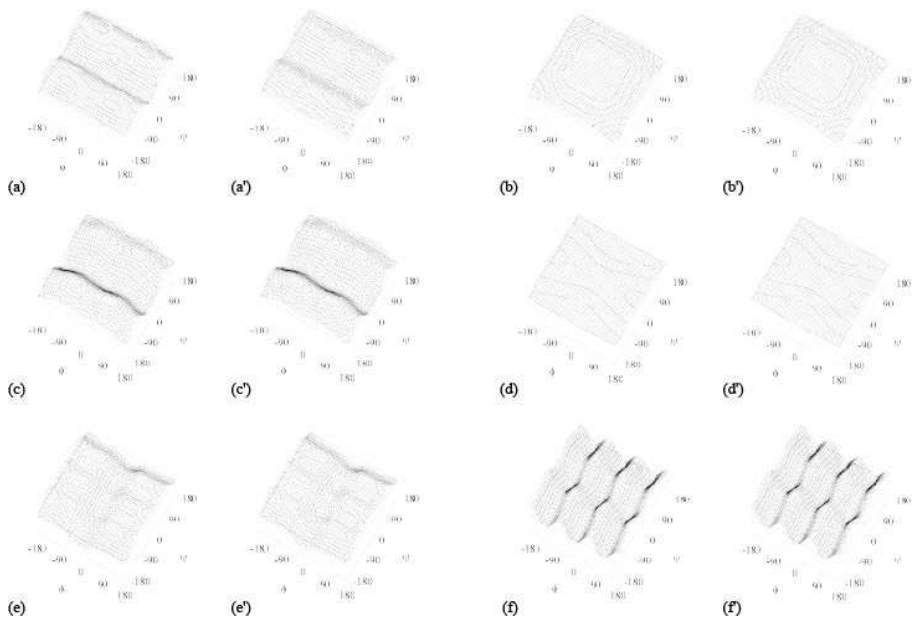}
\caption{Backbone-torsion-energy surfaces of six force fields.  
The backbone dihedral
angles $\tilde{\phi}$ and $\tilde{\psi}$ are in degrees. 
(a), (b), (c), (d), (e), and (f) are those of the original AMBER parm94, 
the original AMBER parm96, 
the original AMBER parm99, the original CHARMM 27, the original OPLS-AA, 
and the original OPLS-AA/L, respectively. 
(a') to (f') are those of (a) to (f), respectively, that were
expressed by the truncated double 
Fourier series in Eq.~(\ref{use_torsion_Fourier}).  The contour lines are drawn every 0.5 kcal/mol.}
\label{fig_ene_surface_org}
\end{center}
\end{figure}

In Fig.~\ref{fig_ene_surface_org}
we compare the six original backbone torsion-energy surfaces with those of
the corresponding double Fourier series in Eq.~(\ref{use_torsion_Fourier}).
Hereafter, the primed labels for figures such as (a') indicate that
the results are those of the double Fourier series.
As can be seen from Fig.~\ref{fig_ene_surface_org}, 
the backbone torsion-energy surfaces are in complete agreement for
all force fields except for AMBER parm94, whereas we see a little difference for AMBER parm94 between
Figs.~\ref{fig_ene_surface_org}(a) and \ref{fig_ene_surface_org}(a'). 
As discussed above, this slight difference for AMBER parm94 reflects the
fact that the $\cos 4 \psi$ term in Eq.~(\ref{ene_torsion_parm94_Fourier})
is missing in the truncated double Fourier series in 
Eq.~(\ref{use_torsion_Fourier}).

We now consider the double Fourier series of non-trigonometric functions.
The functions are those in 
Eqs.~(\ref{mod_surface_Fourier}) and (\ref{mod_surface2_Fourier}).
We try to fine-tune the six original force fields by
subtracting $f(\phi,\psi)$ in Eq.~(\ref{mod_surface2_Fourier})
from the original functions.  The criterion for fine-tuning is, for instance,
whether the refined force fields yield better agreement of the 
secondary-structure-forming tendencies with experimental implications
than the original ones.
For this we need good experimental data.
Because the purpose here is to test whether or not we can control
the secondary-structure-forming tendencies, we simply consider 
extreme cases where we try to modify the existing force fields so that
desired secondary structures may be obtained regardless of the
tendencies of the original force fields.  Note that the six original force fields 
have quite different preferences for $\alpha$-helix and $\beta$-sheet structures \cite{YSO1,YSO2,SO1,SO2,SO3}.

The function $f(\phi,\psi)$ in Eq.~(\ref{mod_surface2_Fourier})
reduces the value of $E(\phi,\psi)$ in a circle of 
radius $r_0$ with the center located at $(\phi_0,\psi_0)$. 
We used $\tilde{r}_0 = 100^\circ$ and $\tilde{B}=5,000$ (degrees)$^2$. 
The coefficient $A$ is calculated by Eq.~(\ref{mod_surface2_Fourier}) from the other parameters 
$f(\tilde{\phi}_0,\tilde{\psi}_0)$, $\tilde{r}_0$, and $\tilde{B}$. 
Namely, we have
\begin{equation}
A = f(\tilde{\phi}_0,\tilde{\psi}_0) \exp \left( \frac{\tilde{B}}{\tilde{r}_0^2} \right) ~.
\label{mod_surface3_Fourier}
\end{equation}

We used
$(\tilde{\phi}_0,\tilde{\psi}_0) = (-57^{\circ},-47^{\circ})$ 
and 
$(\tilde{\phi}_0,\tilde{\psi}_0) = (-130^{\circ},125^{\circ})$ 
in order to enhance
$\alpha$-helix-forming tendency and $\beta$-sheet-forming tendency,
respectively.
The central values $f(\tilde{\phi}_0,\tilde{\psi}_0)$ that we used were
3.0 kcal/mol and 6.0 kcal/mol for enhancing $\alpha$-helix and
$\beta$-sheet, respectively, in the case of AMBER parm94, AMBER parm99, CHARMM27, and OPLS-AA/L.
They were both 3.0 kcal/mol in the case of AMBER parm96 and OPLS-AA.

We remark that the large value of $f(\tilde{\phi}_0,\tilde{\psi}_0)$, 
6.0 kcal/mol, that was necessary to enhance $\beta$-sheet in the case of AMBER parm94, AMBER parm99, 
CHARMM27, and OPLS-AA/L reflects the fact that their original force fields favor $\alpha$-helix.

\begin{figure}
\begin{center}
\includegraphics[width=12cm,keepaspectratio]{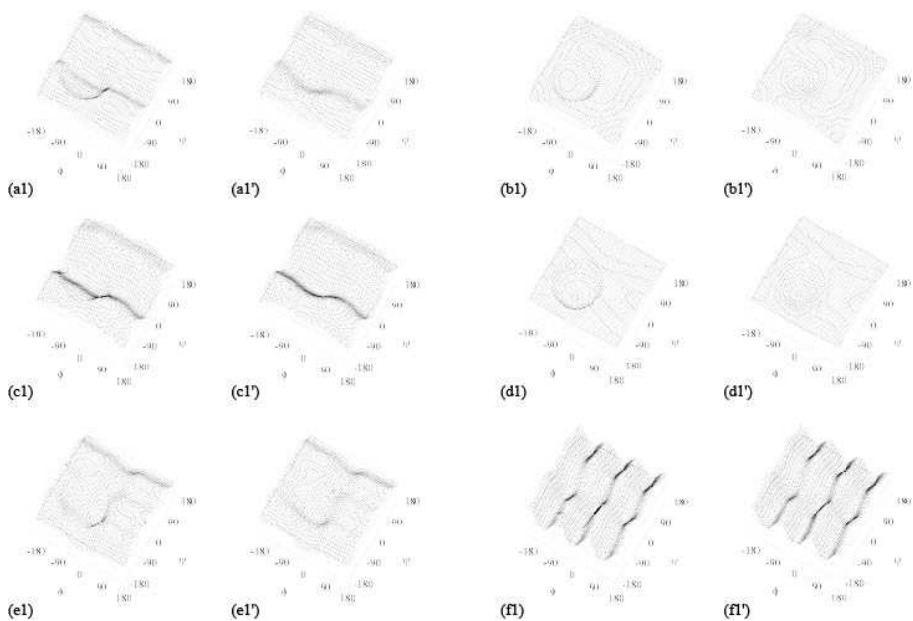}
\caption{Backbone-torsion-energy surfaces of six force fields that were modified by
Eqs.~(\ref{mod_surface_Fourier}), (\ref{mod_surface2_Fourier} and (\ref{use_torsion_Fourier})).
From (a1) to (f1) are those of AMBER parm94, AMBER parm96, AMBER parm99, CHARMM 27, OPLS-AA, and
OPLS-AA/L  force fields that were modified to
enhance $\alpha$-helix structures, respectively.
From (a1') to (f1') are those of AMBER parm94, AMBER parm96, AMBER parm99, CHARMM 27, OPLS-AA, and
OPLS-AA/L  force fields that were expanded by the truncated
double Fourier series in Eq.~(\ref{use_torsion_Fourier}).}
\label{fig_ene_surface_mod}
\end{center}
\end{figure}

\begin{figure}
\begin{center}
\includegraphics[width=12cm,keepaspectratio]{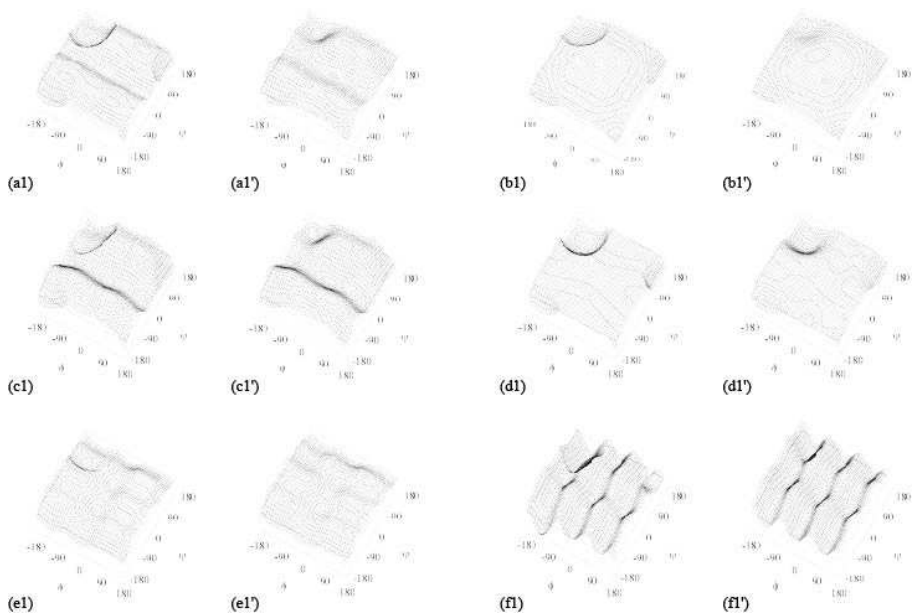}
\caption{Backbone-torsion-energy surfaces of six force fields that were 
modified by
Eqs.~(\ref{mod_surface_Fourier}), (\ref{mod_surface2_Fourier} and 
(\ref{use_torsion_Fourier})).
From (a1) to (f1) are those of AMBER parm94, AMBER parm96, AMBER parm99, 
CHARMM 27, OPLS-AA, and
OPLS-AA/L  force fields that were modified to
enhance $\beta$-sheet structures, respectively.
From (a1') to (f1') are those of AMBER parm94, AMBER parm96, AMBER parm99, 
CHARMM 27, OPLS-AA, and
OPLS-AA/L  force fields that were expanded by the truncated
double Fourier series in Eq.~(\ref{use_torsion_Fourier}).}
\label{fig_ene_surface_mod2}
\end{center}
\end{figure}

In Fig.~\ref{fig_ene_surface_mod}(a1)--(f1)
we compare the six backbone torsion-energy surfaces modified according to Eq.~(\ref{mod_surface_Fourier}), 
which reduced the torsion energy in the $\alpha$-helix region, 
with those of the corresponding double Fourier series in Eq.~(\ref{use_torsion_Fourier}).
In Fig.~\ref{fig_ene_surface_mod}(a1)--(f1), 
$\alpha$-helix is enhanced
from the original AMBER parm94 (a1), AMBER parm96 (b1), AMBER parm99 (c1),
CHARMM27 (d1), OPLS-AA (e1), and OPLS-AA/L (f1).
In Fig.~\ref{fig_ene_surface_mod2}(a1)--(f1)
we show the case of the $\beta$-sheet region, and 
$\beta$-sheet is enhanced
from the original AMBER parm94 (a1), AMBER parm96 (b1), AMBER parm99 (c1),
CHARMM27 (d1), OPLS-AA (e1), and OPLS-AA/L (f1).

These modified backbone torsion-energy functions were expanded by the
truncated double Fourier series in Eq.~(\ref{use_torsion_Fourier}) by evaluating the
corresponding Fourier coefficients from Eq.~(\ref{coe_eqn_Fourier}).
For the numerical integration we again tried two values of 
the bin size $\tilde{\epsilon}$ ($1^{\circ}$ and $10^{\circ}$).
The obtained Fourier coefficients are summarized in
Tables~\ref{coe_table2}, for example, 
in the case of 
AMBER parm94.
For comparisons, the Fourier coefficients of the original AMBER force fields
(before modifications) are also listed.
We see that the two choices of the bin size 
$\tilde{\epsilon}$ gave essentially the same results (agreeing in about 3 digits).

In Figs.~\ref{fig_ene_surface_mod}(a1')--(f1') and \ref{fig_ene_surface_mod2}(a1')--(f1')
we show the backbone torsion-energy surfaces represented by the
truncated double Fourier series.
Comparing these with the original ones in
Fig.~\ref{fig_ene_surface_mod}(a1)--(f1) and \ref{fig_ene_surface_mod2}(a1)--(f1), 
we find that the overall features of the
energy surfaces are well reproduced by the Fourier series.
If more accuracy is desired, we can simply increase the number of 
Fourier terms in the expansion.
As we will see below, the present accuracy of the Fourier series was sufficient for
the purpose of controlling the secondary-structure-forming tendencies
towards $\alpha$-helix or $\beta$-sheet. 

\begin{figure}
\begin{center}
\resizebox*{8cm}{!}{\includegraphics{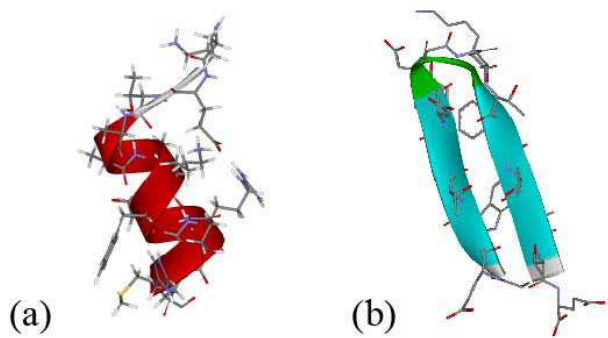}}%
\caption{The structures of C-peptide (a) and G-peptide (b) obtained from 
the experimental results (PDB ID are (a) 1A5P and (b) 1PGA). 
The figures were created with DS Visualizer v1.5\cite{ADS_v2}.}
\label{fig_cpgp}
\end{center}
\end{figure}

We examined the effects of the above modifications of the backbone 
torsion-energy terms
in AMBER parm94, AMBER parm96, AMBER parm99, CHARMM27, OPLS-AA, and OPLS-AA/L 
(towards specific secondary structures) by performing
the folding simulations of two peptides, namely, C-peptide of 
ribonuclease A and the 
C-terminal fragment of the B1 domain of streptococcal protein G, which is 
sometimes referred to as 
G-peptide \cite{gpep3}.
The C-peptide has 13 residues and its amino-acid sequence is
Lys-Glu-Thr-Ala-Ala-Ala-Lys-Phe-Glu-Arg-Gln-His-Met.
This peptide has been extensively studied by experiments and is known
to form an $\alpha$-helix structure \cite{buzz1,buzz2}, as shown 
in Fig.~\ref{fig_cpgp}(a).
Because the charges at peptide termini are known to affect
helix stability \cite{buzz1,buzz2}, we blocked the termini by
a neutral COCH$_3$- group and a neutral -NH$_2$ group.
The G-peptide has 16 residues and its amino-acid sequence is
Gly-Glu-Trp-Thr-Tyr-Asp-Asp-Ala-Thr-Lys-Thr-Phe-Thr-Val-Thr-Glu.
The termini were kept as the usual zwitter ionic states, following the
experimental conditions \cite{gpep3,gpep1,gpep2}.
This peptide is known to form a $\beta$-hairpin structure by experiments
\cite{gpep3,gpep1,gpep2}, as shown 
in Fig.~\ref{fig_cpgp}(b).

Simulated annealing \cite{SA} MD simulations were performed for both
peptides from fully extended initial conformations, where the 12 versions
of the truncated
double Fourier series (which were described in 
Table~\ref{coe_table2} and in
Fig.~\ref{fig_ene_surface_mod}(a1')--(f1') and \ref{fig_ene_surface_mod2}(a1')--(f1')) 
were used for the backbone torsion-energy terms
of AMBER parm94, AMBER parm96, AMBER parm99, CHARMM27, OPLS-AA, and OPLS-AA/L force fields.
For comparisons, the simulations 
with the original force fields
were also performed. 
The unit time step was set to 1.0 fs.
Each simulation was carried out for 1 ns (hence, it consisted of
1,000,000 MD steps).
The temperature during MD simulations was controlled by 
Berendsen's method \cite{berendsen}.
For each run the temperature was decreased exponentially
from 2,000 K to 250 K.
We modified and used the program package
TINKER version 4.1 \cite{tinker_v2} for all the simulations.
As for solvent effects, we used the GB/SA model \cite{gb1,gb2} included in the TINKER program package.
For both peptides, these folding simulations were repeated 60 times 
with different sets of randomly generated initial velocities.

\begin{figure}
\begin{center}
\includegraphics[width=12cm,keepaspectratio]{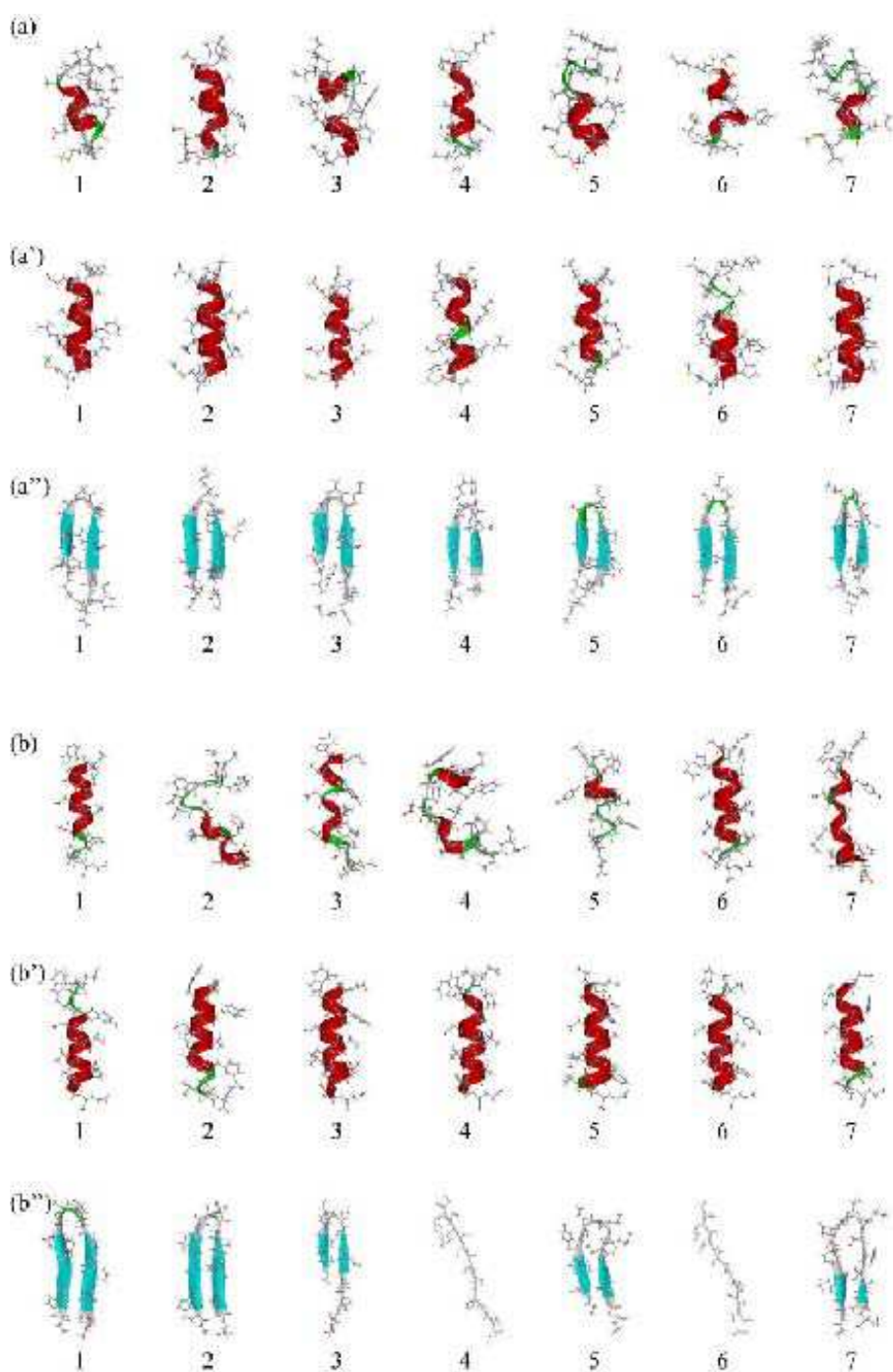}
\caption{Seven lowest-energy final conformations of C-peptide (a)--(a'') and G-peptide (b)--(b'') obtained 
from six sets of 60 simulated annealing MD runs.
(a) and (b) are the results of the original AMBER parm94.
(a') and (b') are the results of AMBER parm94 of the truncated double Fourier series of
six force fields that were modified to
enhance $\alpha$-helix structures.
(a'') and (b'') are the results of AMBER parm94 of the truncated double Fourier series of
six force fields that were modified to
enhance $\beta$-sheet structures.
The conformations are ordered in the increasing order of energy for each case.  
The figures were created with DS Visualizer v1.5\cite{ADS}.}
\label{fig_94_str}
\end{center}
\end{figure}

In Fig.~\ref{fig_94_str}, 
we show seven (out of 60) lowest-energy final conformations of 
C-peptide and G-peptide obtained by the simulated annealing MD simulations, for example, in the case of 
AMBER parm94.

In the Figure, we see that all conformations of 
the original AMBER parm94 (except for 
conformations 2 and 4 of G-peptide) and 
all conformations of its force field modified towards $\alpha$-helix 
are $\alpha$-helix structures 
(conformations 2 and 4 are $3_{10}$-helix structures).
The results show that the original AMBER parm94 favors $\alpha$-helix 
structures, and moreover,  
its force field modified towards $\alpha$-helix favors $\alpha$-helix 
structures more than the original force field 
in the sense that the obtained helices are more extended (and almost 
entirely helical).
On the other hand, AMBER parm94 modified towards $\beta$-sheet 
favors $\beta$ structures strongly.
The results for other force fields were similar.

Therefore, regardless of the secondary-structure-forming tendencies of 
the original force fields, our modifications of the backbone torsion-energy term 
succeeded in enhancing the desired secondary structures.


\subsubsection{Amino-acid-dependent main-chain torsion-energy terms \cite{SO9}}


We present the results of our optimizations of
the force-field parameters 
$V_1(\Phi_{\rm MC}^{(k)})$ for the main-chain angles
$\Phi_{\rm MC}^{(k)}=$ 
$\psi^{(k)}$ (N-C$_{\alpha}$-C-N) and 
$\psi'^{(k)}$ (C$_{\beta}$-C$_{\alpha}$-C-N) 
in Eq.~(\ref{ene_torsion3_eachamino}).
We did this for the case of AMBER ff03 force field.
We determined these
$V_1(\Phi_{\rm MC}^{(k)})$ 
values for the 19 amino-acid residues except for proline.
  
\begin{figure}
\begin{center}
\resizebox*{12cm}{!}{\includegraphics{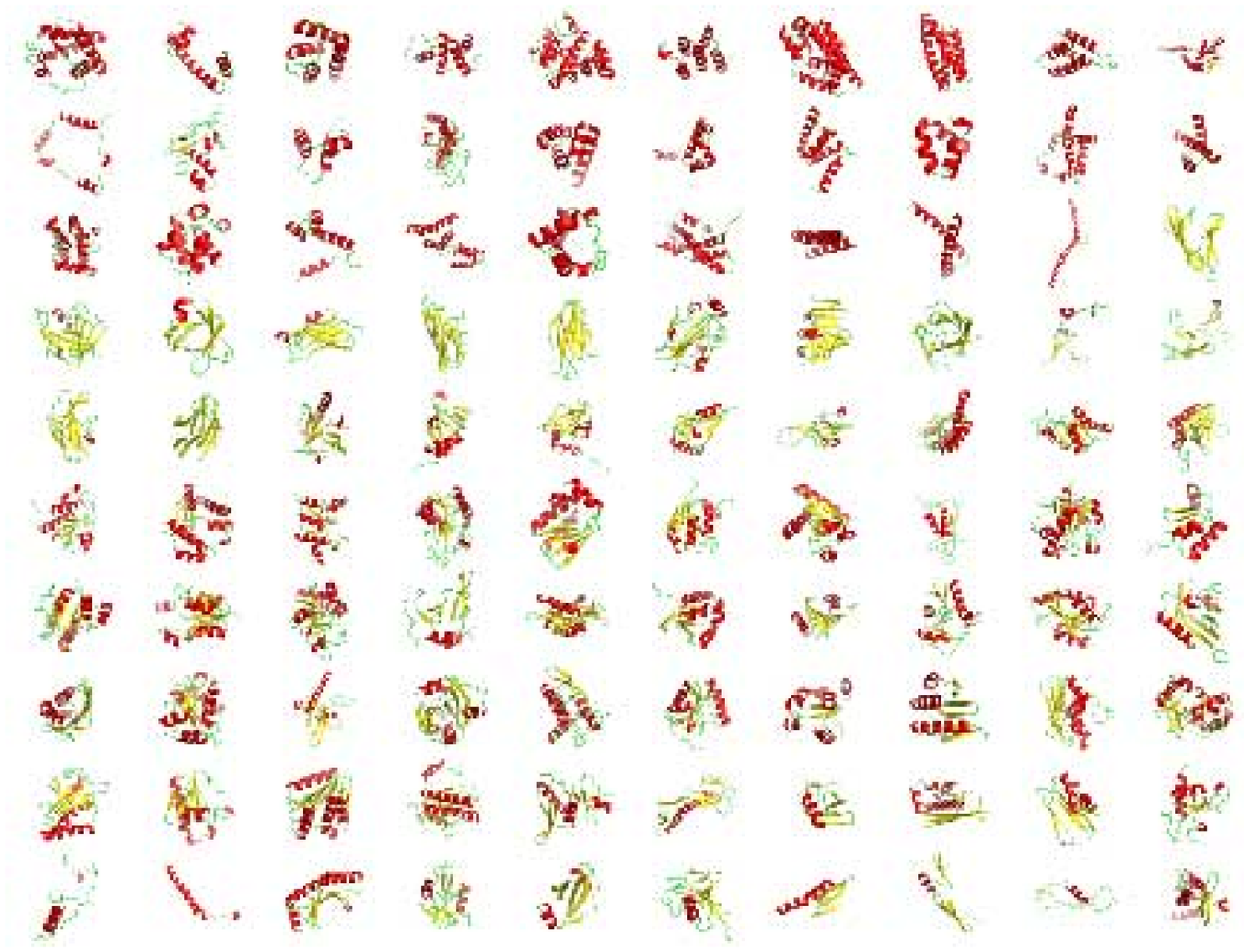}}%
\caption{Structures of 100 proteins in Table~\ref{PDB_list} which
were used in the optimization of force-field parameters.}
\label{fig_100proteins_eachamino}
\end{center}
\end{figure}

At first, we chose 100 PDB files with resolution 2.0 \AA~or better, 
with sequence similarity of amino acid 30.0 \% or lower,  
and with less than 200 residues (the average number of residues is 117.0) from PDB-REPRDB \cite{REPRDB} (see Table~\ref{PDB_list} and Fig.~\ref{fig_100proteins_eachamino}).
We then refined these selected 100 structures.
Generally, data from X-ray experiments do not have coordinates for hydrogen atoms.
Therefore, we have to add hydrogen coordinates.
Many protein simulation software packages provide with routines that add hydrogen atoms to the PDB 
coordinates.
We used the AMBER11 program package \cite{amber_prog11}.   
We thus minimized the total potential 
energy $E_{\rm total} = E_{\rm conf} + E_{\rm solv} + E_{\rm constr}$ 
with respect to the coordinates for each proten conformation, 
where $E_{\rm constr}$ is the harmonic constraint 
energy 
term ($E_{\rm constr} = \sum_{\rm heavy~atom} K_x (\vec{x} - \vec{x}_0)^2$), 
and $E_{\rm solv}$ is the solvation energy term.
Here, $K_x$ is the force constant of the restriction 
and $\vec{x}_0$ are the original coordinate vectors 
of heavy atoms in PDB.
As one can see from $E_{\rm constr}$, the coordinates of 
hydrogen atoms will be mainly adjusted, 
but unnatural heavy-atom coordinates will also be modified.
We performed this minimization for all the 100 protein structures 
separately and obtained 
100 refined structures by using $K_x = 100$ (kcal/mol).
As for the solvation energy term $E_{\rm solv}$, we used the GB/SA solvent 
included in the AMBER program package ($igb = 5$ and $gbsa = 1$)
\cite{gbsa_igb5,gbsa_sa1}. 

\begin{table}
\caption{100 proteins used in the optimization of force-field parameters.}
\label{PDB_list}
\vspace{0.3cm}
\begin{center}
\begin{tabular}{lclclclcl} \hline
fold            &~~ PDB ID ~~&  chain ~&~~ PDB ID  ~~& chain ~&~~ PDB ID  ~~& chain ~&~~ PDB ID ~~& chain  \\ \hline
all $\alpha$    & 1DLW &  A  & 1N1J &  B  & 1U84 &  A  & 1HBK &  A  \\
                & 1TX4 &  A  & 1V54 &  E  & 1SK7 &  A  & 1TQG &  A  \\
                & 1V74 &  B  & 1DVO &  A  & 1HFE &  S  & 1J0P &  A  \\
                & 1Y02 &  A71-114  & 1IJY &  A  & 1I2T &  A  & 1G8E &  A  \\
                & 1VKE &  C  & 1FS1 & A109-149  &  1D9C  &  A  & 1AIL & A  \\
                & 1Q5Z &  A  & 1T8K &  A  & 1OR7 &  C  & 1NG6 & A    \\
                & 1C75 &  A  & 2LIS &  A  & 1NH2 &  B  & 1Q2H &  A  \\
                & 1NKP &  A  &      &     &      &     &      &     \\ \hline
all $\beta$     & 1XAK &  A  & 1T2W &  A  & 1GMU &  C1-70 & 1AYO &  A  \\
                & 1PK6 &  A  & 1NLQ &  C  & 1BEH &  A  & 1UA8 &  A  \\
                & 1UXZ &  A  & 1UB4 &  C  & 1LGP &  A  & 1CQY &  A  \\
                & 1PM4 &  A  & 1OU8 &  A  & 1V76 &  A  & 1UT7 &  B  \\
                & 1OA8 &  D  & 1IFG &  A  &      &     &      &     \\ \hline
$\alpha / \beta$& 1IO0 &  A  & 1U7P &  A  & 1JKE &  C  & 1MXI &  A  \\
                & 1LY1 &  A  & 1NRZ &  A  & 1IM5 &  A  & 1VC1 &  A  \\
                & 1OGD &  A  & 1IIB &  A  & 1PYO &  D  & 1MUG &  A  \\
                & 1H75 &  A  & 1K66 &  A  & 1COZ &  A  & 1D4O &  A  \\ \hline
$\alpha+\beta$  & 1VCC &  A  & 1PP0 &  B  & 1PZ4 &  A  & 1TU1 &  A  \\
                & 1Q2Y &  A  & 1M4J &  A  & 1N9L &  A  & 1LQV &  B  \\
                & 1A3A &  A  & 1K2E &  A  & 1TT8 &  A  & 1HUF &  A  \\
                & 1SXR &  A  & 1CYO &  A  & 1KAF &  A  & 1ID0 &  A  \\
                & 1UCD &  A  & 1F46 &  B  & 1KPF &  A  & 1BYR &  A  \\
                & 1Y60 &  D  & 1SEI &  A  & 1RL6 &  A  & 1WM3 &  A  \\
                & 1FTH &  A  & 1APY &  B  & 1JID &  A  & 1N13 &  E  \\
                & 1LTS &  C  & 1JYO &  F  & 1E87 &  A  & 1UGI &  A  \\
                & 1MWP &  A  & 1PCF &  A  & 1MBY &  A  & 1IHR &  B  \\
                & 1H6H &  A  &      &     &      &     &      &     \\ \hline
\end{tabular}
\end{center}
\end{table}

For these refined protein structures, we performed the 
optimization of force-field parameters $V_1^{(k)}$ of 
$\psi$ and $\psi'$ angles for AMBER ff03 force field by 
using the fucntion $F$ in Eq.~(\ref{F_optF1}) as 
the total potential energy 
function ($E_{\rm total} = E_{\rm conf} + E_{\rm solv}$) for the Monte Carlo simulations in the
parameter space.
Here, we used AMBER11 \cite{amber_prog11} for the force calculations
in Eq.~(\ref{F1_optF1}).  
We have to optimize the 38 ($=2 \times 19$) parameters simultaneously
by the simulations in 38 parameters.  However, here,
for simplicity, we just optimized two 
parameters,  
$V_1(\psi^{(k)})$ and $V_1(\psi'^{(k)})$,  
for each amino-acid residue $k$ separately,
keeping the other $V_1$ values as the original values.
In order to obtain the optimal parameters, we performed Monte Carlo simulations 
of two parameters ($V_1$ of $\psi$ and $\psi'$) for 
the 19 amino-acid residues except for proline. 
In Table \ref{table-opt-parameters}, the optimized parameters 
are listed.

\begin{table}
\caption{Optimized $V_1/2$ parameters for the main-chain dihedral 
angles $\psi$ and $\psi'$ 
for the 19 amino-acid residues (except for proline) 
in Eq.~(\ref{ene_torsion3_eachamino}). 
The rest of the parameters are taken to be the same as
in the original ff03 force field.
The original amino-acid-independent values are also listed
for reference.}
\label{table-opt-parameters}
\vspace{0.3cm}
\begin{center}
\begin{tabular}{lrcr} \hline
                 & $\psi$ (N-C$_{\alpha}$-C-N)  & ~~~ & $\psi'$ (C$_{\beta}$-C$_{\alpha}$-C-N)      \\ \hline
original ff03    & 0.6839      && 0.7784        \\ \hdashline
Ala              & 0.122      && 0.150        \\ 
Arg              & 0.409      && 0.200        \\ 
Asn              & $-0.074$    && $-0.162$       \\ 
Asp              & $-0.137$    && 0.182        \\ 
Cys              & 0.361      && 0.089        \\ 
Gln              & 0.144      && $-0.024$       \\ 
Glu              & 0.180      && 0.152        \\ 
Gly              & 0.258      && -----         \\ 
His              & 0.020      && 0.237        \\ 
Ile              & 0.643      && 0.194        \\ 
Leu              & 0.382      && 0.257        \\ 
Lys              & 0.222      && 0.042        \\ 
Met              & 0.141      && 0.346        \\ 
Phe              & $-0.010$     && 0.553        \\ 
Ser              & $-0.248$     && 0.475        \\ 
Thr              & 0.512      && 0.328        \\ 
Trp              & 0.027      && 0.477        \\ 
Tyr              & 0.082      && 0.652        \\ 
Val              & 0.142      && 0.590        \\ \hline
\end{tabular}
\end{center}
\end{table}

In order to check the force-field parameters obtained by our optimization method, 
we performed the folding simulations using two peptides, namely, C-peptide and G-peptide.

For the folding simulations, we used replica-exchange molecular 
dynamics (REMD) \cite{REMD}.
REMD is one of the generalized-ensemble algorithms, and has high 
conformational sampling efficiency by 
allowing configurations to heat up and cool down while maintaining proper Boltzmann distributions.
We used the AMBER11 program package \cite{amber_prog11}.
The unit time step was set to 2.0 fs, and the bonds involving 
hydrogen atoms were constrained by SHAKE algorithm \cite{SHAKE}.
Each simulation was carried out for 30.0 ns (hence, it consisted of
15,000,000 MD steps) with 16 replicas by using Langevin dynamics.
The exchange procedure for each replica were performed every 3,000 MD steps.
The temperature was distributed exponentially: 
650, 612, 577, 544, 512, 483, 455, 
428, 404, 380, 358, 338, 318, 300, 282, and 266 K.
As for solvent effects, we used the GB/SA model 
in the AMBER program package ($igb = 5$ and $gbsa = 1$)
\cite{gbsa_igb5,gbsa_sa1}. 
The initial conformations for each peptide were fully
extended ones for all the replicas.
The REMD simulations were performed 
with different sets of randomly generated initial velocities
for each replica.

\begin{figure}
\begin{center}
\resizebox*{12cm}{!}{\includegraphics{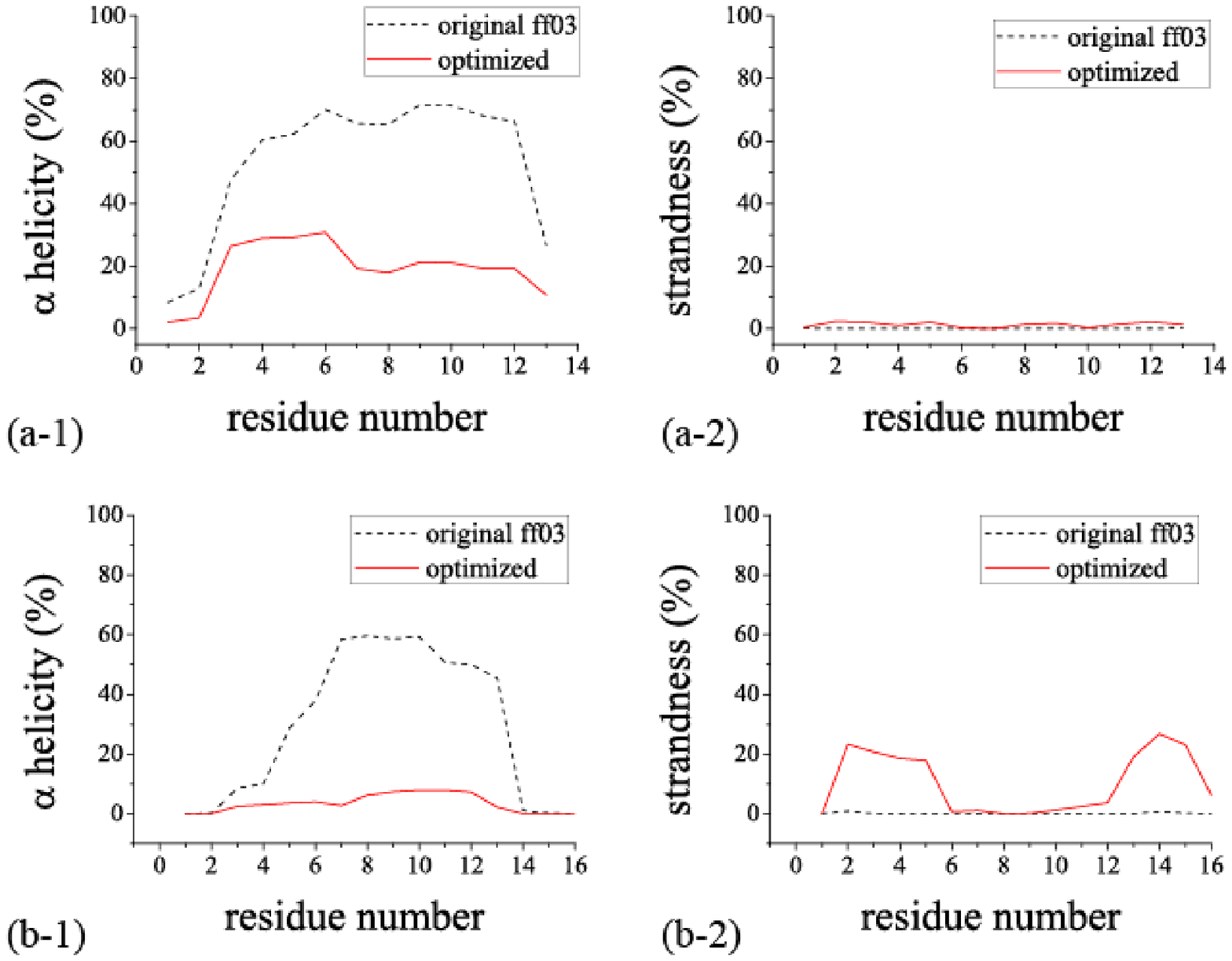}}%
\caption{$\alpha$-helicity (a-1) and $\beta$-strandness (a-2) of C-peptide 
and $\alpha$-helicity (b-1) and $\beta$-strandness (b-2) 
of G-peptide as functions of the residue number at 300 K. These values 
were obtained from the REMD simulations.
Normal and dotted curves stand for the optimized and the
original AMBER ff03 force fields, respectivery.}
\label{fig_secondary_alpha_beta}
\end{center}
\end{figure}

\begin{figure}
\begin{center}
\resizebox*{12cm}{!}{\includegraphics{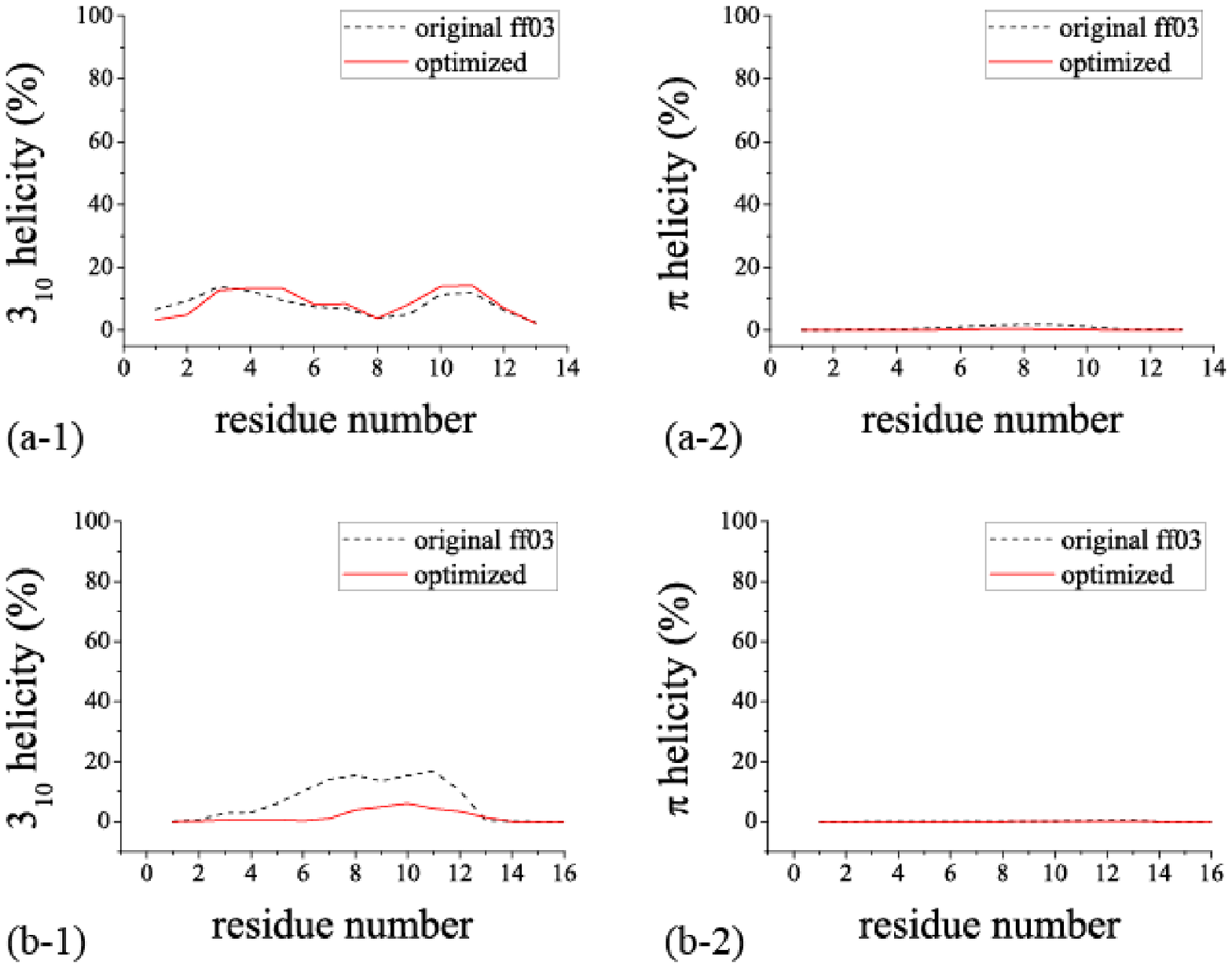}}%
\caption{3$_{10}$-helicity (a-1) and $\pi$-helicity (a-2) of C-peptide 
and 3$_{10}$-helicity (b-1) and $\pi$-helicity (b-2) 
of G-peptide as functions of the residue number at 300 K. These values 
were obtained from the REMD simulations.
Normal and dotted curves stand for the optimized and the
original AMBER ff03 force fields, respectivery.}
\label{fig_secondary_310_pi}
\end{center}
\end{figure}

In Fig.~\ref{fig_secondary_alpha_beta}, $\alpha$-helicity and 
$\beta$-strandness of the two peptides obtained 
from the REMD simulations are shown.
We checked the secondary-structure formations by using the DSSP 
program \cite{DSSP},
which is based on the formations of the intra-main-chain hydrogen bonds.
As is shown in Fig.~\ref{fig_secondary_alpha_beta}, 
for the original AMBER ff03 force field, the $\alpha$-helicity is 
clearly higher than 
the $\beta$-strandness not only in C-peptide but also in G-peptide.
Namely, the original AMBER ff03 force field clearly favors 
$\alpha$-helix and does not favor 
$\beta$-structure.
On the other hand, for the optimized force field, in the case 
of C-peptide, the $\alpha$-helicity is 
higher than the $\beta$-strandness, and in the case of G-peptide, 
the $\beta$-strandness is higher than the $\alpha$-helicity.
We conclude that these results obtained from the optimized force field 
are in better agreement with 
the experimental results in comparison with the original force field.
In Fig.~\ref{fig_secondary_310_pi}, 3$_{10}$-helicity and $\pi$-helicity 
of two peptides obtained from 
the REMD simulations are shown.
For 3$_{10}$ helicity, there is no large difference for both force fields 
in C-peptide, and in the case of G-peptide, 
the value of the optimized force field slightly decreases in comparison 
with the original force field.
$\pi$-helicity has almost no value in the both cases of the original 
and optimized force fields in two peptides.

\begin{figure}
\begin{center}
\resizebox*{12cm}{!}{\includegraphics{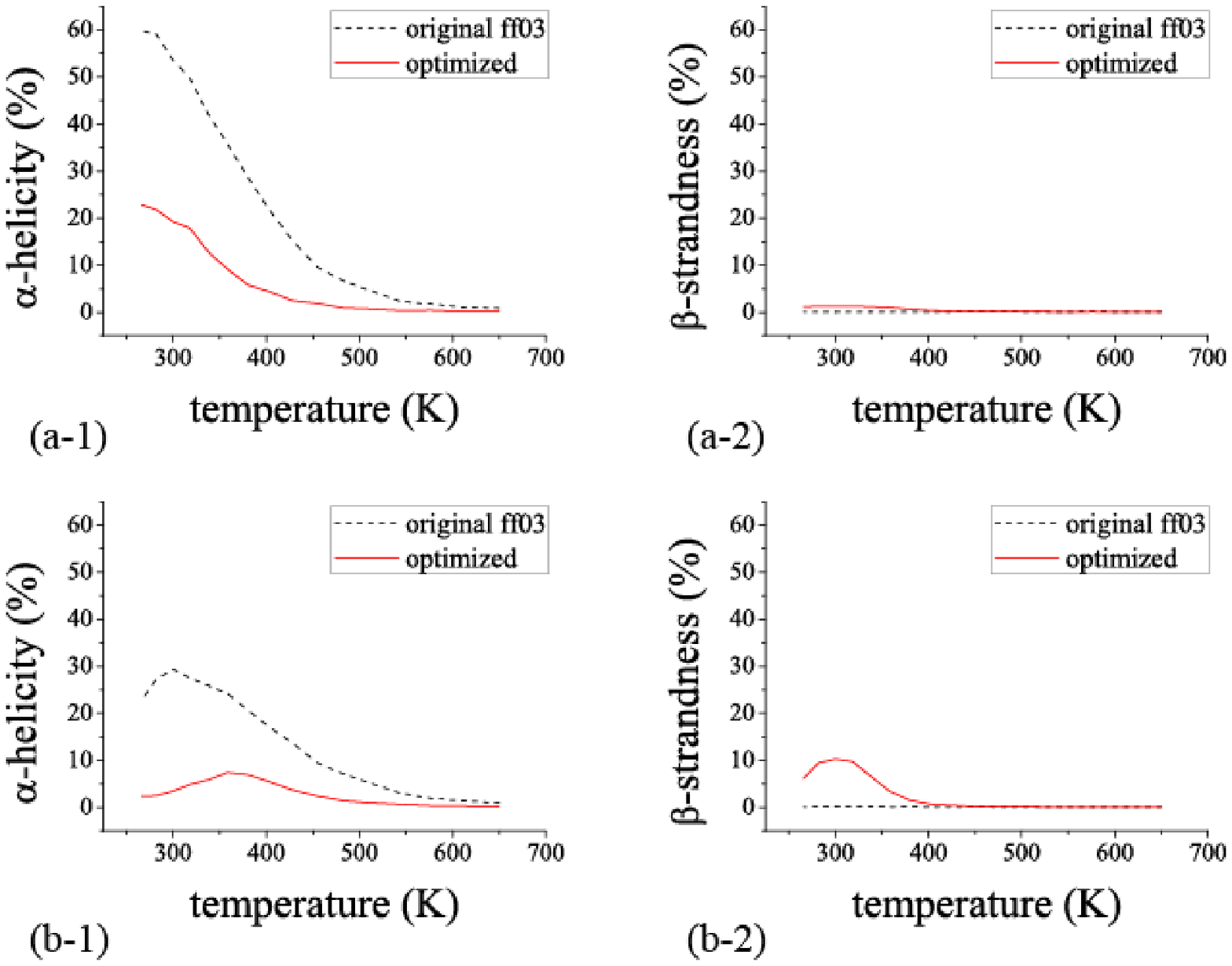}}%
\caption{$\alpha$-helicity (a-1) and $\beta$-strandness (a-2) of C-peptide 
and $\alpha$-helicity (b-1) and $\beta$-strandness (b-2) 
of G-peptide as functions of temperature. These values were
obtained from the REMD simulations.
Normal and dotted curves stand for the optimized and the 
original AMBER ff03 force fields, respectivery.}
\label{fig_temp_secondary_alpha_beta}
\end{center}
\end{figure}

In Fig.~\ref{fig_temp_secondary_alpha_beta}, $\alpha$-helicity and 
$\beta$-strandness as functions of temperature for the
two peptides obtained from the REMD simulations are shown.
For $\alpha$-helicity, the values of both force fields decrease gradually from low temperature 
to high temperature in the case of C-peptide.
On the other hand, in the case of G-peptide, there are small peaks 
at around 300 K and 358 K 
for the original and optimized force fields, respectively.
For $\beta$-strandness, in the case of C-peptide, it is almost zero 
for both force fields.
In the case of G-peptide, for the optimized force field, there is clearly 
a peak around 300 K.

\subsection{Optimization of force-field parameters}

\subsubsection{Use of force acting on each atom in the PDB coordinates \cite{SO1,SO2,SO3,SO8}}



We now present the results of our force-field optimizations.
In Step 1 of the flowchart in Fig.~\ref{fig_flow},
we chose 100 PDB files ($N=100$) from X-ray experiments with resolution 1.8 \AA~ or better and with
less than 200 residues 
(the average number of resiudes is 120.4) from PISCES \cite{pisces}.
Their PDB codes are
2LIS, 1EP0, 
1TIF, 1EB6, 1C1L, 1CCW, 2PTH, 1I6W, 1DBF, 1KPF, 1LRI, 1AAP, 1C75, 1CC8, 
1FK5, 1KQR, 1K1E, 1CZP, 1GP0, 1KOI, 1IQZ, 3EBX, 1I40, 1EJG, 1AMM, 1I07, 
1GK8, 1GVP, 1M4I, 1EYV, 1E29, 1I2T, 1VCC, 1FM0, 1EXR, 1GUT, 1H4X, 1GBS, 
1B0B, 119L, 1IFC, 1DLW, 1EAJ, 1GGZ, 1JR8, 1RB9, 1VAP, 1JZG, 1M55, 1EN2, 
1C9O, 2ERL, 1EMV, 1F41, 1EW6, 2TNF, 1IFR, 1JSE, 1KAF, 1HZT, 1HQK, 1FXL, 
1BKR, 1ID0, 1LQV, 1G2R, 1KR7, 1QTN, 1D4O, 1EAZ, 2CY3, 1UGI, 1IJV, 
3VUB, 1BZP, 1JYR, 1DZK, 1QFT, 1UTG, 2CPG, 1I6W, 1C7K, 1I8O, 1LO7, 1LNI, 
1EQO, 1NDD, 1HD2, 3PYP, 1FD3, 1DK8, 1WHI, 1FAZ, 4FGF, 2MHR, 1JB3, 2MCM, 
1IGD, 1C5E, and 1JIG.
  
In Step 2 of the flowchart, we used the routine in the TINKER package to add hydrogen atoms to the PDB 
coordinates.
The force fields that we optimized are the AMBER parm94 version \cite{parm94}, 
parm96 version \cite{parm96_3}, parm99 version \cite{parm99}, CHARMM version 22 \cite{charmm}, 
and OPLS-AA \cite{opls1}.
We have optimized only two sets of parameters.  The first set is the partial-charge 
parameters ($q_i$ in Eqs.~(\ref{ene_nonbond_optF1}) and (\ref{ene_gbsa3_optF1})).
In order to simplify the constraint-imposing processes on the total charge, 
we did not optimize the charge of one of the hydrogen atoms (HN) in proline when it is 
located at tht N-terminus.
In the original X-ray data, hydrogen coordinates are missing, and in the case of neutral 
histidine whether $N_\delta$ and $N_\epsilon$ are protonated or not is non-trivial to determine.
Because we want to deal with as many as PDB data as possible, we treated all the histidine 
residues as positively charged histidine for simplicity.
Among the five force fields, AMBER has the largest number of remaining partial-charge parameters (602).
We thus optimized these 602 parameters for all the five force fields.
The second set of parameters that we optimized is the backbone torsion-energy parameters
($V_a$, $V_b$, and $V_c$ in Eq.~(\ref{ene_main_optF1})) and there are six such parameters
(three each for $\phi$ and $\psi$).

As explained in detail above, the coodinates of the 100 proteins molecules 
have been prepared (Steps 1 and 2 of the flowchart in Fig. \ref{fig_flow}).
The coordinate refinement in Step 3 of the flowchart was then carried out with 
the constraint in Eq.~(\ref{ene_constr_optF1}) on the heavy atoms.
As for the force constant $K_x$ in Eq.~(\ref{ene_constr_optF1}), we have some freedom for the 
choice of the values.
Our choice is: $K_x$ should be of the same order as $K_l$ in the bond-stretching term in 
Eq.~(\ref{ene_bond_optF1}).
The force constant $K_l$ in AMBER varies from 166 kcal/mol/\AA$^2$ to 656 kcal/mol/\AA$^2$, 
and that in CHARMM varies from 173 kcal/mol/\AA$^2$ to 650 kcal/mol/\AA$^2$.
Hence, in our first trial we set $K_x =$ 100 kcal/mol/\AA$^2$.

\begin{figure}
\begin{center}
\resizebox*{12cm}{!}{\includegraphics{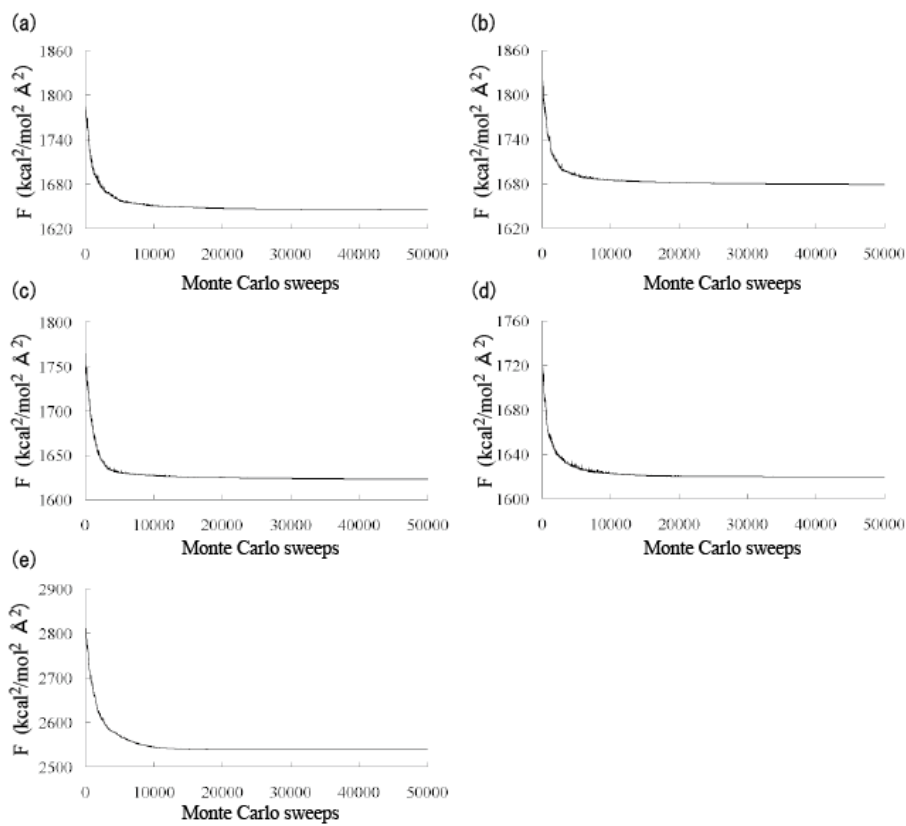}}%
\caption{Time series of MC simulated annealing simulations in force-field parameter 
space of partial charges for AMBER parm94 (a), AMBER parm96 (b), AMBER parm99 (c), CHARMM version 22 (d), 
and OPLS-AA (e).
The ordinate is the value of $F$ in Eq.~(\ref{F_optF1}).}
\label{fig_simcharge2}
\end{center}
\end{figure}

In Step 4 of the flowchart,
we performed the optimization of the 602 partial-charge parameters by 
MC simulated annealing.
Namely, we minimized $F$ in Eq.~(\ref{F_optF1}) by 
MC simulated annealing simulations of these parameters
(the parameters were updated and the updates were accepted or rejected
according to the Metropolis criterion).
For this we introduced an effective ``temperature'' for the parameter space.
The simulation run consisted of 50,000 MC sweeps with the temperature
decreased exponentially from 20 to 0.01.
The simulation was repeated 10 times with different initial random numbers.
The time series of $F$ from these simulations are
shown in Figs.~\ref{fig_simcharge2}(a)--\ref{fig_simcharge2}(e).
We see that $F$ decreases quickly in the beginning until about 5,000
MC sweeps and then it decreases 
very slowly for all force fields; the total number of MC sweeps (50,000) seems sufficient.
The optimized partial charges are taken from those that resulted in
the lowest $F$ value.

In Tables \ref{table-gly2}--\ref{table-glu2},
five examples (glycine, alanine, and glutamic acid)
of the obtained partial
charges together with the original force-field values are listed.
We see from these tables that the values of the partial charges have not changed a lot.
Although the sign of the partial charges remains the same for those with large magnitude,
charges with small magnitude sometimes change their signs 
(see, for example, CA of glycine and CG of glutamic acid).

\begin{table}
\caption{Partial-charge parameters of glycine. 
AMB, CHA, and OPLS respectively stand for the original
AMBER, CHARMM version 22, and OPLS-AA force fields.
Opt(94), Opt(96), Opt(99), Opt(CH), and Opt(OP) are the optimized AMBER parm94, AMBER parm96, 
AMBER parm99, CHARMM version 22, and OPLS-AA, respectively.}
\label{table-gly2}
\vspace{0.3cm}
\begin{center}
\begin{tabular}{lrrrrrrrr} \hline
Atom &  AMB     &   Opt(94)  &   Opt(96) &   Opt(99) &   CHA  & Opt(CH)   &    OPLS   &   Opt(OP)  \\ \hline
N    & $-0.4157$  & $-0.3471$  & $-0.3614$ & $-0.3506$ & $-0.4700$ & $-0.4381$ & $-0.5000$ & $-0.5153$  \\
CA   & $-0.0252$  &   0.0175   &   0.0148  &   0.0166  & $-0.0200$ &   0.0185  &   0.0800  &   0.0909   \\
C    &   0.5973   &   0.5526   &   0.5698  &   0.5577  &   0.5100  &   0.5309  &   0.5000  &   0.6459   \\
HN   &   0.2719   &   0.2492   &   0.2509  &   0.2480  &   0.3100  &   0.3004  &   0.3000  &   0.2615   \\
O    & $-0.5679$  & $-0.5980$  & $-0.5977$ & $-0.5983$ & $-0.5100$ & $-0.5491$ & $-0.5000$ & $-0.5546$  \\
HA   &   0.0698   &   0.0629   &   0.0618  &   0.0633  &   0.0900  &   0.0687  &   0.0600  &   0.0358   \\ \hline
Total&   0.0000   &   0.0000   &   0.0000  &   0.0000  &   0.0000  &   0.0000  &   0.0000  &   0.0000   \\ \hline
\end{tabular}
\end{center}
\end{table}


\begin{table}
\caption{Partial-charge parameters of alanine. See the caption in Table~\ref{table-gly2}.}
\label{table-ala2}
\vspace{0.3cm}
\begin{center}
\begin{tabular}{lrrrrrrrr} \hline
Atom &  AMB     &   Opt(94) &   Opt(96) &  Opt(99)  &   CHA  & Opt(CH)   &  OPLS     &  Opt(OP)   \\ \hline
N    & $-0.4157$  & $-0.3354$ & $-0.3483$ & $-0.3407$ & $-0.4700$ & $-0.3909$ & $-0.5000$ & $-0.5224$  \\
CA   &   0.0337   &   0.0545  &   0.0547  &   0.0511  &   0.0700  &   0.0427  &   0.1400  &   0.1301   \\
C    &   0.5973   &   0.5141  &   0.5240  &   0.5235  &   0.5100  &   0.5215  &   0.5000  &   0.6687   \\
HN   &   0.2719   &   0.2323  &   0.2346  &   0.2317  &   0.3100  &   0.2709  &   0.3000  &   0.2610   \\
O    & $-0.5679$  & $-0.5703$ & $-0.5599$ & $-0.5778$ & $-0.5100$ & $-0.5417$ & $-0.5000$ & $-0.5567$  \\
HA   &   0.0823   &   0.0901  &   0.0912  &   0.0900  &   0.0900  &   0.0741  &   0.0600  &   0.0786   \\
CB   & $-0.1825$  & $-0.0453$ & $-0.0470$ & $-0.0501$ & $-0.2700$ & $-0.2718$ & $-0.1800$ & $-0.0701$  \\
HB   &   0.0603   &   0.0200  &   0.0169  &   0.0241  &   0.0900  &   0.0984  &   0.0600  &   0.0036   \\ \hline
Total&   0.0000   &   0.0000  &   0.0000  &   0.0000  &   0.0000  &   0.0000  &   0.0000  &   0.0000   \\ \hline
\end{tabular}
\end{center}
\end{table}

\begin{table}
\caption{Partial-charge parameters of glutamic acid. See the caption in Table~\ref{table-gly2}.}
\label{table-glu2}
\vspace{0.3cm}
\begin{center}
\begin{tabular}{lrrrrrrrr} \hline
Atom &  AMB     &   Opt(94)  &   Opt(96) &  Opt(99)  &   CHA  & Opt(CH)   &    OPLS   &   Opt(OP)  \\ \hline
N    & $-0.5163$  & $-0.4248$  & $-0.4376$ & $-0.4302$ & $-0.4700$ & $-0.3961$ & $-0.5000$ & $-0.5401$  \\
CA   &   0.0397   &   0.0583   &   0.0553  &   0.0554  &   0.0700  &   0.0423  &   0.1400  &   0.1320   \\
C    &   0.5366   &   0.4728   &   0.4873  &   0.4817  &   0.5100  &   0.5249  &   0.5000  &   0.6538   \\
HN   &   0.2936   &   0.2595   &   0.2620  &   0.2590  &   0.3100  &   0.2845  &   0.3000  &   0.2626   \\ 
O    & $-0.5819$  & $-0.6181$  & $-0.6107$ & $-0.6248$ & $-0.5100$ & $-0.5603$ & $-0.5000$ & $-0.5777$  \\
HA   &   0.1105   &   0.1232   &   0.1232  &   0.1221  &   0.0900  &   0.0837  &   0.0600  &   0.0670   \\
CB   &   0.0560   &   0.1226   &   0.1170  &   0.1217  & $-0.1800$ & $-0.1634$ & $-0.1200$ & $-0.0517$  \\
HB   & $-0.0173$  & $-0.0333$  & $-0.0334$ & $-0.0300$ &   0.0900  &   0.0943  &   0.0600  &   0.0418   \\
CG   &   0.0136   & $-0.0678$  & $-0.0716$ & $-0.0659$ & $-0.2800$ & $-0.2870$ & $-0.2200$ & $-0.2185$  \\
HG   & $-0.0425$  & $-0.0300$  & $-0.0297$ & $-0.0299$ &   0.0900  &   0.1160  &   0.0600  &   0.0437   \\
CD   &   0.8054   &   0.8293   &   0.8340  &   0.8292  &   0.6200  &   0.5465  &   0.7000  &   0.7320   \\
OE   & $-0.8188$  & $-0.8142$  & $-0.8163$ & $-0.8142$ & $-0.7600$ & $-0.7479$ & $-0.8000$ & $-0.8152$  \\ \hline
Total& $-1.0000$  & $-1.0000$  & $-1.0000$ & $-1.0000$ & $-1.0000$ & $-1.0000$ & $-1.0000$ & $-1.0000$  \\ \hline
\end{tabular}
\end{center}
\end{table}

In Step 5 of the flowchart, 
the original coordinates obtained in Step 2 were again refined
with the constraints in Eq.~(\ref{ene_constr_optF1}),
but this time the optimized parameters from Step 4 were used.
This time we used the value $K_x=500$ kcal/mol/\AA$^2$.
For all force fields, the average RMSD of the 100 proteins is 0.012 \AA, and the 
coordinates of heavy atoms have little changed.

In Step 6 of the flowchart, 
we carried out the optimization of the six torsion-energy parameters 
($V_a$, $V_b$, and $V_c$ in Eq.~(\ref{ene_main_optF1}) for both $\phi$ and $\psi$) 
by minimizing $F$ in Eq.~(\ref{F_optF1}) with
MC simulated annealing simulations in this parameter space.
The simulation run consisted of 10,000 MC sweeps with the temperature 
decreasing from 1,000 to 1.0.
The simulation was repeated six times with different random numbers.
We stopped after six trials because the convergence was very good.
The optimized torsion-energy parameters are taken from those that resulted in
the lowest $F$ value.
The obtained torsion-energy parameters 
are listed in Tables~\ref{table-torsion_phi} and \ref{table-torsion_psi}.

\begin{table}
\caption{Torsion parameters of $\phi$ angle. Parm94, Parm96, Parm99, CHARMM, and OPLS are 
AMBER parm94, AMBER parm96, AMBER parm99, CHARMM version 22, and OPLS-AA force fields, respectively.
``Optimized'' stands for the corresponding optimized force field.}
\label{table-torsion_phi}
\vspace{0.3cm}
\begin{center}
\begin{tabular}{lrrrrrrrrr} \hline
Force field &~~~~~$V_a$~~~~~&~~~$n_a$~~~&~~~~~$\gamma_a$~~~~~&~~~~~$V_b$~~~~~&~~~$n_b$~~~&~~~~~$\gamma_b$~~~~~&~~~~~$V_c$~~~~~&~~~$n_c$~~~&~~~~~$\gamma_c$~~~~~ \\ \hline
Parm94      &  0.200  &  2  &  180.0    &  -----  &-----&  -----   &  -----  &-----&  -----     \\
Optimized   &  0.191  &  1  &    0.0    &  0.146  &  2  &  180.0   &$-0.223$ &  3  &    0.0     \\ \hline
Parm96      &  0.850  &  1  &    0.0    &  0.300  &  2  &  180.0   &  -----  &-----&  -----     \\
Optimized   &  1.182  &  1  &    0.0    &  0.359  &  2  &  180.0   &$-0.410$ &  3  &    0.0     \\ \hline
Parm99      &  0.800  &  1  &    0.0    &  0.850  &  2  &  180.0   &  -----  &-----&  -----     \\
Optimized   &  1.380  &  1  &    0.0    &  0.599  &  2  &  180.0   &$-0.330$ &  3  &    0.0     \\ \hline
CHARMM      &  0.200  &  1  &  180.0    &  -----  &-----&  -----   &  -----  &-----&  -----     \\
Optimized   &$-0.047$ &  1  &  180.0    &  0.240  &  2  &  180.0   &$-0.015$ &  3  &    0.0     \\ \hline
OPLS        &$-2.365$ &  1  &    0.0    &  0.912  &  2  &  180.0   &$-0.850$ &  3  &    0.0     \\
Optimized   &  0.502  &  1  &    0.0    &  1.811  &  2  &  180.0   &$-0.567$ &  3  &    0.0     \\ \hline
\end{tabular}
\end{center}
\end{table}

\begin{table}
\caption{Torsion parameters of $\psi$ angle. See the caption in Table~\ref{table-torsion_phi}.}
\label{table-torsion_psi}
\vspace{0.3cm}
\begin{center}
\begin{tabular}{lrrrrrrrrr} \hline
Force field &~~~~~$V_a$~~~~~&~~~$n_a$~~~&~~~~~$\gamma_a$~~~~~&~~~~~$V_b$~~~~~&~~~$n_b$~~~&~~~~~$\gamma_b$~~~~~&~~~~~$V_c$~~~~~&~~~$n_c$~~~&~~~~~$\gamma_c$~~~~~ \\ \hline
Parm94      &  0.750  &  1  & 180.0    &  1.350  &  2  &  180.0   &  0.400  &  4  & 180.0    \\
Optimized   &$-0.368$ &  1  & 180.0    &  1.658  &  2  &  180.0   &  0.265  &  4  & 180.0    \\ \hline
Parm96      &  0.850  &  1  &   0.0    &  0.300  &  2  &  180.0   &  -----  &-----& -----    \\
Optimized   &  0.039  &  1  &   0.0    &  1.011  &  2  &  180.0   &  0.104  &  3  &    0.0   \\ \hline
Parm99      &  1.700  &  1  & 180.0    &  2.000  &  2  &  180.0   &  -----  &-----& -----    \\
Optimized   &  0.228  &  1  & 180.0    &  1.684  &  2  &  180.0   &$-0.031$ &  3  &    0.0   \\ \hline
CHARMM      &  0.600  &  1  &   0.0    &  -----  &-----&  -----   &  -----  &-----& -----    \\
Optimized   &  0.321  &  1  &   0.0    &  0.028  &  2  &  180.0   &  0.251  &  3  &    0.0   \\ \hline
OPLS        &  1.816  &  1  &   0.0    &  1.222  &  2  &  180.0   &  1.581  &  3  &    0.0   \\
Optimized   &  0.880  &  1  &   0.0    &  1.479  &  2  &  180.0   &  0.952  &  3  &    0.0   \\ \hline
\end{tabular}
\end{center}
\end{table}

In the present work, we stopped our process in Step 6 of the flowchart
and did not iterate the optimizations.

\begin{figure}
\begin{center}
\resizebox*{12cm}{!}{\includegraphics{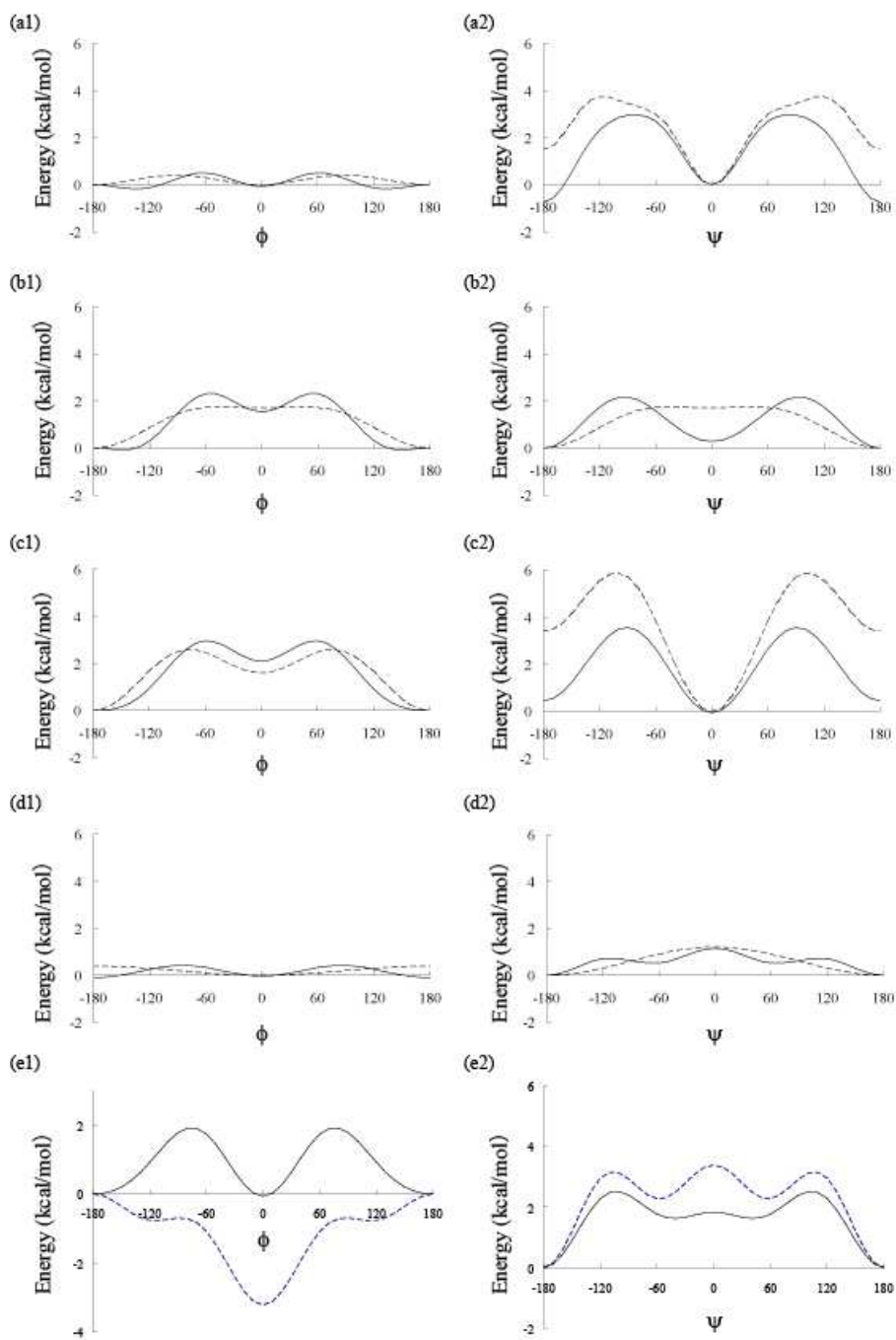}}%
\caption{Backbone torsion-energy curves as functions of 
$\phi$ (in degrees) and $\psi$ (in degrees).
The force fields are
AMBER parm94 (a), AMBER parm96 (b), AMBER parm99 (c), CHARMM version 22 (d), and OPLS-AA (e).
The results for the original force fields are represented by dotted curves, and
those for the optimized force fields are by solid curves.}
\label{fig_tc}
\end{center}
\end{figure}

In order to examine how much the torsion-energy terms have changed after optimizations, we depict them 
in Fig.~\ref{fig_tc} (we remark that the error of factor 2 in the ordinate 
of Fig.~5 (e1) in Ref.~\cite{SO2} is corrected here).
Although the behaviors of the original force fields are quite different, those of the optimized force 
fields are rather similar.
For example, the optimized torsion-energy curves for $\phi$ angles have two maximum peaks around $\phi \sim -60^\circ$ 
and $+60^\circ$ and a local minimum at $\phi = 0^\circ$, while those for $\psi$ angle have two peaks around 
$\psi \sim -100^\circ$ and $+100^\circ$ and a local minimum at $\psi = 0^\circ$ (the exceptions are those 
for CHARMM version 22 and OPLS-AA, which give the global maximum and a local maximum, respectively, 
at $\psi = 0^\circ$).
These results suggest that our optimizations of the torsion-energy term yield a tendency for convergence 
towards a common function.
Some remark is in order.  The case for the optimized CHARMM is the most distinct from other optimized parameters 
in the sense that it gives the global maximum as $\psi = 0^\circ$ whereas that for other cases lie 
around $\psi \sim -100^\circ$ and $+100^\circ$.

\begin{figure}
\begin{center}
\resizebox*{8cm}{!}{\includegraphics{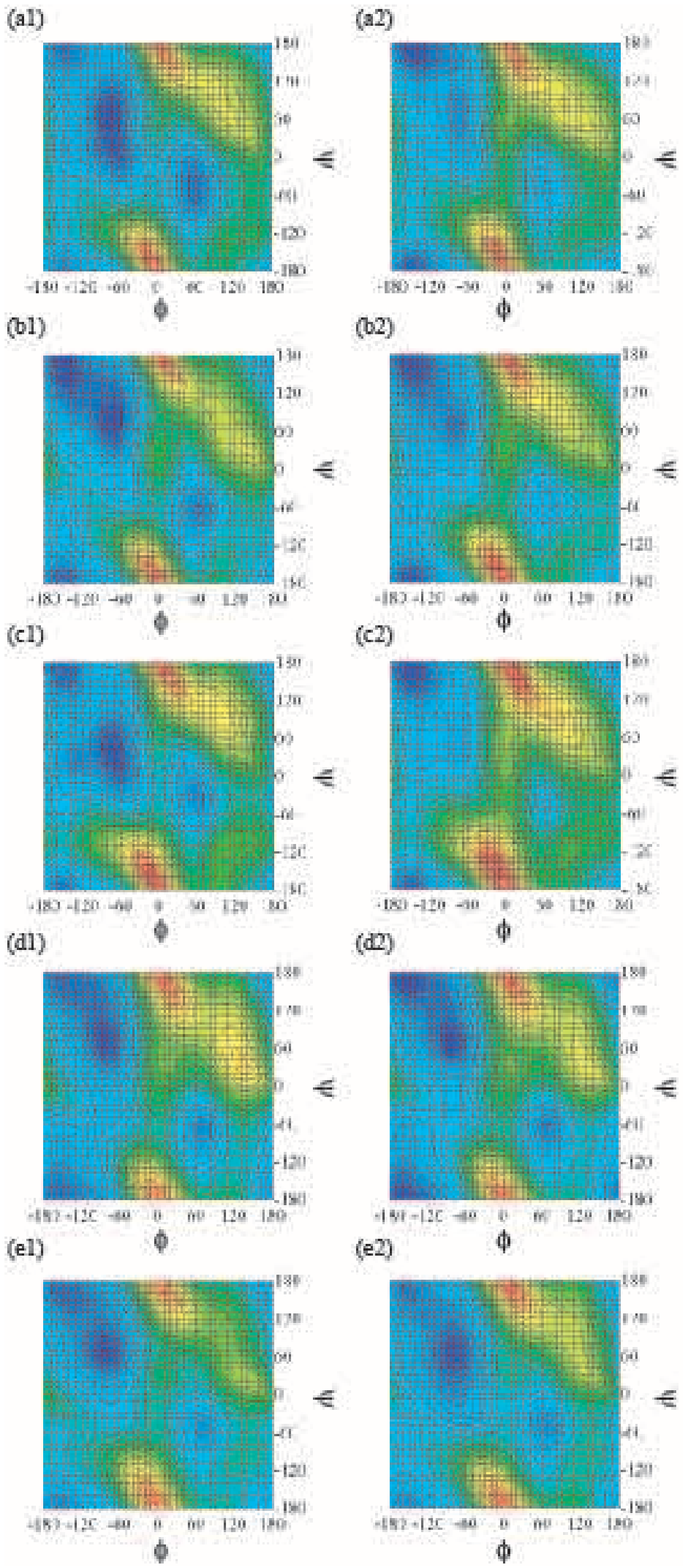}}%
\caption{Potential-energy surfaces of alanine dipeptide.
The force fields are the original
AMBER parm94 (a1), AMBER parm96 (b1), AMBER parm99 (c1), CHARMM version 22 (d1), and OPLS-AA (e1), 
and the corresponding optimized parameters (a2)-(e2). 
The contour maps were evaluated every $10^\circ$ of $\phi$ and $\psi$ angles and
plotted every 1 kcal/mol, 
after minimizing the total potential energy in vacuum with the backbone structures fixed.
The bluer the color is, the lower the potential energy surface is.
As the potential-energy value increases, the color changes from blue to green,
to yellow, and to red.}
\label{fig_rama1}
\end{center}
\end{figure}

In Fig.~\ref{fig_rama1} the potential-energy surfaces of the alanine 
dipeptide (ACE-ALA-NME) are shown for the 10 force-field parameters: the original AMBER parm94, 
AMBER parm96, AMBER parm99, CHARMM version 22, OPLS-AA, and the corresponding optimized parameters.
According to the {\it ab initio} quantum mechanical calculations, there exist three local-minimum states
in the energy surface \cite{parm94}.
They are conformers C$_{7eq}$, C$_5$, and C$_{7ax}$, which correspond to 
$(\phi, \psi) \sim (-80^\circ, +80^\circ)$, $(-160^\circ, +160^\circ)$, and $(+75^\circ, -60^\circ)$, 
respectively (C$_{7eq}$ is the global-minimum state).
We remark that these are the results of quantum chemistry calculations in vacuum, and so it is not
clear how reliable the results are to represent the dipeptide in aqueous solution.
The results of all five original force fields in Figs.~\ref{fig_rama1}(a1)--\ref{fig_rama1}(e1) seem to 
satisfy the above conditions.
Namely, there are three local-minimum states at the locations  of C$_{7eq}$, C$_5$, and C$_{7ax}$, and 
the global-minimum state is C$_{7eq}$.
As for the results of the optimized force fields in Figs.~\ref{fig_rama1}(a2)--\ref{fig_rama1}(e2), 
those for CHARMM version 22 and OPLS-AA also satisfy the above conditions.
Those of the optimized AMBER force fields are less consistent with the quantum mechanical calculations:
C$_{7eq}$ is no longer the global-minimum state, but it is a local-minimum state.
In particular, the optimized AMBER parm99 seems to be in the greatest disagreement in the sense 
that the C$_{7eq}$ state is almost disappearing.

We now present another example of the refinement of our backbone torsion energy
in Eq.~(\ref{new_phipsi_Fourier}). 
We consider the following truncated Fourier series:
\begin{eqnarray}
{\cal E}(\phi,\psi) = a & + & b_{1} \cos \phi + c_{1} \sin \phi + b_{2} \cos 2\phi + c_{2} \sin 2\phi \nonumber\\ 
& + & d_{1} \cos \psi + e_{1} \sin \psi + d_{2} \cos 2\psi + e_{2} \sin 2\psi \nonumber\\ 
& + & f_{11} \cos \phi \cos \psi + g_{11} \cos \phi \sin \psi \nonumber\\ 
& + & h_{11} \sin \phi \cos \psi + i_{11} \sin \phi \sin \psi~.
\label{use_torsion_rmsd3}
\end{eqnarray}
This function has 13 Fourier-coefficient parameters.
We will see below that this number of Fourier terms is sufficient for
the most of our purposes \cite{SO4,SO5}, but that for some cases more number of Fourier terms are preferred.

We optimize the force-field parameters of this double Fourier series by using our optimization method.
At first, we chose 100 PDB files with resolution 2.0 \AA~or better, 
with sequence similarity of amino acid 30.0 \% or lower 
and with less than 200 residues (the average number of residues is 117.0) from PDB-REPRDB \cite{REPRDB}.
Generally, data from X-ray experiments do not have hydrogen atoms.
Therefore, we have to add hydrogen coordinates.
Many protein simulation software packages provide with routines that add hydrogen atoms to the PDB 
coordinates.
We used the TINKER program package  \cite{tinker_v2}.

\begin{figure}
\begin{center}
\resizebox*{8cm}{!}{\includegraphics{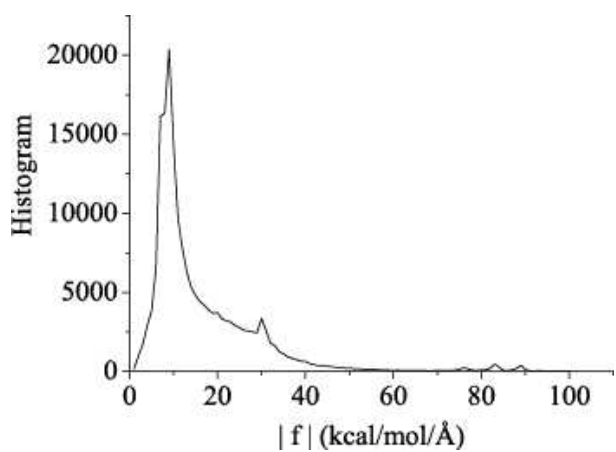}}%
\caption{The distribution of the absolute value of the forces acting on each atom in the 100 protein structures, 
which were obtained from PDB.}
\label{fig_f_hist}
\end{center}
\end{figure}

\begin{figure}
\begin{center}
\resizebox*{8cm}{!}{\includegraphics{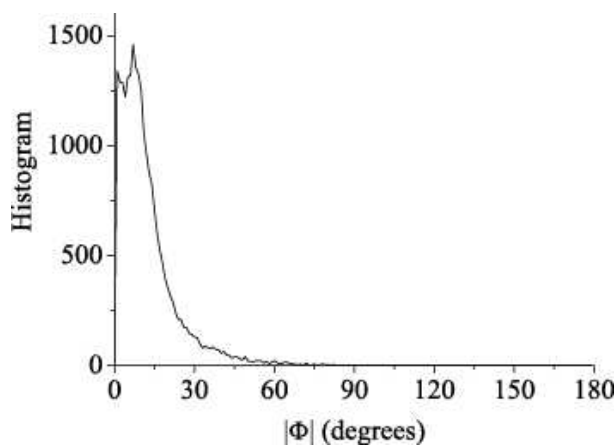}}%
\caption{The distribution of the absolute value of the backbone dihedral angles $\Phi$ ($\phi$ and $\psi$ ) 
in the 100 protein structures, 
which were obtained from PDB.}
\label{fig_dih_hist}
\end{center}
\end{figure}

In our optimization method, the minimizations of $F$ in Eq.~(\ref{F_optF1}) by the Monte Carlo (MC)
simulations of the 13 backbone-torsion-energy parameters
with 3000 MC steps were performed.
The initial values of 13 parameters were all set to be zero.
We performed MC simulations of the optimization for each $f_{\rm cut}$ value 10 times with different seeds for the random numbers.
After that, the minimum $F$ value was selected from the results of the obtained 
10 parameter sets for each case of the $f_{\rm cut}$ value.  The overall
parameter distributions were essentially the same for the 10 runs.
The maximum $f_{\rm cut}$ value was taken to be $f_{\rm cut}^{\rm max} \simeq 9.0$, which was selected from the peak point 
in the distribution of the forces acting on each atom in the 100 protein structures in Fig.~\ref{fig_f_hist}.
For the obtained several parameters, several $\Phi {\rm RMSD}$ were calculated by using Eq.~(\ref{eq_phi_rmsd_rmsd3}).
Here, if a difference between $\Phi_i^{\rm native}$ and $\Phi_i^{\rm min}$ of a backbone dihedral angle in a 
protein was more than 20 degrees,
the value was ignored. 
Because there are about 90\% of differences between $\Phi_i^{\rm native}$ and $\Phi_i^{\rm min}$ including 
less than 20 degrees.
In Fig.~\ref{fig_dih_hist}, the distribution of the backbone dihedral angles in the 100 protein structures is shown. 
Namely, we wanted to consider the majority of the differences of backbone dihedral angles.
After the calculations of several $\Phi {\rm RMSD}$, 
we select $f_{\rm cut} = 8.5$ at the minimum value of $\Phi {\rm RMSD}$ from the several those.

\begin{figure}
\begin{center}
\resizebox*{8cm}{!}{\includegraphics{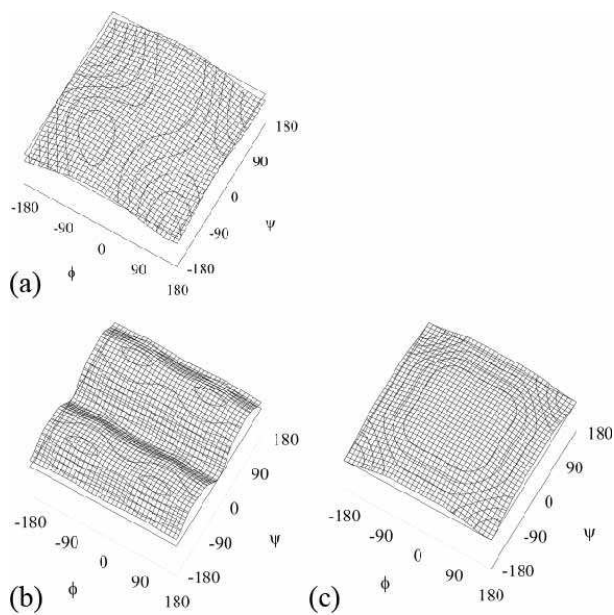}}%
\caption{The backbone-torsion-energy surfaces of the optimized force field (a), the original AMBER ff94 (b), 
and the original AMBER ff96 are shown.}
\label{fig_ene-surfaces}
\end{center}
\end{figure}

In Table.~\ref{table_opt-para}, optimized double Fourier-coefficient parameters and the corresponding 
original AMBER ff94 and ff96 force-field parameters are listed.
Here, the original AMBER ff94 has a Fourier coefficient that the number of waves is four.
Therefore, this coefficient set of the original AMBER ff94 is not complete.
Additionally, in Fig.~\ref{fig_ene-surfaces}, these backbone-torsion-energy surfaces on the Ramachandran space 
are illustrated.

\begin{table}
\caption{Fourier coefficients in Eq.~(\ref{use_torsion_Fourier}) obtained from
the numerical evaluations of the integrals in Eq.~(\ref{coe_eqn_Fourier}).
``org94'' and ``org96'' stand for the original AMBER ff94 and the original AMBER ff96, respectively,
``optimized'' stands for the optimized force field obtained by our optimization method.
Here, the original AMBER ff94 has the Fourier coefficient that the number of waves is four.
Therefore, this coefficient set of the original AMBER ff94 is not complete.}

\label{table_opt-para}
\vspace{0.3cm}
\begin{center}
\begin{tabular}{lrrr} \hline
coefficient & org94    &  org96     &  optimized   \\ \hline
$a$         &  2.700   &  2.300     &  0.000       \\
$b_1$       &  0.000   &  0.850     &  0.835       \\
$b_2$       &$-0.200$  &$-0.300$    &$-0.088$      \\
$c_1$       &  0.000   &  0.000     &$-0.327$      \\
$c_2$       &  0.000   &  0.000     &  0.100       \\
$d_1$       &$-0.750$  &  0.850     &  0.287       \\
$d_2$       &$-1.350$  &$-0.300$    &  0.019       \\
$e_1$       &  0.000   &  0.000     &$-0.160$      \\
$e_2$       &  0.000   &  0.000     &$-0.054$      \\ 
$f_{11}$    &  0.000   &  0.000     &$-0.427$      \\ 
$g_{11}$    &  0.000   &  0.000     &  0.247       \\ 
$h_{11}$    &  0.000   &  0.000     &  0.114       \\ 
$i_{11}$    &  0.000   &  0.000     &  0.603       \\ \hline
\end{tabular}
\end{center}
\end{table}

In order to test the validity of the force-field parameters obtained by our optimization methods, 
we performed folding simulations using two peptides, namely, C-peptide and G-peptide.

For the folding simulations, we used the replica-exchange molecular dynamics (REMD) 
method \cite{REMD}.
We used the TINKER program package \cite{tinker_v2} modified by us for the folding simulations.
The unit time step was set to 1.0 fs.
Each simulation was carried out for 5.0 ns (hence, it consisted of
5,000,000 MD steps) with 32 replicas.
The temperature during MD simulations was controlled by 
Nos\'e-Hoover method \cite{hoover}.
For each replica the temperature was distributed exponentially from 700 K to 250 K.
As for solvent effects, we used the GB/SA model \cite{gb1,gb2} included in the TINKER 
program package \cite{tinker_v2}.

\begin{figure}
\begin{center}
\resizebox*{12cm}{!}{\includegraphics{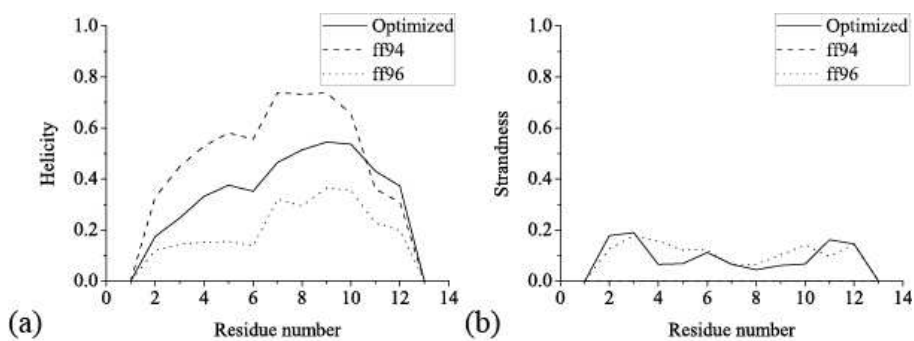}}%
\caption{Helicity (a) and strandness (b) of C-peptide as functions of the residue number. These values are obtained 
from the REMD \cite{REMD} simulations at 300K. Normal, dashed, and dotted lines stand for 
the optimized force field, the original AMBER ff94, and the original AMBER ff96, respectively.
There is only one secondary structural element (an $\alpha$-helix in residues 4 to 12) in the native structure 
(PDB ID: 1A5P). See Fig.~\ref{fig_cpgp}(a).} 
\label{fig_cp_remd_rmsd3}
\end{center}
\end{figure}

We checked the secondary-structure formations, such as the helicity and the strandness, 
 by using the DSSP program \cite{DSSP},
which is based on the formations of the intra-backbone hydrogen bonds.
Strandness means that there are $\beta$-bridge or extended strand in the
corresponding amino acid.
In Fig.~\ref{fig_cp_remd_rmsd3}, the helicity and strandness of C-peptide 
which were obtained with
the optimized force field, the original AMBER ff94 and ff96 are shown.
In comparison with the helicity of the original AMBER ff94, 
the helicity of the optimized force field decreases and in comparison with that of the original 
AMBER ff96, that of the optimized force field increases.
For the strandness, the original AMBER ff94 is almost zero, and 
both the optimized force field and the original AMBER ff96 have the low strandness.

\begin{figure}
\begin{center}
\resizebox*{12cm}{!}{\includegraphics{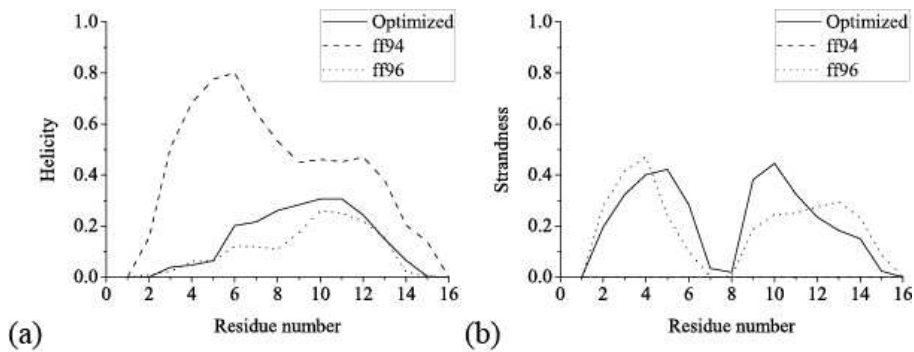}}%
\caption{Helicity (a) and strandness (b) of G-peptide as functions of the residue number. These values are obtained 
from the REMD \cite{REMD} simulations at 300K. Normal, dashed, and dotted lines stand for 
the optimized force field, the original AMBER ff94, and the original AMBER ff96, respectively.
There is only one secondary structural element (a $\beta$-hairpin; $\beta$-strands are 
in residues 2 to 6 and residues 11 to 15) in the native structure 
(PDB ID: 1PGA). See Fig.~\ref{fig_cpgp}(b).} 
\label{fig_gp_remd_rmsd3}
\end{center}
\end{figure}

In Fig.~\ref{fig_gp_remd_rmsd3}, the helicity and strandness of G-peptide 
which were obtained with
the optimized force field, the original AMBER ff94 and ff96 are shown.
The helicity of the original AMBER ff94 obviously has high value the same as the case of C-peptide.
On the other hand, the helicity of both the optimized force field and the original AMBER ff96 
decrease in comparison with the case of the original AMBER ff94.
However, in comarison with the original AMBER ff96, the optimized force field slightly favors 
the helix structure in the region around amino-acid residues 6--8.
In the experimental results, there is a turn region around residues 7--10 in G-peptide, and 
the backbone-torsion angles of the turn conformation are similar to that of the helix structure.
Therefore, we consider that this tendency is not disagreement with the experimental results.
For the strandness, the original AMBER ff94 is also almost zero the same as the case of C-peptide, 
and both the optimized force field and the original AMBER ff96 have higher values of the strandness 
than those ot the helicity.
In Fig.~\ref{fig_gp_remd_rmsd3}(b), the strandness decreases in the region around 7--8 residues
in agreement with the experiments.

These secondary-structure-forming tendencies of the optimized force field for two peptides 
agree with experimental implications in comparison with those of the original AMBER ff94 and ff96 force field.
Therefore, our improvement methods succeeded in enhancing the accuracy of the AMBER force field.

\subsubsection{Use of RMSD I \cite{SO7}}



We now present the results of the applications of our 
optimization method in Subsection 2.3.2, which we refer
to as Method 2, as well as that in Subsection 2.3.1, which we
refer to as Method 1.

At first, we chose 100 PDB files with resolution 2.0 \AA~or better, 
with sequence similarity of amino acid 30.0 \% or lower 
and with less than 200 residues (the average number of residues is 117.0) from PDB-REPRDB \cite{REPRDB}.
Next, we refine these selected 100 structures.
Generally, data from X-ray experiments do not have hydrogen atoms.
Therefore, we have to add hydrogen coordinates.
Many protein simulation software packages provide with routines that add hydrogen atoms to the PDB 
coordinates.
We used the TINKER program package  \cite{tinker_v2}.   
We thus minimize the total potential energy $E_{\rm total} = E_{\rm conf} + E_{\rm solv} + E_{\rm constr}$ 
with respect to the coordinates for each proten conformation, where $E_{\rm constr}$ is the constraint 
energy term in Eq.~(\ref{ene_constr_optF1}).
Here, $K_x$ is the force constant of the restriction and $\vec{x}_0$ are the original coordinate vectors 
of heavy atoms in PDB.
As one can see from Eq.~(\ref{ene_constr_optF1}), the coordinates of hydrogen atoms will be mainly adjusted, 
but unnatural heavy-atom coordinates will also be modified.
We performed this minimization for all the 100 protein structures separately and obtained 
100 refined structures.

We focused on the parameters of torsion-energy term, which we believe to be an important
force-field term that influences the backbone conformational preferences 
such as $\alpha$-helix structure and $\beta$-sheet structure.
For example, AMBER parm94 \cite{parm94} and AMBER parm96 \cite{parm96_3} have very different behaviors about 
the secondary-structure-forming tendencies, although these force fields differ only 
in the backbone torsion-energy terms for rotations of the backbone $\phi$ and $\psi$ angles.
Recently, new force-field parameters of the backbone torsion-energy term about $\phi$ and $\psi$ angles 
have been developed, which are, e.g., AMBER ff99SB \cite{parm99SB}, AMBER ff03 \cite{parm03}, and CHARMM 22/CMAP \cite{CMAP}.

The force field that we optimized is the OPLS-UA \cite{oplsua}.
The torsion-energy term $E_{\rm torsion} (\Phi)$ for this force field is 
given by Eq.~(\ref{ene_torsion_optF1}).
We performed 
the force-field parameter optimizations that correspond to the following torsion angles 
by Methods 1 and/or 2.

\begin{enumerate}
\item N--C$_\alpha$--C$_\beta$--C$_\gamma$ and C--C$_\alpha$--C$_\beta$--C$_\gamma$ ($\chi_1$) by Method 2 \\

\item C--N--C$_\alpha$--C ($\phi$), N--C$_\alpha$--C--N ($\psi$), C--N--C$_\alpha$--C$_\beta$ and N--C--C$_\alpha$--C$_\beta$ by Methods 1 and 2 \\

\item C--N--C$_\alpha$--C$_\beta$ by Method 2 \\

\item N--C$_\alpha$--C--N by Method 2 \\

\item C$_\alpha$--C$_\beta$--C$_\gamma$--C$_\delta$ ($\chi_2$ of Glu) by Methods 1 and 2 \\
\end{enumerate}

Here, we also optimized the force-field parameters of $\chi_2$ of Glu.
The reason is given below.

\begin{figure}
\begin{center}
\resizebox*{7cm}{!}{\includegraphics{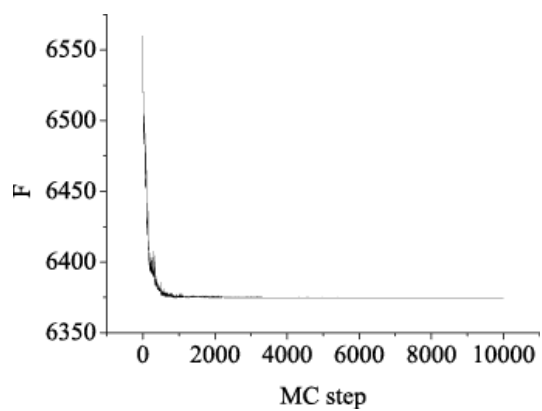}}%
\caption{Time series of Monte Carlo simulated annealing simulations in force-field parameter space of 
torsion-energy for OPLS-UA. The ordinate is the value of $F$ in Eq.~(\ref{F_optF1}).}
\label{fig_sim_method1}
\end{center}
\end{figure}

In Method 1, the minimizations of $F$ in Eq.~(\ref{F_optF1}) by the Monte Carlo (MC)
simulated annealing simulations of the torsion-energy parameters
with 10000 MC steps were performed 10 times.
Here, we neglected the improper-torsion-energy contributions to $E_{\rm conf}$ in Eq.~(\ref{ene_tot_optF1}).
In order to make a better force field, we have to optimize many force-field parameters.
However, we ignored the uncertainty of improper-torsion-energy parameters with 
this optimization, because we wanted to focus on 
the torsion-energy parameters and Method 1 is very sensitive for the energy of dihedral angles.
For example, one of the results of the simulations of Method 1 above is shown in Fig.~\ref{fig_sim_method1}.

In Method 2, the lowest $R$ value was selected from about 10--30 optimization runs with different initial conditions.
In order to calculate $R$, the minimizations of 100 proteins were performed using these new parameter sets.
In Table~\ref{opt_para}, all the optimized torsion-energy parameters are listed.
As one can see in Table~\ref{opt_para}, the original parameters of OPLS-UA force field 
for the optimization are almost zero. 

\begin{table}
\caption{Original and optimized torsion-energy parameters of OPLS-UA.}
\label{opt_para}
\vspace{0.3cm}
\begin{center}
\begin{tabular}{lcccccccccccc} \hline
                                       & \multicolumn{2}{c}{$V_1/2$}   &  $\gamma_1$ &~ & \multicolumn{2}{c}{$V_2/2$}   & $\gamma_2$ &~  &  \multicolumn{2}{c}{$V_3/2$}   & $\gamma_3$ &~ \\ \hline
                                       & org   & opt    &          &~ & org   & opt   &         &~  & org  & opt  &     &~ \\ \hline
N--C$_\alpha$--C$_\beta$--C$_\gamma$ ($\chi_1$) &       &          &             &&       &         &             && 0.5 or 1.0     &   1.950     &   0.0    &  \\
C--C$_\alpha$--C$_\beta$--C$_\gamma$ ($\chi_1$) &       &          &             &&       &         &             && 0.5 or 1.0     &   1.950     &   0.0    &  \\
C--N--C$_\alpha$--C ($\phi$)                    & 0.0 & -0.662   &   0.0       && 0.0 & 0.277   & $\pi$       &&  0.0    &   -0.050    &   0.0    &  \\
N--C$_\alpha$--C--N ($\psi$)                    & 0.0 & 0.974    &   0.0       && 0.0 & 0.576   & $\pi$       &&  0.0    &   -0.083    &   0.0    &  \\
C--N--C$_\alpha$--C$_\beta$                     & 0.0 & 0.811    &   0.0       && 0.0 & 0.328   & $\pi$       &&  0.0    &    0.155    &   0.0    &  \\
N--C--C$_\alpha$--C$_\beta$                     & 0.0 & 0.215    &   0.0       && 0.0 & 0.036   & $\pi$       &&  0.0    &    0.015    &   0.0    &  \\
C$_\alpha$--C$_\beta$--C$_\gamma$--C$_\delta$ ($\chi_2$ of Glu)  & 0.0  &  0.565    &   0.0       && 0.0 & 0.177   & $\pi$       &&  2.0 & -0.025    &   0.0    &  \\ \hline
\end{tabular}
\end{center}
\end{table}

In comparison with Method 1, Method 2 can optimize force-field 
parameters appropriately even if there are some errors in PDB structures.
However, the computational cost of Method 2 is much larger than that of 
Method 1.
Therefore, we could not apply Method 2 to the global optimization in  
the force-field-parameter space.
The force-field parameters of the backbone-torsion 
angles need the global optimization, 
because we consider that these parameters are the most problematic. 
Thus, at first, we performed the global optimization of the 
backbone-torsion parameters by using  Method 1.
After that, Method 2 was applied only on the local region of the 
parameter space, 
which was identified as relevant by Method 1.


In order to test the validity of the force-field parameters obtained by 
our optimization methods, 
we performed folding simulations using two peptides, namely, 
C-peptide and G-peptide.

Only Glu amino acid appears twice in each of the two peptides.
Therefore, we consider that Glu amino acid is the most important, 
and the $\chi_2$ parameters were optimized for this amino acid.
(Of cource, we expect that it becomes a better force field if the remaining  
force-field parameters of other amino acids are also optimized.)

For the folding simulations, we used the replica-exchange molecular dynamics (REMD) 
method \cite{REMD}.
REMD is one of the generalized-ensemble simulation algorithms and has high 
conformational sampling efficiency by 
allowing configurations to heat up and cool down while maintaining proper Boltzmann distributions.
We used the TINKER program package \cite{tinker_v2} modified by us for the folding simulations.
The unit time step was set to 1.0 fs.
Each simulation was carried out for 10 ns (hence, it consisted of
10,000,000 MD steps) with 16 replicas.
The temperature during MD simulations was controlled by 
Nos\'e-Hoover method \cite{hoover}.
For each replica the temperature was distributed exponentially: 700, 662, 625, 591, 558, 528, 499, 
471, 446, 421, 398, 376, 355, 336, 317, and 300 K.
As for solvent effects, we used the GB/SA model \cite{gb1,gb2} included in the TINKER 
program package \cite{tinker_v2}.
These folding simulations were repeated 10 times 
with different sets of randomly generated initial velocities.

\begin{figure}
\begin{center}
\resizebox*{12cm}{!}{\includegraphics{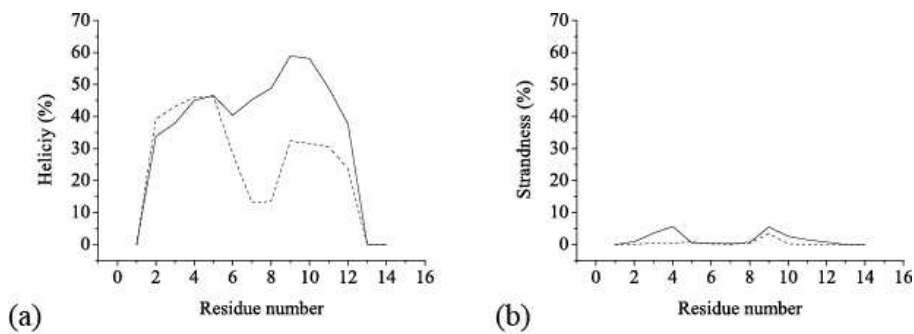}}%
\caption{Helicity (a) and strandness (b) of C-peptide as functions of the residue number. These values are the average of the 
10 independent REMD \cite{REMD} simulations at 300 K. Normal and dotted lines stand for the optimized and original OPLS-UA force fields, respectively.}
\label{fig_cp_remd}
\end{center}
\end{figure}

In Fig.~\ref{fig_cp_remd}, the helicity and strandness of C-peptide 
which were obtained with
the original OPLS-UA and its optimized force field are shown.
These values are the averages of the 10 REMD simulations at 300 K. 
In comparison with the helicity of the original OPLS-UA, 
the helicity of the optimized force field increases at the amino-acid sequence 
between 6 and 12.
For the strandness, both the original and optimized OPLS-UA force fields are almost zero.

\begin{figure}
\begin{center}
\resizebox*{12cm}{!}{\includegraphics{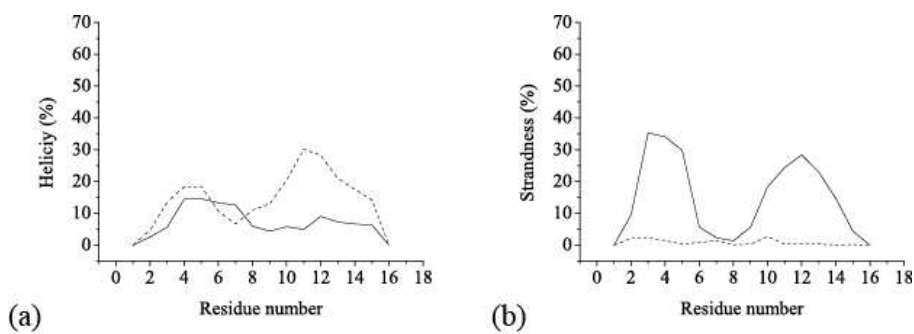}}%
\caption{Helicity (a) and strandness (b) of G-peptide as functions of the residue number. These values are the average of the 
10 independent REMD \cite{REMD} simulations at 300 K. Normal and dotted lines stand for the optimized and original OPLS-UA force fields, respectively.}
\label{fig_gp_remd}
\end{center}
\end{figure}

In Fig.~\ref{fig_gp_remd}, the helicity and strandness of G-peptide at
the original OPLS-UA and its optimized force field are shown.
In comparison with the helicity of the original OPLS-UA, 
the helicity of the optimized force field decreases at the area of amino-acid 
sequence between 8 and 15, and 
in comparison with the strandness of 
the original OPLS-UA, the strandness of the optimized force field clearly increases at the two areas of amino-acid sequences 2--6 and 9--15.
We checked the secondary-structure formations by using the DSSP program \cite{DSSP},
which is based on the formations of the intra-backbone hydrogen bonds.
Strandness means that there are $\beta$-bridge or extended strand in the
corresponding amino acid.
In the experimental results, there is a turn region around residues 7--10  
and there are five intra-backbone hydrogen bond pairs, namely, between
residue pairs 2--15, 3--14, 4--13, 5--12, and 6--11 in G-peptide.
In Fig.~\ref{fig_gp_remd}(b), the strandness decreases in the region around 7--8 residues
in agreement with the experiments.

These results show that the optimized force field favors helix structures more than 
the original OPLS-UA in the case of C-peptide 
and favors $\beta$ structures more than the original OPLS-UA in the case of G-peptide.
We see that these secondary-structure-forming-tendencies of the optimized force field are 
better than those of the original OPLS-UA, because these results are consistent with the native 
structures of the two peptides.

\begin{figure}
\begin{center}
\resizebox*{12cm}{!}{\includegraphics{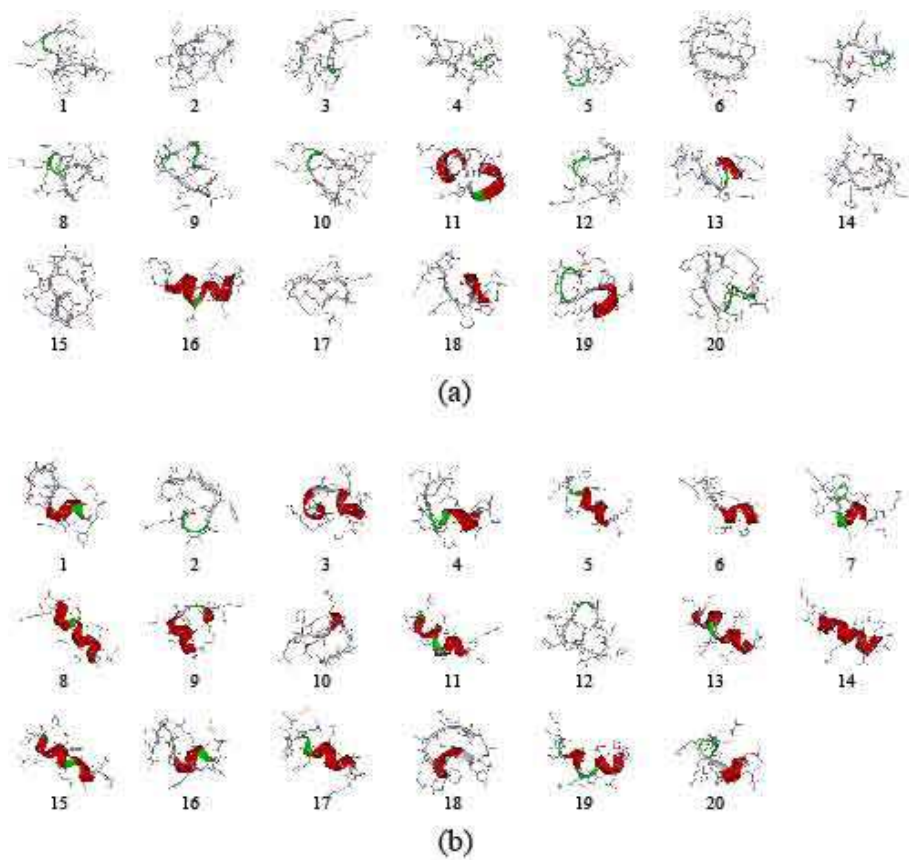}}%
\caption{Twenty lowest-energy conformations of C-peptide obtained from 10 sets of REMD \cite{REMD} simulation runs. (a) and (b) are   
the results of the original and optimized OPLS-UA force field, respectively. 
The conformations are ordered in the increasing order of energy for each case. 
The figures were created with DS Visualizer v1.5\cite{ADS_v2}.}
\label{fig_cp_str}
\end{center}
\end{figure}

\begin{figure}
\begin{center}
\resizebox*{12cm}{!}{\includegraphics{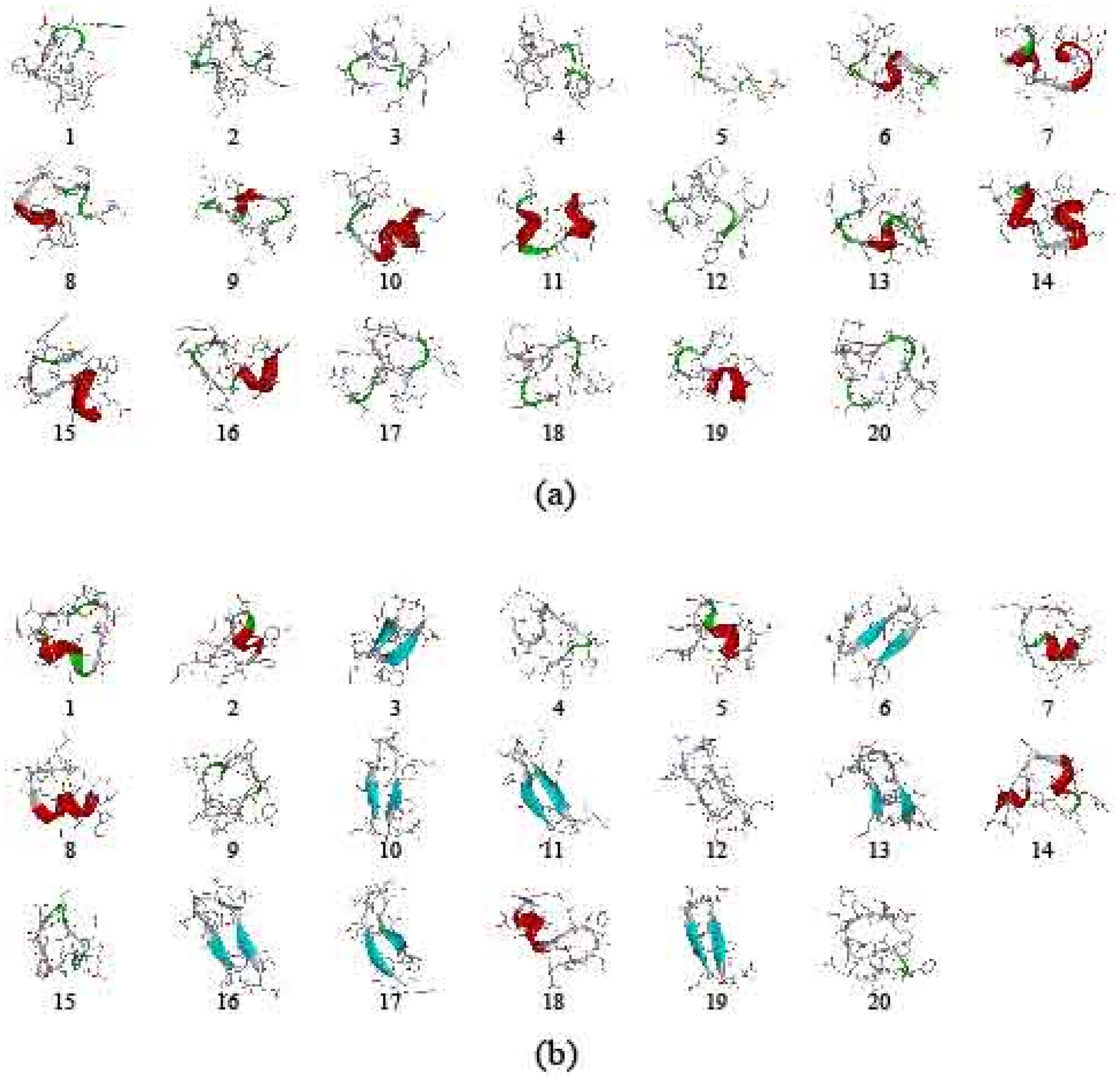}}%
\caption{Twenty lowest-energy conformations of G-peptide obtained from 10 sets of REMD \cite{REMD} simulation runs. (a) and (b) are   
the results of the original and optimized OPLS-UA force field, respectively. 
The conformations are ordered in the increasing order of energy for each case. 
The figures were created with DS Visualizer v1.5\cite{ADS_v2}.}
\label{fig_gp_str}
\end{center}
\end{figure}

In Figs.~\ref{fig_cp_str} and \ref{fig_gp_str}, we show the 20 lowest-energy conformations of 
C-peptide and G-peptide obtained by 
the REMD simulations in the case of the original and optimized OPLS-UA force fields, respectively.
In Fig.~\ref{fig_cp_str}(a), five conformations (Nos. 11, 13, 16, 18, and 19) have $\alpha$-helix structures for the original 
OPLS-UA in the case of C-peptide.
In Fig.~\ref{fig_cp_str}(b), 18 conformations (all conformations except for Nos. 2 and 12) have $\alpha$-helix structures for the 
optimized OPLS-UA in the case of C-peptide.
>From these results, we can see that the optimized OPLS-UA force field favor $\alpha$-helix structure 
more than the original OPLS-UA force field in the case of C-peptide.
In Fig.~\ref{fig_gp_str}(a), 11 conformations have $\alpha$-helix structures for the original OPLS-UA in the case of G-peptide.
In Fig.~\ref{fig_gp_str}(b), seven conformations have $\alpha$-helix structures, and eight conformations have $\beta$-hairpin structures
 for the optimized OPLS-UA in the case of G-peptide.
In Fig.~\ref{fig_gp_str}(b), two conformations (Nos. 3 and 16) out of the eight $\beta$-hairpin conformations
have the right hydrogen bond formations that are inferred by the experiments.
Namely, conformation No.3 has three native-like hydrogen bonds between residue pairs 
3--14, 4--13, and 5--12, and conformation No.16 has two native-like hydrogen bonds between 
residue pairs  3--14 and 4--13. 
These results for G-peptide show that the optimized OPLS-UA force field does not favor $\alpha$-helix structure  
and clearly favors $\beta$-hairpin structure more than the original OPLS-UA force field.

These secondary-structure-forming tendencies of the optimized OPLS-UA force field for two peptides 
agree with experimental implications in comparison with those of the original OPLS-UA force field.
Therefore, our optimization methods succeeded in enhancing the accuracy of the OPLS-UA force field.

\subsubsection{Use of RMSD II \cite{SO6}}


We now present the results of the applications of our new optimization method of force-field parameters.

At first, we chose 100 PDB files with resolution 2.0 \AA~or better, 
with sequence similarity of amino acid 30.0 \% or lower, 
and with less than 200 residues (the average number of residues is 122.2) from PDB-REPRDB \cite{REPRDB}.
We selected the number of each fold (all $\alpha$, all $\beta$, $\alpha / \beta$, and $\alpha+\beta$)
in 100 proteins based on the number of folds given by SCOP (version 1.73 at November 2007) \cite{SCOP}.
Namely, we used 29 all $\alpha$, 18 all $\beta$, 16 $\alpha / \beta$, and 37 ($\alpha+\beta$) proteins 
(the list is slightly different from that in Table~\ref{PDB_list}).

The force field that we optimized is the AMBER parm96 version \cite{parm96_3}.
The backbone-torsion-energy term $E_{\rm torsion} (\Phi, \Psi)$ for this force field is given by
\begin{equation}
E_{\rm torsion} (\Phi, \Psi) = \frac{V^{\phi}_1}{2}[1+\cos\phi] + 
\frac{V^{\phi}_2}{2}[1-\cos2\phi] + \frac{V^{\psi}_1}{2}[1+\cos\psi] + 
\frac{V^{\psi}_2}{2}[1-\cos2\psi],
\end{equation}
where we have 
$V^{\phi}_1=1.7$, $V^{\phi}_2=0.6$, $V^{\psi}_1=1.7$, and $V^{\psi}_2=0.6$.
Here, we have optimized only two parameters in the backbone-torsion-energy 
term, namely, $V^{\psi}_1$ and $V^{\psi}_2$ for $\psi$ angle. 
As described above, AMBER parm94 and AMBER parm96 have quite different secondary-structure-forming-tendencies, 
although these force fields differ only 
in the backbone torsion-energy terms for rotations of the $\phi$ and $\psi$ angles.
Moreover, we can easily imagine that force-field parameters $V^{\psi}_1$ 
and $V^{\psi}_2$ for $\psi$ angle are 
important for the secondary-structure-forming-tendencies,
because the energy surface in the Ramachandran space is  quite sensitive 
to this energy term in the helix and $\beta$-sheet regions.
Namely, if the torsion-energy term for the $\psi$ angle changes, 
the stabilities of helix structure region and $\beta$-sheet region on 
the Ramachandran space change.
Therefore, we considered some trial force-field parameters for
$V^{\psi}_1$ and $V^{\psi}_2$, which are given by the following equations: 
\begin{equation}
V_1^{\rm trial} = 1.7 \cdot 0.2 i = 0.34 i,
\label{v_1_rmsd2}
\end{equation}
\begin{equation}
V_2^{\rm trial} = 0.6 \cdot 0.2 i = 0.12 i.
\label{v_2_rmsd2}
\end{equation}
Here, $i$ is any real number.
When $i$ is 5, the force-field parameters $V_1^{\rm trial}$ and $V_2^{\rm trial}$ of $\psi$ angle are equal to those of the original AMBER parm96.
>From our experience, if $i$ has a small number ($i < 5$), the force field favors helix structure, and if $i$ has a large number ($i > 5$), 
the force field favors $\beta$-sheet structure (see also Figs.~\ref{fig_cp_rep} and \ref{fig_gp_rep} below).
We calculated $\Phi {\rm RMSD}_{\rm 2ndly}$ values in Eq.~(\ref{helix_plus_beta_rmsd2}) about some trial force-field parameters obtained by changing $i$ 
in Eqs.~(\ref{v_1_rmsd2}) and (\ref{v_2_rmsd2}).

\begin{figure}
\begin{center}
\resizebox*{10cm}{!}{\includegraphics{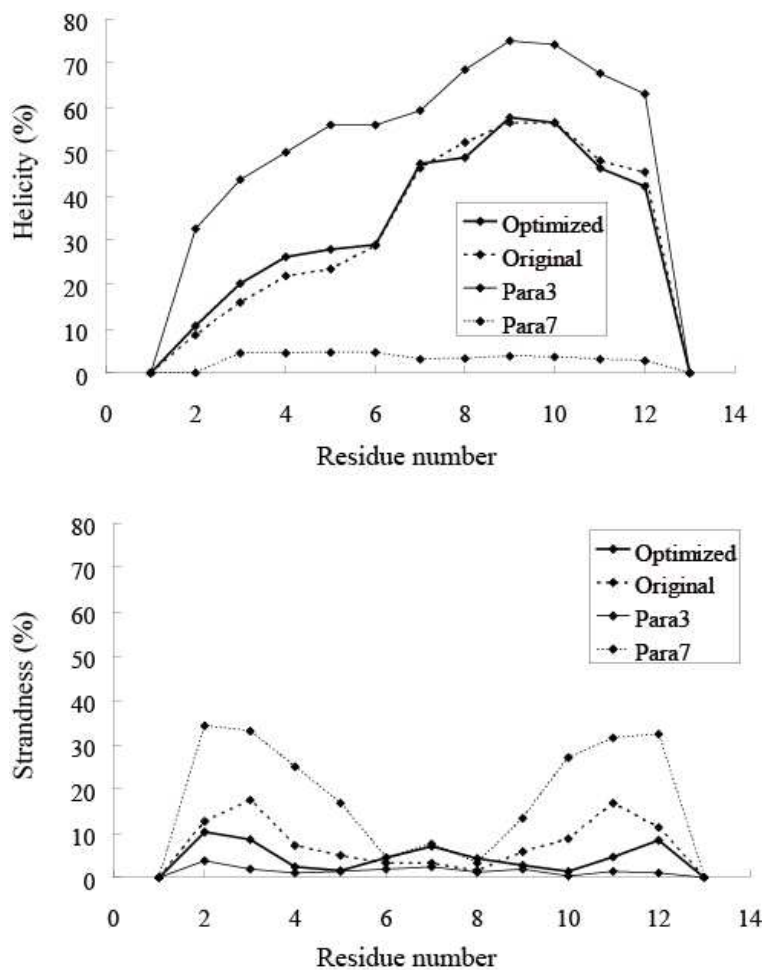}}%
\caption{Helicity (a) and strandness (b) of C-peptide as functions of the residue number.
These values are the averages of the 10 independent REMD \cite{REMD} simulations at 300 K. 
Optimized, original, para3, and para7 stand for the optimized AMBER parm96 ($i = 4.7$), original AMBER parm96 ($i = 5.0$), 
trial force field para3 ($i = 3.0$), and trial force field para7 ($i = 7.0$), respectively.}
\label{fig_cp_rep}
\end{center}
\end{figure}

\begin{figure}
\begin{center}
\resizebox*{10cm}{!}{\includegraphics{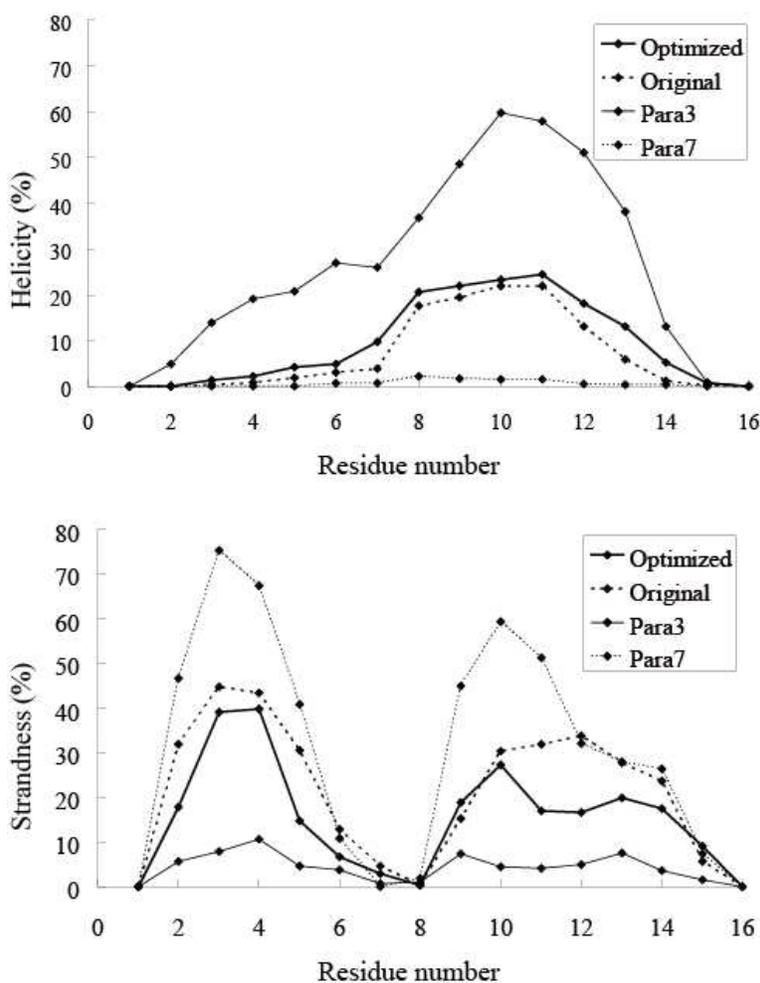}}%
\caption{Helicity (a) and strandness (b) of G-peptide as functions of the residue number.
These values are the averages of the 10 REMD \cite{REMD} simulations at 300 K. 
Optimized, original, para3, and para7 stand for the optimized AMBER parm96 ($i = 4.7$), original AMBER parm96 ($i = 5.0$), 
trial force field para3 ($i = 3.0$), and trial force field para7 ($i = 7.0$), respectively.}
\label{fig_gp_rep}
\end{center}
\end{figure}

We performed the minimization, which was terminated when the root-mean-square (RMS) potential energy 
gradients were less than 0.1 (kcal/mol/\AA) by using TINKER program package \cite{tinker_v2}.
For solvent effects, we used GB/SA solvent model in TINKER.

\begin{figure}
\begin{center}
\resizebox*{12cm}{!}{\includegraphics{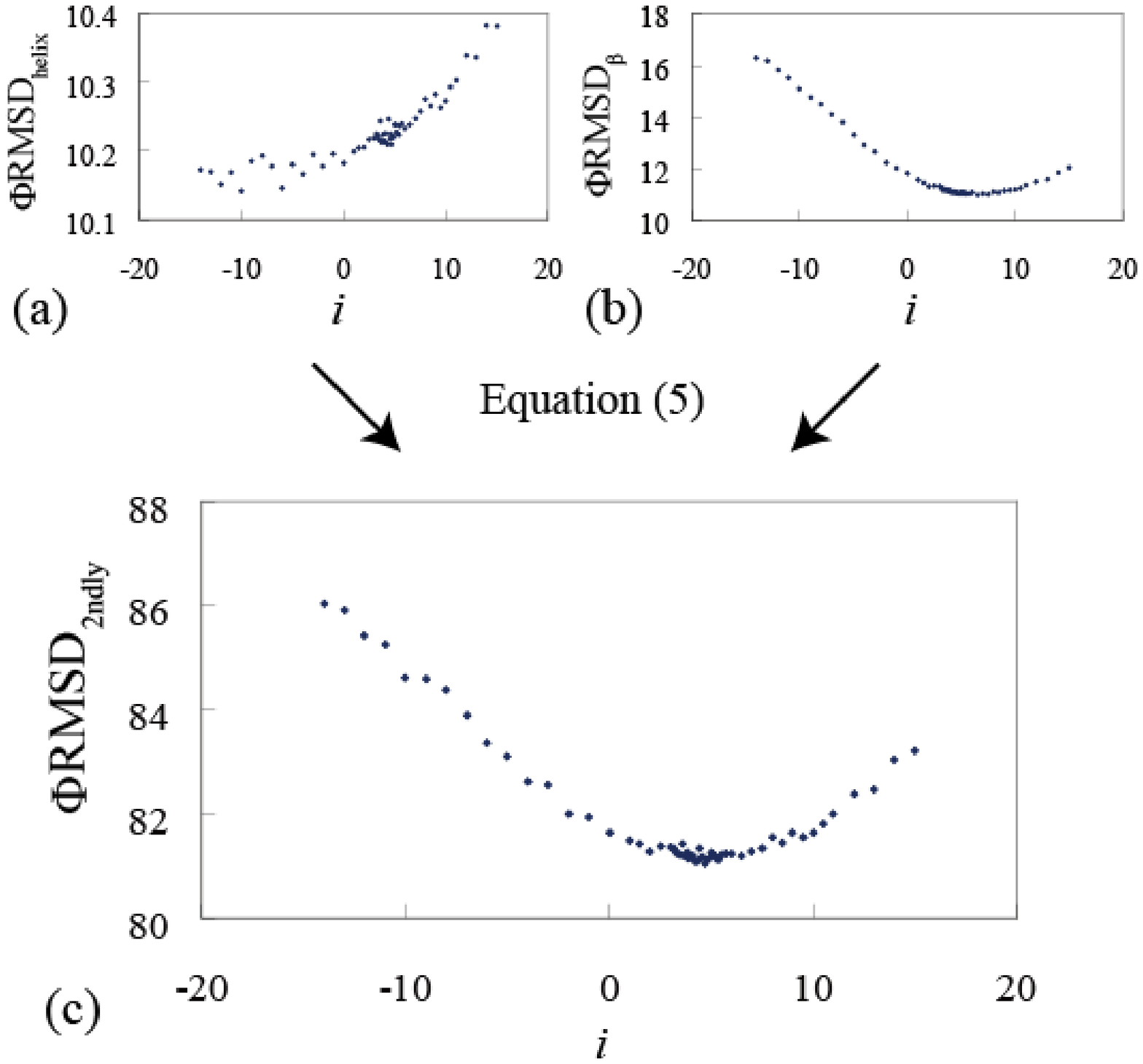}}%
\caption{Distributions of $\Phi {\rm RMSD}_{\rm helix}$ (a), $\Phi {\rm RMSD}_{\rm \beta}$ (b), 
and $\Phi {\rm RMSD}_{\rm 2ndly}$ (c)} obtained from 
the minimization of 100 proteins using the trial force-field parameters $V_1^{\rm trial}$ and $V_2^{\rm trial}$ 
depending on the number $i$.
\label{fig_phi_rmsd_bunpu}
\end{center}
\end{figure}

The results of $\Phi {\rm RMSD}_{\rm helix}$ and $\Phi {\rm RMSD}_{\rm \beta}$ 
are shown in Fig.~\ref{fig_phi_rmsd_bunpu}(a) and Fig.~\ref{fig_phi_rmsd_bunpu}(b), recpectively.
In these calculations, if the differences of the backbone-dihedral angles between $\Phi_i^{\rm native}$ and $\Phi_i^{\rm min}$  in Eq.~(\ref{eq_e_dif_rmsd2}) are more than 30 degrees, they were ignored, 
assuming that the uncertaintties in those angles are too large.
We see that $\Phi {\rm RMSD}_{\rm helix}$ decreases gradually with a decrease in $i$.
If $i$ decreases, the torsion energy of the helix structure region in the Ramachandran space also decreases.
On the other hand, $\Phi {\rm RMSD}_{\rm \beta}$ decreases gradually with an increase in $i$.
If $i$ increases, the torsion energy of the $\beta$ structure region in the Ramachandran space decreases.
Hence, this result is reasonable.
However, $\Phi {\rm RMSD}_{\rm \beta}$ reaches the global minimium, when $i$ is 6.5.
If $i$ is larger than 6.5, $\Phi {\rm RMSD}_{\rm \beta}$ increases gradually.
This result implies that the $\Phi {\rm RMSD}_{\rm \beta}$ does not correspond to 
the parameters $V_1^{\rm trial}$ and $V_2^{\rm trial}$ completely.

For $\Phi {\rm RMSD}_{\rm helix}$ and $\Phi {\rm RMSD}_{\rm \beta}$ in Fig.~\ref{fig_phi_rmsd_bunpu} (a) and (b), 
we can see the difference clearly.
The noteworthy point obtaind from these results is that $\Phi {\rm RMSD}$ can distinguish 
between helix structure and $\beta$ structure.

We combined $\Phi {\rm RMSD}_{\rm helix}$ and $\Phi {\rm RMSD}_{\rm \beta}$ 
by Eq.~(\ref{helix_plus_beta_rmsd2}).
Here, in order to have roughly equal contributions from both terms, we can 
set the value of the scaling factor $\lambda$ to be, for example, the 
coefficients of variations:
\begin{equation}
\lambda = \frac{\displaystyle{\frac{\sigma_{\rm \beta}}{\mu_{\rm \beta}}}}
{\displaystyle{\frac{\sigma_{\rm helix}}{\mu_{\rm helix}}}}. 
\label{lambda_rmsd2}
\end{equation}
Here, $\mu_{\rm helix}$ and $\mu_{\rm \beta}$ are the averages and $\sigma_{\rm helix}$ and $\sigma_{\rm \beta}$ are 
the corresponding standard deviations for $\Phi {\rm RMSD}_{\rm helix}$ and $\Phi {\rm RMSD}_{\rm \beta}$.
For the calculations, we have chosen a small number of $i$ values in a range $i_{\rm min} \leq i \leq i_{\rm max}$.
For $i_{\rm min} = 0$ and $i_{\rm max} = 10$, we obtained $\lambda = 6.857$, and this fixied value was used for all 
the calculations in the present work.


In Fig.~\ref{fig_phi_rmsd_bunpu}(c), the combined result is shown.
The smallest $\Phi {\rm RMSD}_{\rm 2ndly}$ is obtained value $i = 4.7$, namely, the obtained force-field parameters are $V_1^{\rm trial} = 1.598$ 
and $V_2^{\rm trial} = 0.564$.
These values are slightly smaller than those of the original AMBER parm96, which corresponds to $i = 5$.
We can easily expect the new obtained force-field parameters slightly favor helix structure more and  
$\beta$-sheet structure less than the original AMBER parm96.

In order to check the force-field parameters obtained by our optimization method, 
we performed the folding simulations using two peptides, namely, C-peptide and G-peptide.

For the folding simulations, we used replica-exchange molecular dynamics (REMD) \cite{REMD}.
We used the TINKER program package \cite{tinker_v2} modified by us for the folding simulations.
The unit time step was set to 1.0 fs.
Each simulation was carried out for 2 ns (hence, it consisted of
2,000,000 MD steps) with 16 replicas and repeated 10 times.
The temperature during MD simulations was controlled by 
Berendsen's method \cite{berendsen}.
For each replica the temperature was distributed exponentially: 700, 662, 625, 591, 558, 528, 499, 
471, 446, 421, 398, 376, 355, 336, 317, and 300 K.
As for solvent effects, we used the GB/SA model \cite{gb1,gb2} included in the TINKER program package \cite{tinker_v2}.
These folding simulations were performed 
with different sets of randomly generated initial velocities.

In Fig.~\ref{fig_cp_rep}, the helicity and strandness of C-peptide 
which were obtained with
the original AMBER parm96 and its optimized force field are shown.
These values are the averages of the 10 REMD simulations at 300 K. 
In comparison with the helicity of the original AMBER parm96, 
the helicity of the optimized force field is similler.
However, the helicity of Thr3, Ala4, and Ala5 of the optimized force field slightly increases.
In comparison with the strandness of 
the original AMBER parm96, the strandness of the optimized force field decreases except for
those at Ala6, Lys7, and Phe8.

In Fig.~\ref{fig_gp_rep}, the helicity and strandness of G-peptide at
the original AMBER parm96 and its optimized force field are shown.
In comparison with the helicity of the original AMBER parm96, 
the helicity of the optimized force field slightly increases, and 
in comparison with the strandness of 
the original AMBER parm96, the strandness of the optimized force field slightly decreases.
For trial force fields of para3 and para7, the scondary-structure-forming-tendencies are simillar to the case of 
C-peptide.

These results clearly show that the optimized force field favors helix structures 
and does not favor $\beta$ structures than the original AMBER parm96.
We can see that these secondary-structure-forming-tendencies of the optimized force field are 
better than those of the original AMBER parm96, becasue it is known that the AMBER parm96 slightly favors 
the $\beta$ structure too much \cite{YSO1,YSO2,SO1,SO2,SO3}.

We also performed the folding simulations with two extreme cases of the trial force fields, namely, 
para3 ($i = 3.0$) and para7 ($i = 7.0$) (see Figs.~\ref{fig_cp_rep} and \ref{fig_gp_rep})
for comparisons.
The trial force field para3 favors helix structure strongly and does not 
favors $\beta$ structure clearly.
On the other hand, the trial force field para7 has the tendency that is quite reverse to para3.
According to the results of $\Phi {\rm RMSD}_{\rm helix}$ and $\Phi {\rm RMSD}_{\rm \beta}$ in 
Fig.~\ref{fig_phi_rmsd_bunpu}(a)(b), 
$\Phi {\rm RMSD}_{\rm helix}$ decreases gradually with a decrease in $i$, and 
$\Phi {\rm RMSD}_{\rm \beta}$ reaches the global minimum, when $i$ is 6.5.
Namely, we can see that the values of $\Phi {\rm RMSD}_{\rm helix}$ and $\Phi {\rm RMSD}_{\rm \beta}$ are 
related to the stabilities of helix structure and $\beta$ structure well.


\subsubsection{Use of short MD simulations \cite{SO10}}

\begin{figure}
\begin{center}
\resizebox*{12cm}{!}{\includegraphics{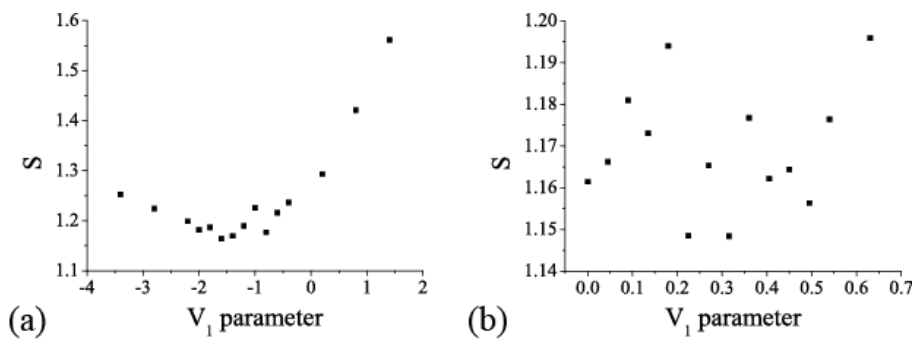}}%
\caption{$S$ values (defined in Eq.~(\ref{su_num_MD})) obtained from 
MD simulations of 31 proteins with the force fields which  have 
different $V_1$ parameter values for $\psi'$ (C$_{\beta}$-C$_{\alpha}$-C-N) 
(a) and $\psi$ (N-C$_{\alpha}$-C-N) (b) angles.}
\label{fig_S_values}
\end{center}
\end{figure}

We present the results of the applications of our optimization method 
in Subsection 2.3.4 to the AMBER ff99SB force field.
At first, we chose 31 PDB files ($M=31$) with resolution 2.0 \AA~or better, 
with sequence similarity of amino acid 30.0 \% or lower 
and with from 40 to 111 residues (the average number of residues is 86.7) from PDB-REPRDB \cite{REPRDB}.
Namely, the PDB IDs of these 31 proteins are 1LDD, 1HBK, 1Y02, 1I2T, 1U84, 2ERL, 1TQG, 1O82, 1V54, 1XAK, 
1GMU, 1O5U, 1NLQ, 1WHO, 1CQY, 1H75, 1GMX, 1IIB, 1VC1, 1AY7, 1KAF, 1KPF, 1BM8, 1MK0, 1EW4, 1OSD, 1VCC, 
1OPD, 1CYO, 1CTF, and 1N9L.
Generally, data from X-ray experiments do not have hydrogen atoms.
Therefore, we have to add hydrogen coordinates.
Many protein simulation software packages provide with routines that add hydrogen atoms to the PDB 
coordinates.
After adding the hydrogen atoms, we performed the short potential energy 
minimizations while restraining the heavy atoms.
We use the obtained conformations as the initial structures (experimental 
structures).
We performed MD simulations for these proteins.
Each simulation was carried out for 40.0 ps (hence, it consisted of
20,000 MD steps, and the unit time step was set to 2.0 fs and the bonds involving hydrogen were constrained 
by SHAKE algorithm \cite{SHAKE}) by using Langevin dynamics at 300 K.
The nonbonded cutoff of 20 \AA~ were used.
As for solvent effects, we used the GB/SA model \cite{gbsa_igb5} included 
in the AMBER program package ($igb = 5$).
These simulations were performed 
with different sets of the same generated initial velocities of atoms 
in 31 proteins.
For all the process, we used the AMBER11 program package \cite{amber_prog11}.
As trial force-field parameters, we used the parameters $V_1$ of 
$\psi$ (N-C$_{\alpha}$-C-N) and $\psi'$ (C$_{\beta}$-C$_{\alpha}$-C-N) 
angles for torsion-energy term in Eq.~(\ref{ene_torsion_optF1}).
We performed the simulations by using 14 and 15 values of the 
$V_1$ parameters 
of $\psi$ and $\psi'$, respectively, and these simulations 
with each set of parameter values were 
performed five times by changing the initial velocities of atoms in 
the 31 proteins. 
Namely, we calculated $n_i^{\rm S \to U}$ and $n_i^{\rm U \to S}$ 
in Eq.~(\ref{su_num_MD}) as 
the average numbers of $n_i^{\rm S \to U}$ and $n_i^{\rm U \to S}$ 
of 10 trajectories from 20.0 ps to 40.0 ps of the five simulations.
These results are shown in Fig.~\ref{fig_S_values}.
We determined the optimized force-field parameters in order 
of $\psi'$ and $\psi$, by searching the minimum value of $S$ 
in Fig.~\ref{fig_S_values}.
$V_1$ parameter for $\psi$ changed from 0.45 to 0.31, and $V_1$ parameter 
for $\psi'$ changed from 0.20 to $-1.60$.

In order to test the validity of the force-field parameters obtained 
by our optimization method, 
we performed the folding simulations using two peptides, namely, 
C-peptide and G-peptide.

For test simulations, we used replica-exchange molecular dynamics (REMD) \cite{REMD}.
We used the AMBER11 program package \cite{amber_prog11}.
The unit time step was set to 2.0 fs, and the bonds involving hydrogen were constrained by SHAKE algorithm \cite{SHAKE}.
Each simulation was carried out for 30.0 ns (hence, it consisted of
15,000,000 MD steps) with 32 replicas by using Langevin dynamics.
The replica exchange was tried every 3,000 steps.
The temperature was distributed exponentially: 
600, 585, 571, 557, 544, 530, 517, 505, 492, 480, 469, 457, 446, 435, 425, 414, 404, 394, 385, 375, 366, 
357, 348, 340, 332, 324, 316, 308, 300, 293, 286, and 279 K.
As for solvent effects, we used the GB/SA model \cite{gbsa_igb5} included in the AMBER program package ($igb = 5$).
These simulations were performed 
with different sets of randomly generated initial velocities.

\begin{figure}
\begin{center}
\resizebox*{12cm}{!}{\includegraphics{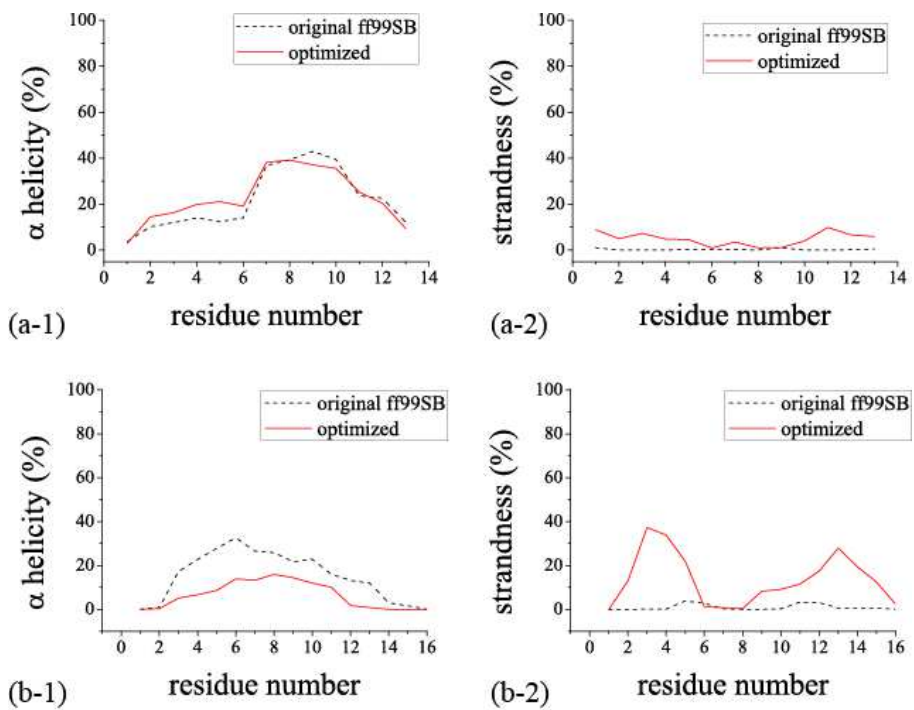}}%
\caption{$\alpha$ helicity (a-1) and strandness (a-2) of C-peptide and $\alpha$ helicity (b-1) and strandness (b-2) 
of G-peptide as functions of the residue number. These values are obtained from REMD \cite{REMD} simulations at 300 K.
Normal and dotted lines stand for the optimized and original AMBER ff99SB force field, respectively.}
\label{fig_secondary_alpha_beta_mdopt}
\end{center}
\end{figure}

In Fig.~\ref{fig_secondary_alpha_beta_mdopt}, $\alpha$ helicity and strandness of two peptides obtained 
from the test simulations are shown.
We checked the secondary-structure formations by using the DSSP program \cite{DSSP},
which is based on the formations of the intra-backbone hydrogen bonds.
For the original AMBER ff99SB force field, the $\alpha$ helicity is clearly larger than 
the strandness in not only C-peptide but also G-peptide.
Namely, the original AMBER ff99SB force field clearly favors $\alpha$-helix structure, and does not favor 
$\beta$ structure.
On the other hand, for the optimized force field, in the case of C-peptide, the $\alpha$ helicity is 
larger than the strandness, and in the case of G-peptide, the strandness is larger than the $\alpha$ helicity.
We can see that these results obtained from the optimized force field are in better agreement with 
the experimental results in comparison with the original force field.

\section{Conclusions}
In this Chapter we reviewed our works on force fields for molecular
simulations of protein systems.  
We first discussed the functional forms of
the force fields and present some extensions of the conventional ones.
Because the main-chain torsion-energy
terms are the most problematic among the force-field terms in the
existing force fields, we mainly considered the main-chain torsion-energy
terms.  We have generalized them into the double Fourier series in
$\phi$ and $\psi$.  We have also introduced the amino-acid dependence
on these terms.

Given the functional forms, we then presented various methods for 
force-field parameter optimizations.
Some of our methods use the coordinates from PDB,
which were determined by experiments.  We tried to minimize the
effects of systematic experimental errors by considering many
protein structures.
Other methods rely on short molecular dynamics simulations
with the native conformations from PDB as initial ones for the simulations.
   
Some examples of our applications of these parameter optimization
methods were given and they were compared with the results from
the existing force-fields.  It turned out that all the examples
resulted in improvement of the existing force fields.  We thus believe
that we are at least on the right track.

Our optimization methods for the force-field parameters are
quite general and they can be readily applied to any new energy
terms whenever they are introduced in the future.

\begin{acknowledgement}
The computations were performed on the computers at the Research Center for Computational Science, 
Institute for Molecular Science, Information Technology Center, Nagoya University, and Center for 
Computational Sciences, University of Tsukuba.
This work was supported, in part, by 
the Grants-in-Aid for the Academic Frontier Project, ``Intelligent Information Science'', 
for Scientific Research on Innovative Areas (``Fluctuations and Biological Functions'' ),
and for the Next Generation Super Computing Project, Nanoscience Program
and Computational Materials Science Initiative 
from the Ministry of Education, Culture, Sports, Science 
and Technology (MEXT), Japan.
\end{acknowledgement}
%
%
%


\bibliographystyle{spphys}

\end{document}